\documentclass[preprint]{aastex62}
\usepackage{amsmath}
\usepackage{cases}
\usepackage{natbib}
\usepackage{subdepth}
\usepackage{tipa}
\usepackage{units}
\usepackage{hyperref}
\usepackage{epstopdf}

\newcommand{\D}{\mathrm{d}}
\newcommand{\delr}[1]{\frac{\partial #1}{\partial r}}
\newcommand{\dr}[1]{\frac{\D #1}{\D r}}
\newcommand{\delth}[1]{\frac{\partial #1}{\partial \theta}}
\newcommand{\delphi}[1]{\frac{\partial #1}{\partial \phi}}
\newcommand{\delthsq}[1]{\frac{\partial^2 #1}{\partial \theta^2}}
\newcommand{\xr}{\xi^r}
\newcommand{\xrh}{\hat{\xi}^r}
\newcommand{\xh}{\xi^h}
\newcommand{\xhh}{\hat{\xi}^h}
\newcommand{\sinth}{\sin\theta}
\newcommand{\cotth}{\cot\theta}
\newcommand{\gsharm}[3]{Y_{#1}^{#2,#3}}
\newcommand{\sharm}[2]{Y_{#1}^{#2}}
\newcommand{\om}[2]{\Omega_{#1}^{#2}}
\newcommand{\gl}{\gamma_l}
\newcommand{\glp}{\gamma_{l\p}}
\newcommand{\Dr}[1]{\frac{\D #1}{\D r}}
\newcommand{\wtj}[6]{\begin{pmatrix} #1 & #2 & #3 \\ #4 & #5 & #6 \end{pmatrix}}
\newcommand{\p}{'}

\defcitealias{Broomhall2017}{B17}
\shorttitle{Toroidal Magnetic Fields and Solar Oscillation Frequencies}
\shortauthors{Kiefer \& Roth}

\begin{document}

\title{The Effect of Toroidal Magnetic Fields on Solar Oscillation Frequencies}
\submitjournal{The Astrophysical Journal}
\author{Ren\'e Kiefer}
\affil{Kiepenheuer-Institut f\"ur Sonnenphysik, Sch\"oneckstra\ss e 6, 79104, Freiburg, Germany}

\author{Markus Roth}
\affiliation{Kiepenheuer-Institut f\"ur Sonnenphysik, Sch\"oneckstra\ss e 6, 79104, Freiburg, Germany}

\begin{abstract}
Solar oscillation frequencies change with the level of magnetic activity. Localizing subsurface magnetic field concentrations in the Sun with helioseismology will help us to understand the solar dynamo. Because the magnetic fields are not considered in standard solar models, adding them to the basic equations of stellar structure changes the eigenfunctions and eigenfrequencies. We use quasi-degenerate perturbation theory to calculate the effect of toroidal magnetic fields on solar oscillation mean multiplet frequencies for six field configurations. In our calculations, we consider both the direct effect of the magnetic field, which describes the coupling of modes, and the indirect effect, which accounts for changes in stellar structure due to the magnetic field. We limit our calculations to self-coupling of modes. We find that the magnetic field affects the multiplet frequencies in a way that depends on the location and the geometry of the field inside the Sun. Comparing our theoretical results with observed shifts, we find that strong tachocline fields cannot be responsible for the observed frequency shifts of p modes over the solar cycle. We also find that part of the surface effect in helioseismic oscillation frequencies might be attributed to magnetic fields in the outer layers of the Sun. The theory presented here is also applicable to models of solar-like stars and their oscillation frequencies.
\end{abstract}


%%%%%%%%%%%%%%%%%%%%%%%%%%%%%%%%%%%%%%%%%%%%%%%%%%%%%%%%%%%%%%%%%%%%%%%%%%%%%%%%%%%%%%%%%%%%%%%%%%%%%%%%%%%%%%%%%%%%%%%%%%%%%%%%%%%%%%%%%%%%%%%%%%%%
%%%%%%%%%%%%%%%%%%%%%%%%%%%%%%%%%%%%%%%%%%%%%%%%%%%%%%%%%%%%%%%%%%%%%%%%%%%%%%%%%%%%%%%%%%%%%%%%%%%%%%%%%%%%%%%%%%%%%%%%%%%%%%%%%%%%%%%%%%%%%%%%%%%%

\section{Introduction}\label{sec:1}
The mapping of subsurface magnetic field concentrations and with it the pinpointing of the region in the Sun, where its dynamo operates can surely be regarded as one of the outstanding open issues of solar physics. In this article, we develop the theory for the forward calculation of the effect of a superposition of zonal toroidal magnetic fields on solar oscillation frequencies in the framework of quasi-degenerate perturbation theory. The toroidal component of the global solar magnetic field is assumed to be responsible for the bulk of phenomena associated with magnetic activity \citep[see][and references therein]{ Fan2009, Charbonneau2010, Hathaway2015}. In simulations of flux-transport solar dynamos, the energy that is stored in the toroidal component of the large-scale magnetic field is orders of magnitude larger than the energy in the poloidal field \citep[e.g.,][]{Miesch2016}. Hence, the restriction to purely toroidal magnetic field configurations in the present work is adequate.

From observations, it is well known that solar p-mode frequencies vary in phase with the solar activity cycle. \citet{Woodard1985} measured these changes in frequencies for oscillation of low harmonic degree. Later, the frequency shifts over the solar cycle were confirmed and thoroughly investigated by, e.g., \citet{Libbrecht1990}, \citet{Jimenez-Reyes1998}, and \citet{Broomhall2017}. In the Sun, these shifts are larger for modes of higher frequency \citep{Jimenez-Reyes1998}. As, e.g., \citet{Basu2012} showed, modes of higher frequency have their maximum sensitivity to structural changes in the solar interior in shallower layers than modes of low frequency. This can be used to study the change of subsurface solar activity as a function of time and depth \citep{Basu2012}. 

It has been shown that the frequencies of acoustic oscillations of solar-like stars undergo changes similar to those observed on the Sun. The first time that such changes were reported for a star other than the Sun was done by \citet{Garcia2010}. They also showed that these changes are correlated with stellar magnetic activity. Since then this behavior has been observed for several more stars by \citet{Salabert2016a} and \citet{Kiefer2017}. Gaining insight about the underlying magnetic field causing these changes in stars would supplement the simulation of solar and stellar dynamos greatly. 

Until now, the attempts to determine the magnetic field in the Sun with helioseismology have been rather sparse. \citet{Gough1990} calculated multiplet shifts for low degree modes for an axisymmetric buried magnetic field in a perturbational approach. Building on their framework, \citet{Antia2000} analyzed splitting coefficients to probe the solar acoustic asphericity and magnetic fields in the convection zone. They were able to limit the magnetic field strength at the base of the convection zone to $\unit[300]{kG}$. \citet{Baldner2009} extended this work and matched simulated splitting coefficients to their observed counterparts. They found that a superposition of two very shallow toroidal magnetic fields in the upper 1\% of the convection zone and a poloidal component could explain the observed even splitting coefficients best. \citet{Dziembowski2005} analyzed the behavior of centroid multiplet frequencies between the minimum and maximum of cycle 23. They found that the frequency increase can be explained by a less than 2\% decrease in the radial component of the turbulent velocity in the convection zone. They explain the p-mode frequency increase with rising levels of activity by the inhibiting effect of the magnetic field on convection. Recently, \citet{Hanasoge2017} developed a formalism to calculate sensitivity kernels of mode coupling to Lorentz stresses in the Sun. 

In this article, we describe our forward calculations of the effect of subsurface toroidal magnetic fields on solar oscillation multiplet frequencies. For this, we use an ansatz from quasi-degenerate perturbation theory to calculate the strength of the coupling between solar oscillation modes \citep{Lavely1992}. The coupling of initially independent modes leads to changes in mode frequencies and eigenfunctions. The distortions of the solar mode eigenfunctions by flows have previously been exploited by \citet{Schad2013a} to infer a double cell profile of the meridional circulation. In the ansatz, we use here, the total perturbation by the magnetic field can be separated in a direct and an indirect contribution. The direct contribution couples the modes and thus changes their frequencies and eigenfunctions. An analytical derivation of the general matrix element of this direct contribution caused by a superposition of zonal toroidal magnetic field was recently presented by \citet{Kiefer2017b}. The indirect contribution is due to the perturbation to the equilibrium stellar structural quantities, which is caused by the magnetic field. The effect of a magnetic field on stellar structure was studied by, e.g., \citet{Mestel1977, Mathis2005, Duez2008, Duez2010}.

We start by introducing the necessary theoretical background of quasi-degenerate perturbation theory in Section~\ref{sec:2}. The general matrix element for the indirect effect of toroidal magnetic fields is then derived in Section~\ref{sec:3}. We present the six magnetic field configurations we tested and the resulting multiplet shifts in Section~\ref{sec:4}. These results are discussed in Section ~\ref{sec:5} and we end the paper with our conclusions in Section~\ref{sec:6}. We include mathematical supplements in Appendix~\ref{app:sec:decompose}, a detailed derivation of the projection of the Lorentz force onto spherical harmonics in Appendix~\ref{app:sec:project}, the sensitivity kernels in Appendix~\ref{app:sec:kernels}, derivations of the perturbations in stellar structural quantities in Appendix~\ref{app:sec:perturb}, and additional figures for the modeled magnetic fields and the resulting frequency shifts in Appendix~\ref{app:sec:figures}.

%%%%%%%%%%%%%%%%%%%%%%%%%%%%%%%%%%%%%%%%%%%%%%%%%%%%%%%%%%%%%%%%%%%%%%%%%%%%%%%%%%%%%%%%%%%%%%%%%%%%%%%%%%%%%%%%%%%%%%%%%%%%%%%%%%%%%%%%%%%%%%%%%%%%
%%%%%%%%%%%%%%%%%%%%%%%%%%%%%%%%%%%%%%%%%%%%%%%%%%%%%%%%%%%%%%%%%%%%%%%%%%%%%%%%%%%%%%%%%%%%%%%%%%%%%%%%%%%%%%%%%%%%%%%%%%%%%%%%%%%%%%%%%%%%%%%%%%%%
\section{Perturbation Theory}\label{sec:2}
The equations that describe the equilibrium state of a star --- without flows, rotation, or magnetic field --- can be solved to give a system of eigenfunctions \citep{Christensen-Dalsgaard2008}. These eigenfunctions, with their respective eigenfrequencies, are perturbed when flows \citep{Lavely1992}, rotation, or a magnetic field \citep[e.g.,][]{Gough1990} are added to the star. If the perturbation is small enough, the perturbed eigenfunctions and eigenfrequencies can be obtained from the unperturbed eigenfunctions and eigenfrequencies with techniques from standard perturbation theory. As the spectrum of the eigenfrequencies of a solar-like star is dense, the techniques from quasi-degenerate perturbation theory can be adopted. The solutions to the eigenvalue problem 
\begin{align}
\mathbf{ZC} = \boldsymbol{\Lambda} \mathbf{C},\label{eq:sec1:1}
\end{align}
define the perturbed eigenvectors and eigenfrequencies of the system. Here, $\mathbf{Z}$ is called the supermatrix with entries
\begin{align}\label{eq:sec2:5}
Z_{k\p k}&= \begin{cases}
H_{k\p k}-\left(\omega_{\text{ref}}^2-\omega^2_{k}\right)\delta_{k\p k} &\text{for }k\p,k\in K,\\
0 & \text{otherwise,}
\end{cases}
\end{align}
where $H_{k\p k}$ is the general matrix element, which is discussed in detail below, and the indices $k=\left(n,l,m\right)$, $k'=\left(n',l',m'\right)$ define the considered eigenmodes with radial orders $n,n'$, harmonic degrees $l,l'$, and azimuthal orders $m,m'$. The coupling set $K$ is made up of those modes, which satisfy the following two conditions: Firstly, their frequencies obey the quasi-degeneracy condition
\begin{align}
\left|\omega_{\text{ref}}^2-\omega_k^2\right|<\Delta\omega^2,\label{eq:sec2:3}
\end{align}
where the reference frequency $\omega_{\text{ref}}$ is typically chosen equal or close to the central frequency of a multiplet $k$, $\omega_k$ is the frequency of the considered mode, and $\Delta\omega^2$ is the width of this range. Second, the geometry of the modes, determined by their harmonic degree and azimuthal order, complies with angular momentum selection rules imposed by the configuration of the perturbation. These selection rules are given in Section~5.1 of \citet{Kiefer2017b} for the magnetic field configurations considered in this work.

The eigenvector $\mathbf{C}$ holds the expansion coefficients of the perturbed eigenfunction $\boldsymbol{\xi}_j$:
\begin{align}
\boldsymbol{\xi}_j &= \sum\limits_{k\in K}{c_{jk}\boldsymbol{\xi}_k^0}\label{eq:sec2:2a},
\end{align}
where $c_{jk}$ is the $k$'th component of $\mathbf{C}$ and $\boldsymbol{\xi}_k^0$ is an unperturbed eigenfunction. 
The matrix $\boldsymbol{\Lambda}$ is a diagonal matrix with the frequency perturbations $\delta\omega^2_k$ as entries.

Detailed discussions of quasi-degenerate perturbation theory with application to solar and stellar problems can be found in, e.g., \citet{Lavely1992}, \citet{Roth2002}, \citet{Schad2013}, \citet{Herzberg2016}, and \citet{Herzberg2017}.

The complete general matrix element $H_{k'k}$ is calculated as
\begin{align}
H_{k'k} = D_{k\p k} - I_{k\p k},\label{fullGME}
\end{align}
where $D_{k\p k}$ is the general matrix element, which accounts for the mode coupling due to the magnetic field. An analytical expression of $D_{k\p k}$ for a superposition of zonal toroidal magnetic fields was recently derived by \citet{Kiefer2017b}. It is given by
\begin{align}
D_{k\p k} = -\frac{1}{4\pi}\sum_{s,s\p}\int_V \overline{\boldsymbol{\xi}}_{k\p} \cdot &\left[ \left(\left(\boldsymbol{\nabla}\times\left(\boldsymbol{\nabla}\times\left(\boldsymbol{\xi}_k\times\mathbf{B}_{s}\right)\right)\right)\times\mathbf{B}_{s\p}\right)+\left(\left(\boldsymbol{\nabla}\times\mathbf{B}_s\right)\times\left(\boldsymbol{\nabla}\times\left(\boldsymbol{\xi}_k\times\mathbf{B}_{s\p}\right)\right)\right)\right.\notag\\
&\left.+\left(\boldsymbol{\xi}_k\cdot\boldsymbol{\nabla}\right)\left(\left(\boldsymbol{\nabla}\times\mathbf{B}_s\right)\times\mathbf{B}_{s\p}\right)+\left(\boldsymbol{\nabla}\cdot\boldsymbol{\xi}_k\right)\left(\left(\boldsymbol{\nabla}\times\mathbf{B}_s\right)\times\mathbf{B}_{s\p}\right)\right]\,\D V,\label{eq:directGME}
\end{align}
where $\mathbf{B}_{s},\mathbf{B}_{s'}$ are toroidal magnetic fields with harmonic degrees $s$ and $s'$, $\boldsymbol{\xi}_{k}$ and $\boldsymbol{\xi}_{k\p}$ are eigenfunctions, and the integral extends over the stellar volume $V$. The sum in Equation~(\ref{eq:directGME}) extends over the values of harmonic degrees $s$ and $s'$ of the investigated magnetic field model, see Equation~(\ref{eq:sec2:2}). The cross-terms between configurations of unequal harmonic degree are thus included in the calculation of the matrix element. The full analytical expression of $D_{k\p k}$ is given in Equations~(31), and (G62)--(G86) of \citet{Kiefer2017b}.
$I_{k\p k}$ in Equation~(\ref{fullGME}) accounts for the modal interactions due to the perturbations in stellar structural quantities, which arise as a result of the magnetic field. We derive $I_{k\p k}$ for the case of a superposition of zonal toroidal magnetic fields in Section~\ref{sec:3}.

The shift in angular frequency of a mode can be approximated by
\begin{align}
\delta\omega_k = \sqrt{\omega_k^2 +\delta\omega^2_k}-\omega_k,\label{eq:sec1:10}
\end{align}
where $\delta\omega^2_k$ is the perturbation of the squared angular frequency:
\begin{align}
\delta\omega^2_k \approx H_{kk}+\sum_{k'\in K}\frac{H_{k'k}H_{kk'}}{\omega_k^2-\omega_{k'}^2}+\dots,\label{eq:sec1:9}
\end{align}
which is expanded up to second order \citep{Schad2011}. In this article, we do not consider perturbations to the eigenfunctions but focus on the more accessible quantity -- the shifts in mode frequency. However, the perturbation of the eigenfunctions by a superposition of zonal toroidal magnetic fields can be calculated with the theory presented here: To second order, the perturbed eigenfunction $\boldsymbol{\xi}_k$ is given by
\begin{align}\label{eq:sec1:11}
\boldsymbol{\xi}_k \approx \boldsymbol{\xi}^0_{k} + 
\sum_{k'\in K\setminus k}\frac{H_{k'k}}{\omega_k^2-\omega_{k'}^2}\boldsymbol{\xi}_{k'}^0
+\sum_{k',k''\in K\setminus k}\frac{H_{k'k''}H_{k''k}}{\left(\omega_k^2-\omega_{k'}^2\right)\left(\omega_k^2-\omega_{k''}^2\right)}\boldsymbol{\xi}_{k'}^0
- \sum_{k'\in K\setminus k}\frac{H_{k'k}H_{kk}}{\left(\omega_k^2-\omega_{k'}^2\right)^2}\boldsymbol{\xi}_{k'}^0,
\end{align}
where $\boldsymbol{\xi}^0_{k}$ is the unperturbed eigenfunction. Equations~(\ref{eq:sec1:9}) and (\ref{eq:sec1:11}) are basic results of non- or quasi-degenerate perturbation theory, see, e.g., \citet{Sakurai2014}.

In the present work, we concentrate on the self-coupling of modes. This reduces the computational effort to obtain $H_{k'k}$ by a factor of ${\vert K\vert}^2$, i.e., from the square of the number of coupling modes to only one general matrix element $H_{kk}$. This approximation is justified, as the coupling strength decreases with increasing frequency difference and the radial and horizontal eigenfunction become less similar for modes of different radial order and harmonic degree. We expect the error introduced by this approximation to be of the order of a few percent. \citet{Schad2013} found that for differential rotation, the self-coupling matrix elements are typically two orders of magnitude larger than the cross-coupling matrix elements. We expect this to be similar for a magnetic field as the perturbing agent. The eigenfunctions are not perturbed in the self-coupling limit, as can be seen from Equation~(\ref{eq:sec1:11}). In this approximation, we find from Equations~(\ref{eq:sec1:10}) and (\ref{eq:sec1:9}) that the shift in mode frequency can be calculated by
\begin{align}
\delta\nu_k = \frac{\sqrt{\omega_k^2 +H_{kk}}-\omega_k}{2\pi}.\label{eq:sec1:12}
\end{align}

%%%%%%%%%%%%%%%%%%%%%%%%%%%%%%%%%%%%%%%%%%%%%%%%%%%%%%%%%%%%%%%%%%%%%%%%%%%%%%%%%%%%%%%%%%%%%%%%%%%%%%%%%%%%%%%%%%%%%%%%%%%%%%%%%%%%%%%%%%%%%%%%%%%%
%%%%%%%%%%%%%%%%%%%%%%%%%%%%%%%%%%%%%%%%%%%%%%%%%%%%%%%%%%%%%%%%%%%%%%%%%%%%%%%%%%%%%%%%%%%%%%%%%%%%%%%%%%%%%%%%%%%%%%%%%%%%%%%%%%%%%%%%%%%%%%%%%%%%
\section{The Indirect Effect}\label{sec:3}
If a magnetic field is present, the Lorentz force has to be added to the equation of motion. Hence, the structure of the star is slightly changed compared to the equilibrium. We treat this change as a small perturbation to the nonmagnetic star. The Lorentz force for a superposition of axisymmetric zonal toroidal magnetic fields is given by
\begin{align}
\mathbf{F}_{\text{tor}}(r,\theta) = \frac{1}{4\pi}\left(\boldsymbol{\nabla}\times\mathbf{B}_{\text{tor}}(r,\theta)\right)\times\hat{\mathbf{B}}_{\text{tor}}(r,\theta)\label{eq:sec2:1}
\end{align}
with the toroidal magnetic field
\begin{align}
\mathbf{B}_{\text{tor}}(r,\theta) &= \sum_{s}\mathbf{B}_{\text{s}}(r,\theta) = \sum_{s}-a(r) \delth{} \sharm{s}{0}(\theta)\mathbf{e}_\phi\label{eq:sec2:2},\\
\hat{\mathbf{B}}_{\text{tor}}(r,\theta) &= \sum_{s'}\mathbf{B}_{\text{s}'}(r,\theta)= \sum_{s'}-\hat{a}(r) \delth{} \sharm{s'}{0}(\theta)\mathbf{e}_\phi.
\end{align}
The second magnetic field $\hat{\mathbf{B}}_{\text{tor}}$ is introduced to keep track of the individual contributions in a superposition of components of distinct harmonic degree. Here, $a(r)$ and $\hat{a}(r)$ are the radial profiles of the component of the magnetic field with harmonic degrees $s$ and $s'$, respectively. $\sharm{s}{0}(\theta)$ is a spherical harmonic function of degree $s$ and azimuthal order $0$. In this article, we will be referring to relations and properties of the spherical harmonic functions and the generalized spherical harmonic functions $\gsharm{l}{N}{m}$, which can be found in Appendix~(D) of \citet{Kiefer2017b}. Many useful relations for the spherical harmonic functions can also be found in, e.g., \citet{Dahlen1998}. The spherical harmonics are a special case of the generalized spherical harmonics with $\sharm{l}{m}=\gsharm{l}{0}{m}$ in the convention that is used in this work. 

The Lorentz force for a magnetic field of the form presented in Equation~(\ref{eq:sec2:2}) can be written as 
\begin{align}
\mathbf{F}_{\text{tor}}(r,\theta)=& \sum_{s,s'}\sum_{\lambda=0}^{s+s'}\left(\mathcal{R}_{s,s',\lambda}(r){Y}_{\lambda}^{0}(\theta)\mathbf{e}_r + \mathcal{S}_{s,s\p,\lambda}(r)\delth{}Y_{\lambda}^{0}(\theta)\mathbf{e}_{\theta}\right),\label{app:project:result:main}
\end{align}
where the radial functions of the radial component and the colatitudinal component are given by
\begin{align}
\mathcal{R}_{s,s\p,\lambda}(r) &= \om{0}{s}\om{0}{s'}\gamma_{\lambda}\gamma_{s}\gamma_{s'}\left(\frac{a\hat{a}}{r}+\hat{a}\delr{a}\right)
\wtj{s}{s'}{\lambda}{0}{0}{0}\wtj{s}{s'}{\lambda}{1}{-1}{0},\label{project:R}\\
\mathcal{S}_{s,s\p,\lambda}(r) &=	
-\om{0}{s}\om{0}{s}\om{0}{s'}\gamma_{\lambda}\gamma_{s}\gamma_{s'}\frac{a\hat{a}}{\sqrt{2}r}\wtj{s}{s'}{\lambda}{0}{0}{0}\wtj{s}{s'}{\lambda}{0}{-1}{1},\label{project:S}
\end{align}
respectively. The coefficients are $\om{l}{N}= \sqrt{(l+N)(l-N+1)/2}$ and $\gamma_{l}= \sqrt{(2l+1)/4\pi}$. The last two factors of both equations are Wigner 3j symbols. Extensive lists of their properties can be found in, e.g., \citet{Edmonds}, \citet{Regge1958} and \citet{Dahlen1998}. We will be referring to only those properties of the Wigner 3j symbols that are listed in Appendix~(E) of \citet{Kiefer2017b}. Equations~(\ref{project:R}) and (\ref{project:S}) are the vector spherical harmonic expansion coefficients of the Lorentz force for a toroidal magnetic field specified in Equation~(\ref{eq:sec2:2}). They are derived in detail in Appendix~\ref{app:sec:project}.

We expand aspherical perturbations to all stellar structural quantities in spherical harmonics \citep{Woodhouse1978, Lavely1992, Dahlen1998}:
\begin{align}
\delta Q\left(r,\theta\right) &= \sum\limits_{s,s\p,\lambda}\delta Q_{s,s\p}^{\lambda}(r)\sharm{\lambda}{0}(\theta),\label{expand:perturbation}
\end{align}
where the quantity $Q$ can be the gravitational potential $\phi$, density $\rho$, pressure $p$, or squared sound speed $c^2$. We consider only zonal toroidal magnetic fields as perturbations. Hence, the azimuthal order of the spherical harmonic in the expansion (\ref{expand:perturbation}) is set to $0$.
The ranges of the summation indices are determined by the configuration of the considered magnetic field: the indices $s$ and $s\p$ take the values of the model magnetic field. The index $\lambda$ extends over all even values between $0$ and $s+s'$. This can be seen from properties of the Wigner 3j symbols (E29) and (E30c) in \citet{Kiefer2017b} and the fact that we only consider magnetic fields with even harmonic degree. Restricting the magnetic field to even harmonic degrees ensures antisymmetry about the equator which is generally observed for the Sun \citep{Hathaway2015}.

The aspherical perturbation to the gravitational potential $\delta\phi_{s,s\p}^{\lambda}(r)$ can be calculated by solving
\begin{align}
\frac{1}{r^2}\Dr{}\left(r^2\Dr{\delta\phi_{s,s\p}^{\lambda}(r)}\right) - \left[\frac{\lambda\left(\lambda+1\right)}{r^2}+\frac{4\pi G}{g_0}\Dr{\rho_0}\right]\delta\phi_{s,s\p}^{\lambda}(r) = \frac{4\pi G}{g_0}\left[\mathcal{R}_{s,s\p,\lambda}(r)+\Dr{}\left(r\mathcal{S}_{s,s\p,\lambda}(r)\right)\right].\label{eq:sweet}
\end{align}
where $G$ is the constant of gravitation, $g_0$ is the unperturbed gravitational acceleration, $\rho_0$ is unperturbed density, and $\mathcal{R}_{s,s\p,\lambda}$ and $\mathcal{S}_{s,s\p,\lambda}$ are the vector spherical harmonic coefficients of the radial and colatitudinal component of the Lorentz force. This equation is derived in \citet{Sweet1950}. Having obtained $\delta\phi_{s,s\p}^{\lambda}$ by numerically integrating Equation~(\ref{eq:sweet}), we can calculate the perturbations in density, pressure, and squared sound speed:
\begin{align}
\delta\rho_{s,s\p}^{\lambda}(r) &= \frac{1}{g_0}\left[\Dr{\rho_0}\delta\phi_{s,s\p}^{\lambda}(r)+\mathcal{R}_{s,s\p,\lambda}(r)+\Dr{}\left(r\mathcal{S}_{s,s\p,\lambda}(r)\right)\right],\label{perturb:rho}\\
\delta p_{s,s\p}^{\lambda}(r) &= -\rho_0\delta\phi_{s,s\p}^{\lambda}(r)-r\mathcal{S}_{s,s\p,\lambda}(r),\label{perturb:p}\\
\delta {c^2}_{s,s\p}^{\lambda}(r) &= \left[\left(\frac{\partial \ln\Gamma_1}{\partial\ln p}\right)_\rho+1\right]\frac{\delta p_{s,s\p}^{\lambda}(r)}{p_0}c_0^2+ \left[\left(\frac{\partial \ln\Gamma_1}{\partial\ln \rho}\right)_p-1\right]\frac{\delta \rho_{s,s\p}^{\lambda}(r)}{\rho_0}c_0^2\label{perturb:soundspeed},
\end{align}
where $\Gamma_1$ is the first adiabatic exponent and $c_0^2$ is unperturbed sound speed. The derivatives of $\ln \Gamma_1$ are supplied in the solar model that was used for this work \citep{Christensen-Dalsgaard1996}. We present brief derivations of Equations~(\ref{perturb:rho})-(\ref{perturb:soundspeed}) in Appendix~(\ref{app:sec:perturb}).

The general matrix element for the indirect effect is derived in Appendix (E) of \citet{Lavely1992}. We neglect terms due to rotation and ellipticity and transform the general matrix element from perturbations in bulk modulus $\kappa_0$ and density $\rho_0$ into perturbations in squared sound speed $c_0^2$ and density $\rho_0$, see Equation~(\ref{app:transform:perturbations}). This yields
\begin{align}
I_{k\p k} = 4\pi\glp\gl\left(-1\right)^{m\p}\sum\limits_{s,s',\lambda}\gamma_{\lambda}\wtj{l\p}{\lambda}{l}{-m\p}{0}{m}\int\limits_{0}^{R_{\odot}}&\left(\rho_0(r)\delta {c^2}_{\,s,s\p}^{\lambda}(r)K_{\lambda}\left(r\right) + c_0^2\delta \rho_{s,s\p}^{\lambda}(r)K_{\lambda}\left(r\right)\right.\notag\\&+\left.\delta\rho_{s,s'}^{\lambda}\left(r\right)R_{\lambda}^{(2)}(r)\right)r^2\D r,\label{matrixelement_indirect}
\end{align}
where $K_{\lambda}$ is the bulk modulus perturbation kernel and $R_{\lambda}^{(2)}$ is the density perturbation kernel, which are listed in Appendix~(\ref{app:sec:kernels}).

%%%%%%%%%%%%%%%%%%%%%%%%%%%%%%%%%%%%%%%%%%%%%%%%%%%%%%%%%%%%%%%%%%%%%%%%%%%%%%%%%%%%%%%%%%%%%%%%%%%%%%%%%%%%%%%%%%%%%%%%%%%%%%%%%%%%%%%%%%%%%%%%%%%%
%%%%%%%%%%%%%%%%%%%%%%%%%%%%%%%%%%%%%%%%%%%%%%%%%%%%%%%%%%%%%%%%%%%%%%%%%%%%%%%%%%%%%%%%%%%%%%%%%%%%%%%%%%%%%%%%%%%%%%%%%%%%%%%%%%%%%%%%%%%%%%%%%%%%
\begin{table}
	\centering
	\caption{Computed models of the toroidal magnetic field.}
	\begin{tabular}{ccccccc}
		\hline\hline 
		Model&Degree&$\mu$&$\sigma$&$B_{\text{max}}$&$\beta(B_{\text{max}})$&$\beta(\text{photosphere})$\\
		&& $\left(R_{\odot}\right)$& $\left(R_{\odot}\right)$& $\left(\unit{kG}\right)$ & $\times 10^{-3}$ &\\
		\hline
		A&2&0.9&0.04&50&21.3&0.39\\
		B&2&0.9&0.04&40&33.3&0.62\\
		C&2&0.72&0.05&300&14.5&$9\times 10^{8}$\\
		D&4&0.9&0.04&50&21.3&0.39\\
		E&2&0.9&0.04&50&15.5&0.29\\
		&4&0.9&0.04&-30&&\\
		F&2&0.97&0.01&10&16.0&154\\
		\hline
	\end{tabular} \label{table:1}
\end{table}

\section{Disturbing the Sun}\label{sec:4}
\begin{figure}
	\centering{
		\includegraphics[width=0.35\textwidth]{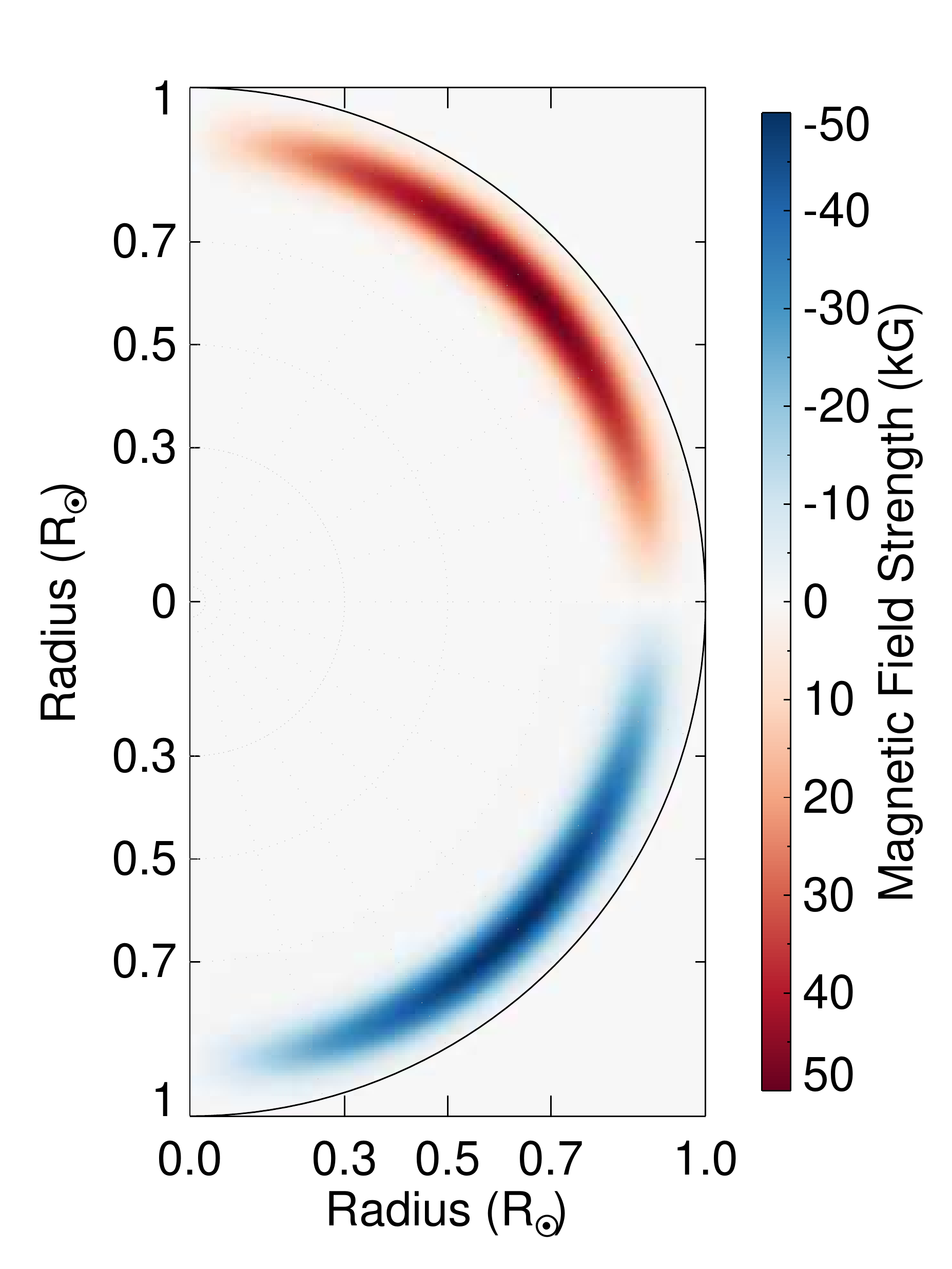}\\
		\includegraphics[width=0.6\textwidth]{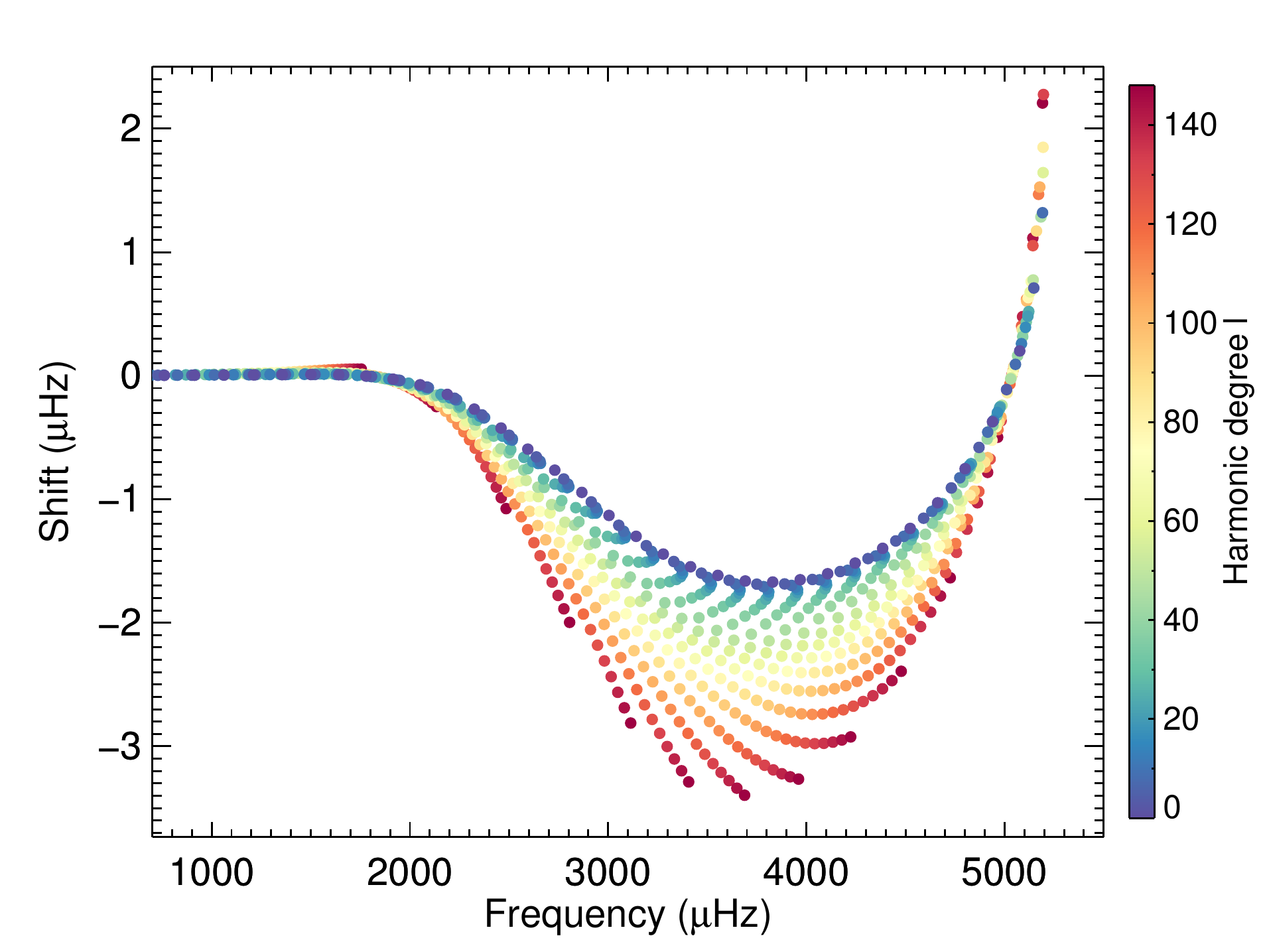}
		\caption{\textit{Top panel:} visualization of magnetic field model A. \textit{Bottom panel:} multiplet frequency shifts for model A as a function of unperturbed mode frequency. Every fourth harmonic degree is shown.}
		\label{fig:1}}
\end{figure}

We modeled six different toroidal magnetic field distributions. Their radial profiles were modeled with Gaussians 
\begin{align}
a(r)= \frac{B_{\text{scale}}}{\sigma \sqrt{2\pi}} \exp\left(-0.5 \left(\frac{r-\mu}{\sigma}\right)^2\right),\label{eq:sec:4:1}
\end{align}
where $B_{\text{scale}}$ is a factor to scale the distribution to the desired maximum value. In Table~\ref{table:1} the locations of the maximum of the field distribution $\mu$ and the width in terms of the Gaussian standard deviation $\sigma$ are given in the third and fourth columns.\footnote{To avoid confusion, the center of the Sun is at a value of $\unit[0]{R_{\odot}}$ and the photosphere is located at $\unit[1]{R_{\odot}}$. This places the tachocline around $\unit[0.72]{R_{\odot}}$.} The maximum values of the distribution are given in the fifth column. The sixth and seventh columns of Table~\ref{table:1} give the values of the plasma beta
\begin{align}
\beta = \frac{8\pi p }{B^2}\label{eq:sec:4:2}
\end{align}
at the location of the maximum of the field distribution and at the photospheric level above the maximum of the distribution. Here, $p$ is the gas pressure and $B$ is the magnetic field strength. The plasma beta is the ratio of gas pressure to magnetic pressure. It must be noted that the values for $\beta$ change with latitude as the spherical harmonics, with which the radial distribution are multiplied, see Equation~(\ref{eq:sec2:2}), have a latitudinal dependence. This latitudinal dependence of the field distributions can be appreciated in the top panel of Figure~\ref{fig:1}, where magnetic field model A is shown in a meridional cut. The $\beta$ values in Table~\ref{table:1} can thus be seen as minimal values.

Perturbations to the gravitational potential were neglected, i.e., we applied the Cowling approximation \citep{Cowling1941}. This affects the sensitivity kernel $R_{\lambda}^{(1)}$ (Equation~(\ref{app:kernels:6})) and the perturbations to the structural quantities (Equations~(\ref{eq:sweet})-(\ref{perturb:soundspeed})).\footnote{If the Cowling approximation is not applied, the eigenfunctions of the gravitational potential $\delta\phi$ have to be calculated because they are required in Equation~(\ref{app:kernels:6}). The ADIPLS code \citep{Christensen-Dalsgaard2008} that was used for the computation of the set of eigenmodes can provide these functions.}

In our calculations, we used the standard solar model Model S by \citet{Christensen-Dalsgaard1996}. Included in the version of the model we used is an extended set of variables, e.g., the derivatives $\left(\frac{\partial \ln\Gamma_1}{\partial\ln p}\right)_\rho$ and $\left(\frac{\partial \ln\Gamma_1}{\partial\ln \rho}\right)_p$ which are needed for the computation of the perturbation of the squared sound speed.

Model A is of harmonic degree $s=2$, has its maximum at $\mu=0.90\,R_{\odot}$ with a width of $\sigma=0.04\,R_{\odot}$, and has a maximum field strength of $B_{\text{max}}=\unit[50]{kG}$, see the top panel of Figure~\ref{fig:1}. The maximum field strength is located at latitudes of $\pm 45^{\circ}$. The bottom panel of Figure~\ref{fig:1} shows the resulting frequency shifts as a function of unperturbed mode frequency for modes with $4\le l \le 148$. To enhance clarity, we show only every fourth harmonic degree starting at $l=4$. The frequency shifts are averaged over the azimuthal order $m$ and are thus the mean shift of each multiplet. This shift is usually reported in studies of solar oscillation frequencies as a function of the level of activity \citep[e.g.,][]{Broomhall2017}. 

\begin{figure}
	\centering{
		\includegraphics[width=0.6\textwidth]{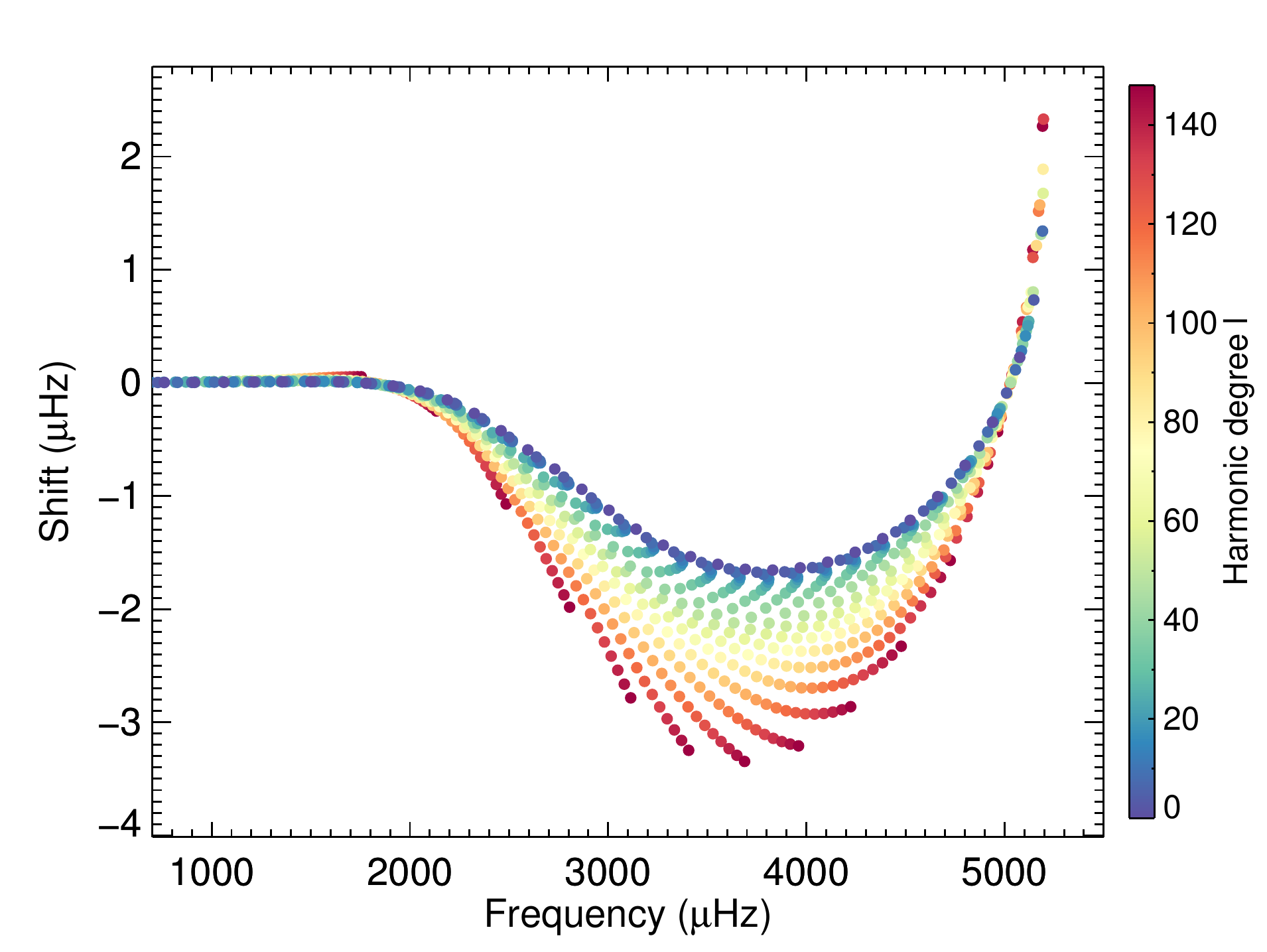}
		\includegraphics[width=0.6\textwidth]{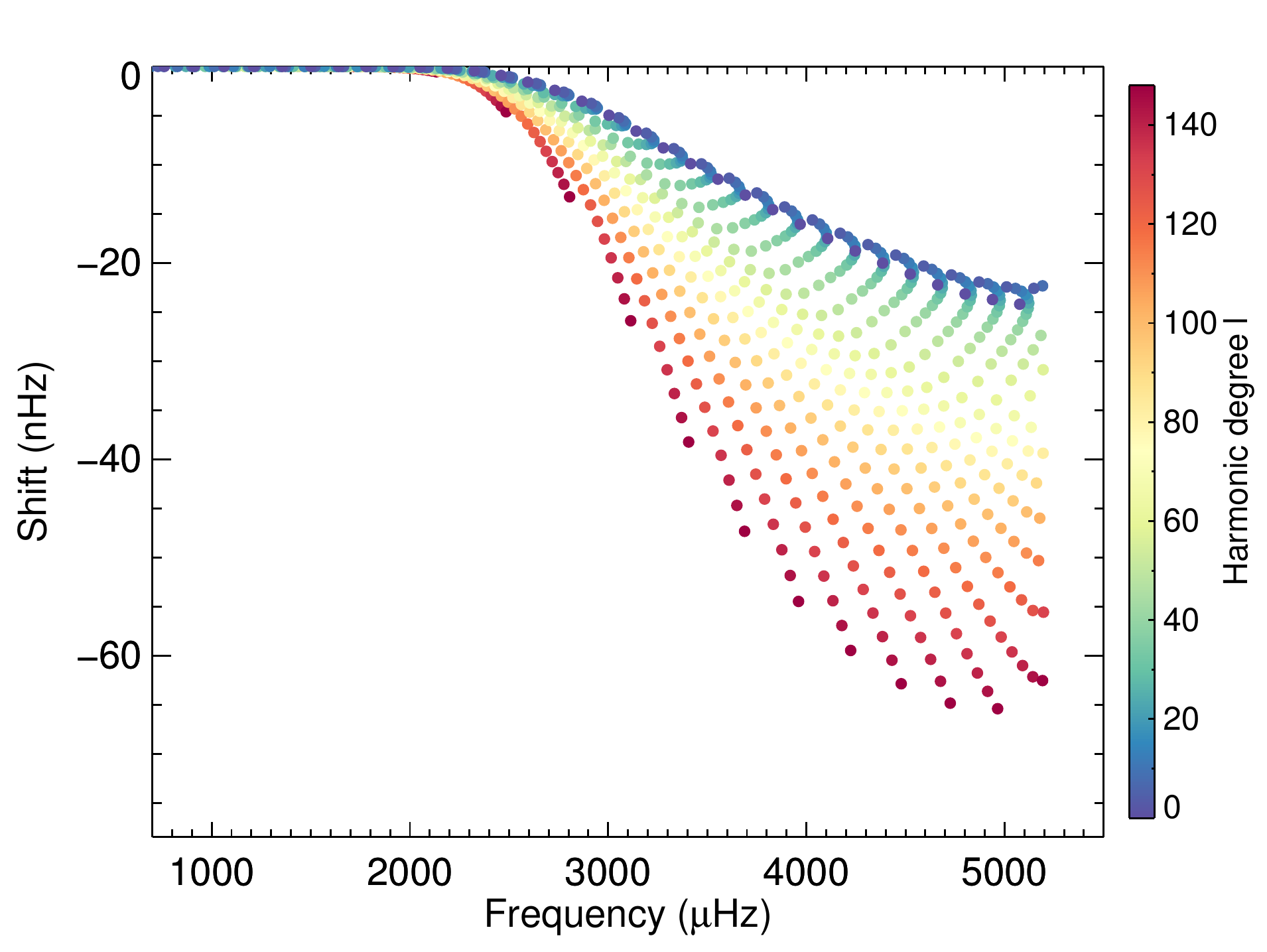}
		\caption{Multiplet frequency shifts caused by the direct (\textit{top panel}) and indirect effect (\textit{bottom panel}) for magnetic field model A. Notice the different magnitudes of the shifts in the top and bottom panels.}
		\label{fig:2}}
\end{figure}

We calculated the general matrix elements for the direct and indirect effects separately. Thus, we can also examine the shifts caused by the two effects separately. In the top panel of Figure~\ref{fig:2}, the mean multiplet shifts of model A are shown for only the direct effect. In the bottom panel, the same is shown for the indirect effect. Note the different orders of magnitude of the shifts in the two panels.

\begin{figure}
	\centering{
		\includegraphics[width=0.35\textwidth]{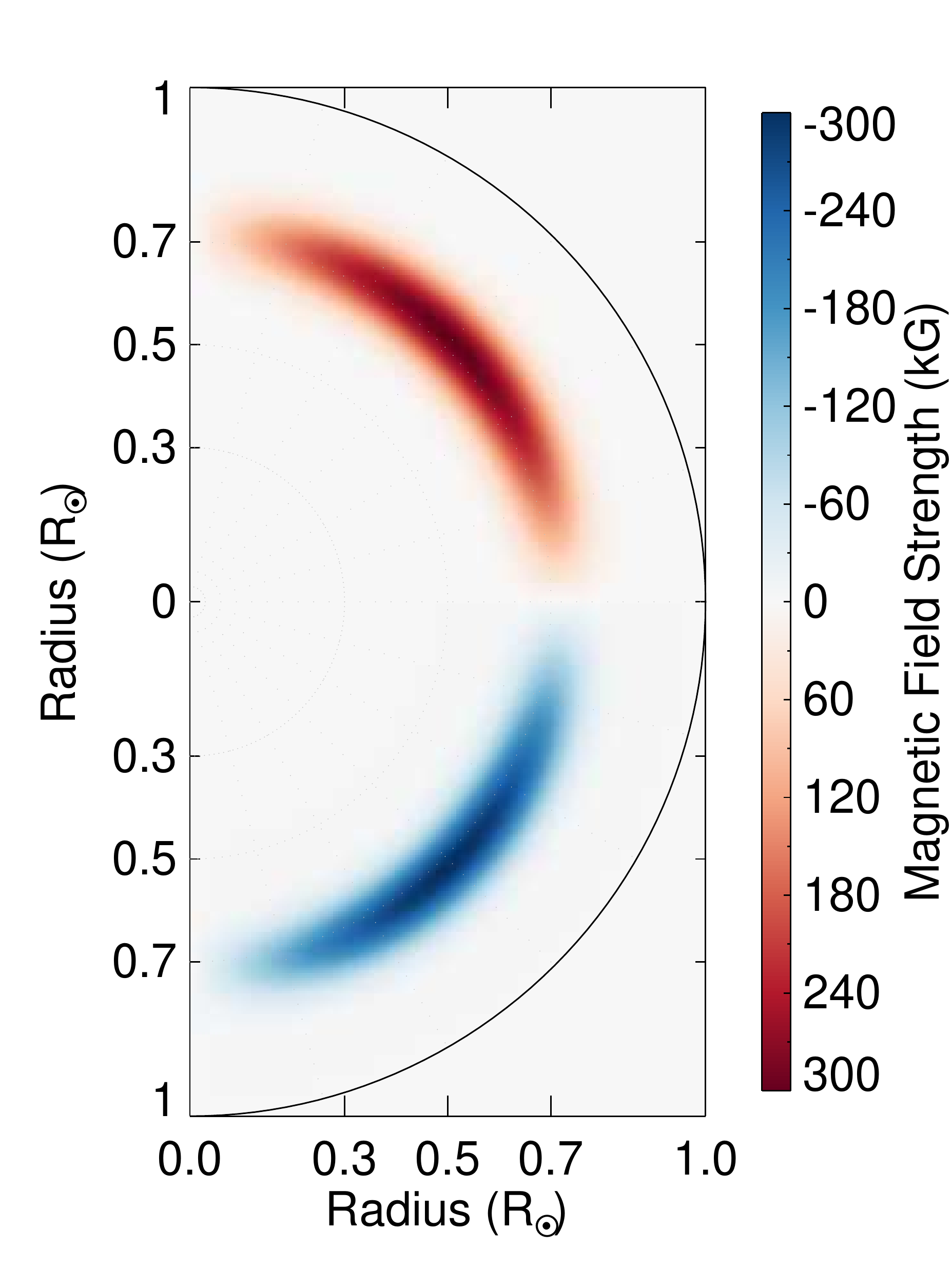}\\
		\includegraphics[width=0.6\textwidth]{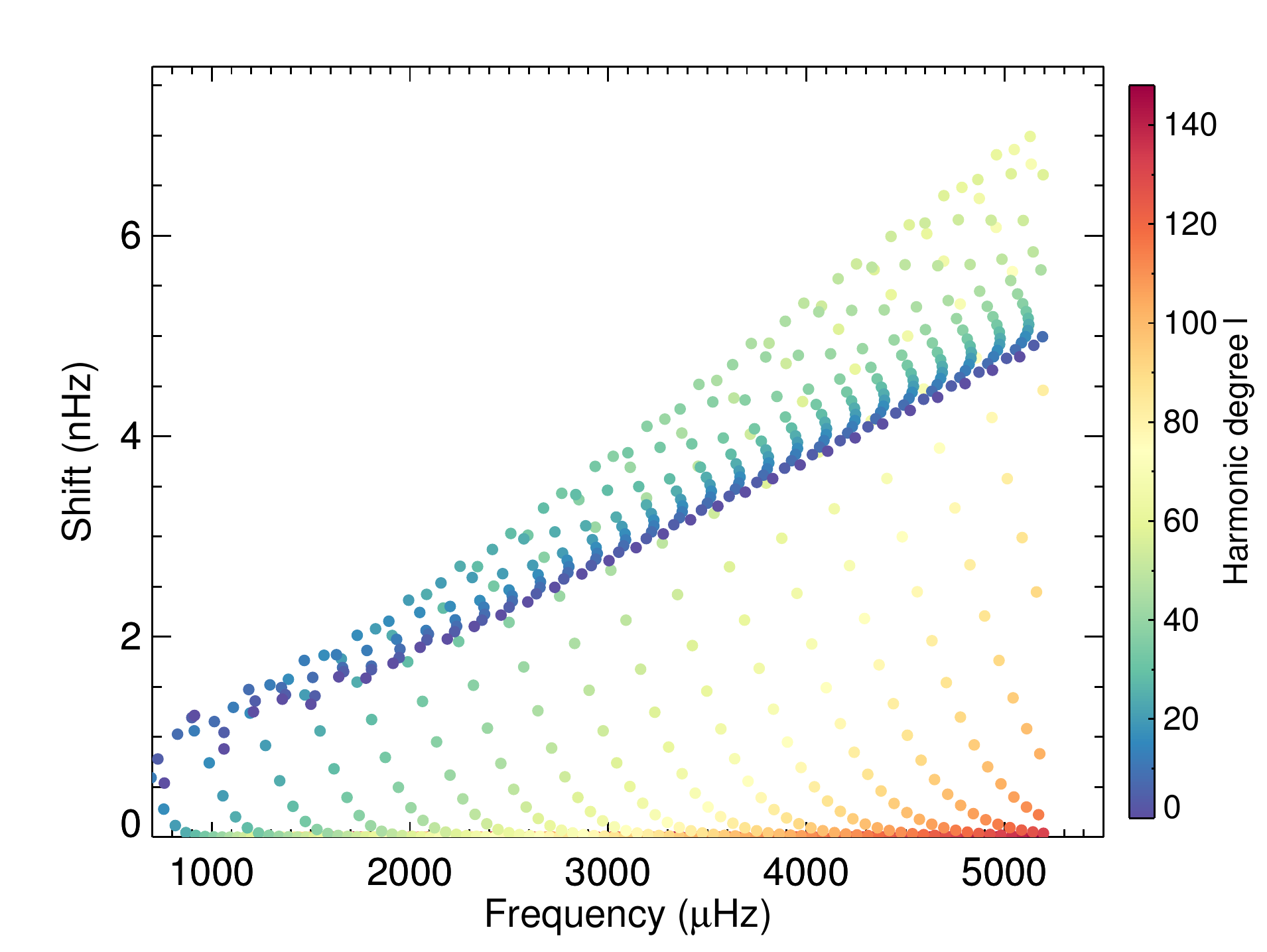}
		\caption{\textit{Top panel:} visualization of magnetic field model C. \textit{Bottom panel:} multiplet frequency shifts for model C as a function of unperturbed mode frequency. Every fourth harmonic degree is shown. }
		\label{fig:4}}
\end{figure}

Magnetic field model C, which is a strong field in the tachocline region with $B_{\text{max}}=\unit[300]{kG}$, is depicted in the top panel of Figure~\ref{fig:4}. We modeled this field with a harmonic degree of $s=2$, at a depth of $\mu=0.72\,R_{\odot}$, and with a width of $\sigma=0.04\,R_{\odot}$. The resulting frequency shifts as a function of unperturbed mode frequency for modes of harmonic degree $4\le l \le 148$ are shown in the bottom panel of Figure~\ref{fig:4}. Notice the small magnitude of the shifts compared to the field model A.

We chose the strength of this model because \citet{Antia2000} put an upper limit of $\unit[300]{kG}$ on magnetic fields in the tachocline region from analyses of mode splitting coefficients.

Model B is the same as model A, but with a lower maximum field strength. We will use this model in the next section to investigate differences between two configurations of the same geometry and depth but different strengths. Field configuration D has the same parameters for the radial function $a(r)$ as models A and B, but has a harmonic degree of $s=4$. Model E is a superposition of two fields, one with $s=2$ and one with $s=4$. Both fields are located at $\mu=0.9\,R_{\odot}$ with a width of $\sigma=0.04\,R_{\odot}$. The $s=2$ component has a maximum field strength of $B_{\text{max}}=\unit[50]{kG}$, while the $s=4$ component has a strength of $B_{\text{max}}=\unit[-30]{kG}$. The cross-terms between the two configurations are included in the calculation of the general matrix elements. As can be seen in Figure~\ref{app:fig:3.1} in Appendix~\ref{app:sec:figures}, due to the superposition of the two fields, the maximum of the field is closer to the equator compared to model A. 

Model F is a shallow field located at $\mu=0.97\,R_{\odot}$ with a width of $\sigma=0.01\,R_{\odot}$, with a maximum field strength of $B_{\text{max}}=\unit[10]{kG}$, and has harmonic degree $s=2$. The multiplet shifts for this model are only of the order of some tens of $\unit{nHz}$ as can be seen in Figure~\ref{app:fig:4.2} in Appendix~\ref{app:sec:figures}. The shifts caused by the indirect effect are much smaller still, being at the level of a few $\unit[10^{-15}]{Hz}$. The shape of the frequency shifts as a function of mode frequency is rather different compared to the other models.

Figures~\ref{app:fig:1.2}--\ref{app:fig:4.2} in Appendix~\ref{app:sec:figures} show visualizations of the magnetic fields, the resulting frequency shifts as functions of mode frequency and of lower turning point, as well as the shifts as a function of mode frequency separated for the direct and indirect effect for field models D, E, and F.

In Figure~\ref{fig:3} the frequency shifts for models A and C are shown as functions of lower turning point of the modes. As can be seen in the bottom panel, only modes with lower turning point below or in the magnetized region around the tachocline experience a significant shift. Modes with turning points above the magnetic field are no longer disturbed by the direct effect. They can, however, still experience a weak shift due to the indirect effect, see lower row of panels in Figure~\ref{app:fig:1.2} in Appendix~\ref{app:sec:figures}. As can be seen in the top panel of Figure~\ref{fig:3}, for model A, the maximum shifts are experienced by the modes of the highest degree we calculated, as they have their lower turning point in the magnetized region.

%%%%%%%%%%%%%%%%%%%%%%%%%%%%%%%%%%%%%%%%%%%%%%%%%%%%%%%%%%%%%%%%%%%%%%%%%%%%%%%%%%%%%%%%%%%%%%%%%%%%%%%%%%%%%%%%%%%%%%%%%%%%%%%%%%%%%%%%%%%%%%%%%%%%
%%%%%%%%%%%%%%%%%%%%%%%%%%%%%%%%%%%%%%%%%%%%%%%%%%%%%%%%%%%%%%%%%%%%%%%%%%%%%%%%%%%%%%%%%%%%%%%%%%%%%%%%%%%%%%%%%%%%%%%%%%%%%%%%%%%%%%%%%%%%%%%%%%%%
\section{Discussion}\label{sec:5}
The six magnetic field configurations we tested produce very distinct patterns in the frequency shifts as a function of unperturbed mode frequency. While models A, B, and D lead to a decrease of the multiplet frequency for a wide range of mode frequencies, the shift is generally positive for models C, E, and F. 

The multiplet shifts for field model A are presented in the bottom panel of Figure~\ref{fig:1}. The shifts for this model decrease for mode frequencies higher than $\nu\approx\unit[2]{mHz}$ and have their minimum at around $\nu\approx\unit[3.8]{mHz}$. For even higher frequencies they increase and become positive for modes with $\nu\gtrsim\unit[5]{mHz}$. Ridges of modes with the same radial order are apparent. The pattern seen here is determined by the direct effect. As seen in Figure~\ref{fig:2}, the shifts caused by the indirect effect (bottom panel) are much smaller than those caused by the direct effect (top panel).

The mean multiplet shift caused by the indirect effect is negative for all models and all modes. The dependence of the shift as a function of frequency resembles the surface effect, see, e.g., \citet{Ball2014} or \citet{Basu2016}. For modes of low degree, as can be seen for the $l=4$ modes, the lowest harmonic degree we considered here, the multiplet shift caused by the indirect effect is lower than for the next harmonic degrees we considered for models D and E. This peculiarity may help to rule out some field configurations if it is not observed for the Sun. We shall study the behavior of the low degree mode frequencies in a separate article as these are of special importance for asteroseismic studies of the magnetic activity of a star for which only modes up to $l=3$ can be observed. We speculate that magnetic fields in the upper part of the solar convection zone can explain at least part of the discrepancy between model frequencies and observed frequencies. 

\begin{figure}
	\centering{
		\includegraphics[width=0.6\textwidth]{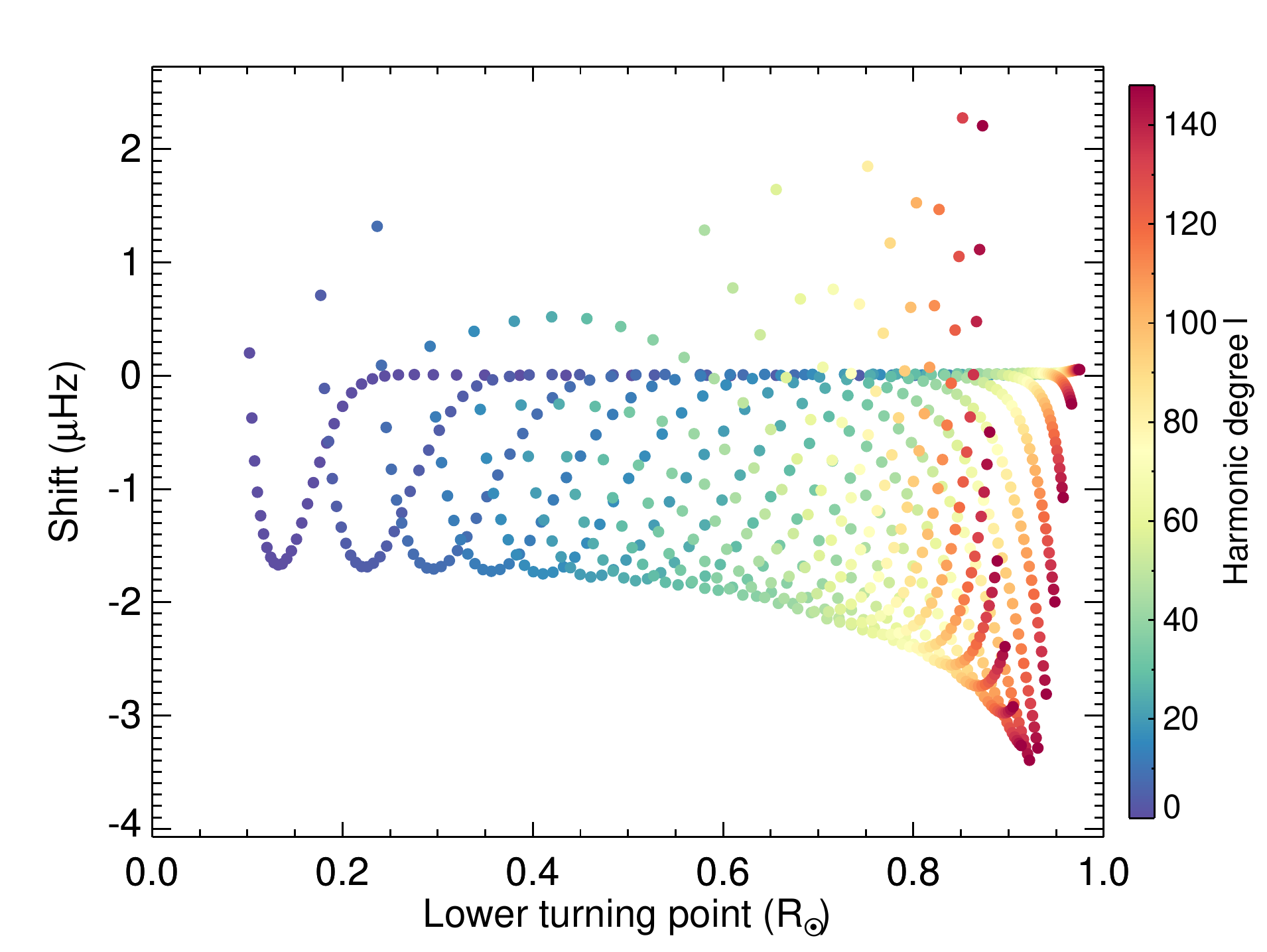}
		\includegraphics[width=0.6\textwidth]{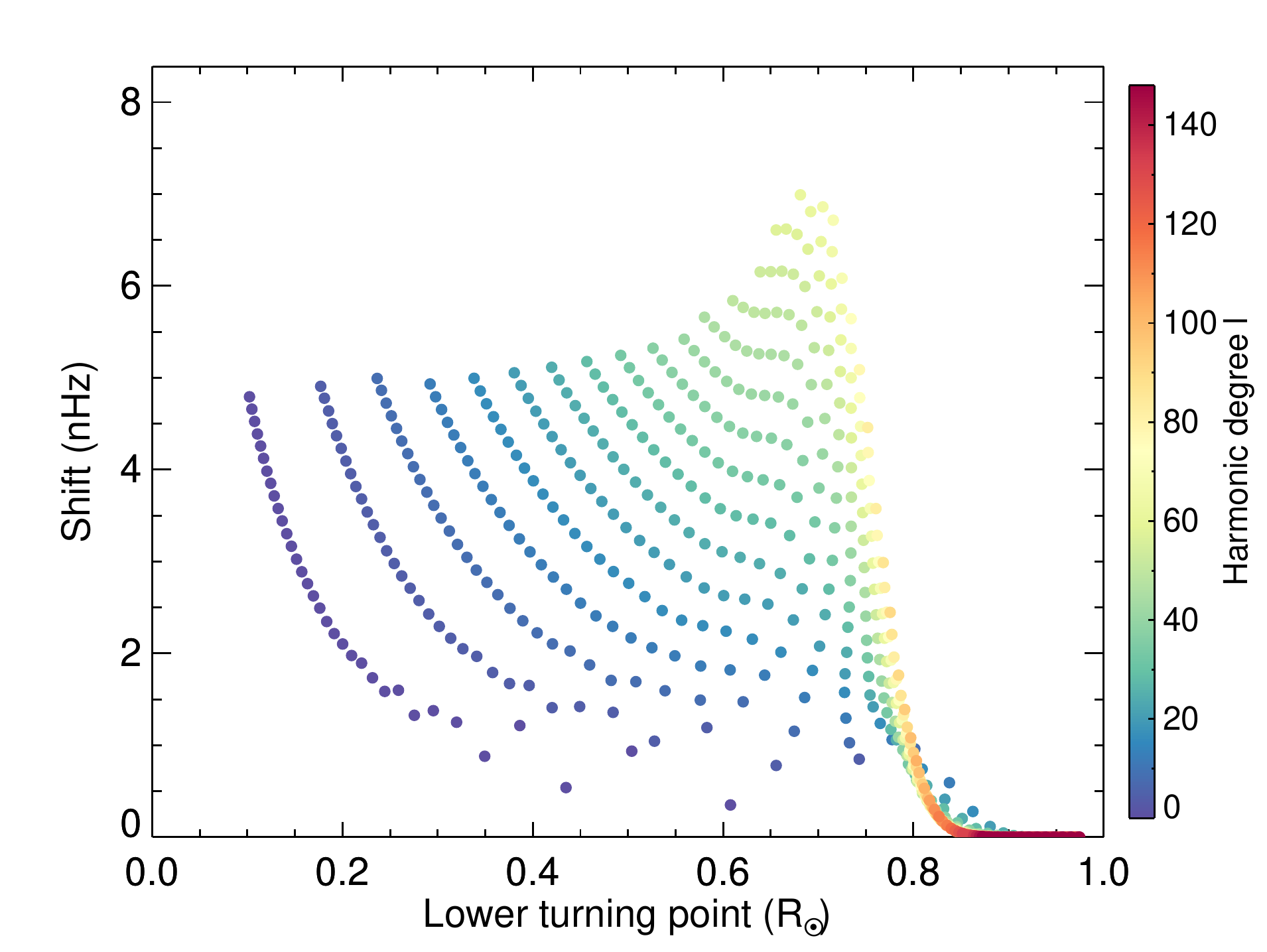}
		\caption{Multiplet frequency shifts for model A \textit{(top panel)} and model C \textit{(bottom panel)} as a function of lower turning point. Notice the different magnitudes of the shifts in the top and bottom panels.}
		\label{fig:3}}
\end{figure}

The direct effect is the dominant contribution to the total shift for all model fields we investigated. The indirect effect is only important for more complex field configurations and when the magnetic field strength near the photosphere is so strong as to give a plasma beta $\beta<1$. The shifts caused by the indirect effect are of the same order of magnitude as those caused by the direct effect for models D and E, as can be seen from Figures~\ref{app:fig:2.2} and \ref{app:fig:3.2} in Appendix~\ref{app:sec:figures}. For the other models, which all have harmonic degree $s=2$, the contribution from the indirect effect to the mean multiplet shift can well be neglected.

Generating plots for observed shifts as a function of the lower turning point of the modes, analogously to Figure~\ref{fig:3}, might give an indication of the location of magnetic field concentrations in the Sun. For this, the shifts of modes that have their lower turning point rather close to the surface have to be determined. If the shifts begin to decrease at a certain harmonic degree, as they do for the shifts for model C, their turning point can then be interpreted as the location of maximal magnetic field strength in this region. We note that such a location is not necessarily the location of the maximum field strength in the Sun, as deeper seated magnetic fields result in a smaller frequency shift even if they are stronger than more shallow magnetic fields. This can be seen by comparison of the top and bottom panels of Figure~\ref{fig:3}: model A, which has a maximum magnetic field strength of $\unit[50]{kG}$ located at $\unit[0.9]{R_{\odot}}$, produces shifts at the \unit{$\mu$Hz} level. Whereas model C, which has a maximum magnetic field strength of six times that of model A, produces shifts that are only at the $\unit{nHz}$ level.

\begin{figure}
	\centering{
		\includegraphics[width=0.6\textwidth]{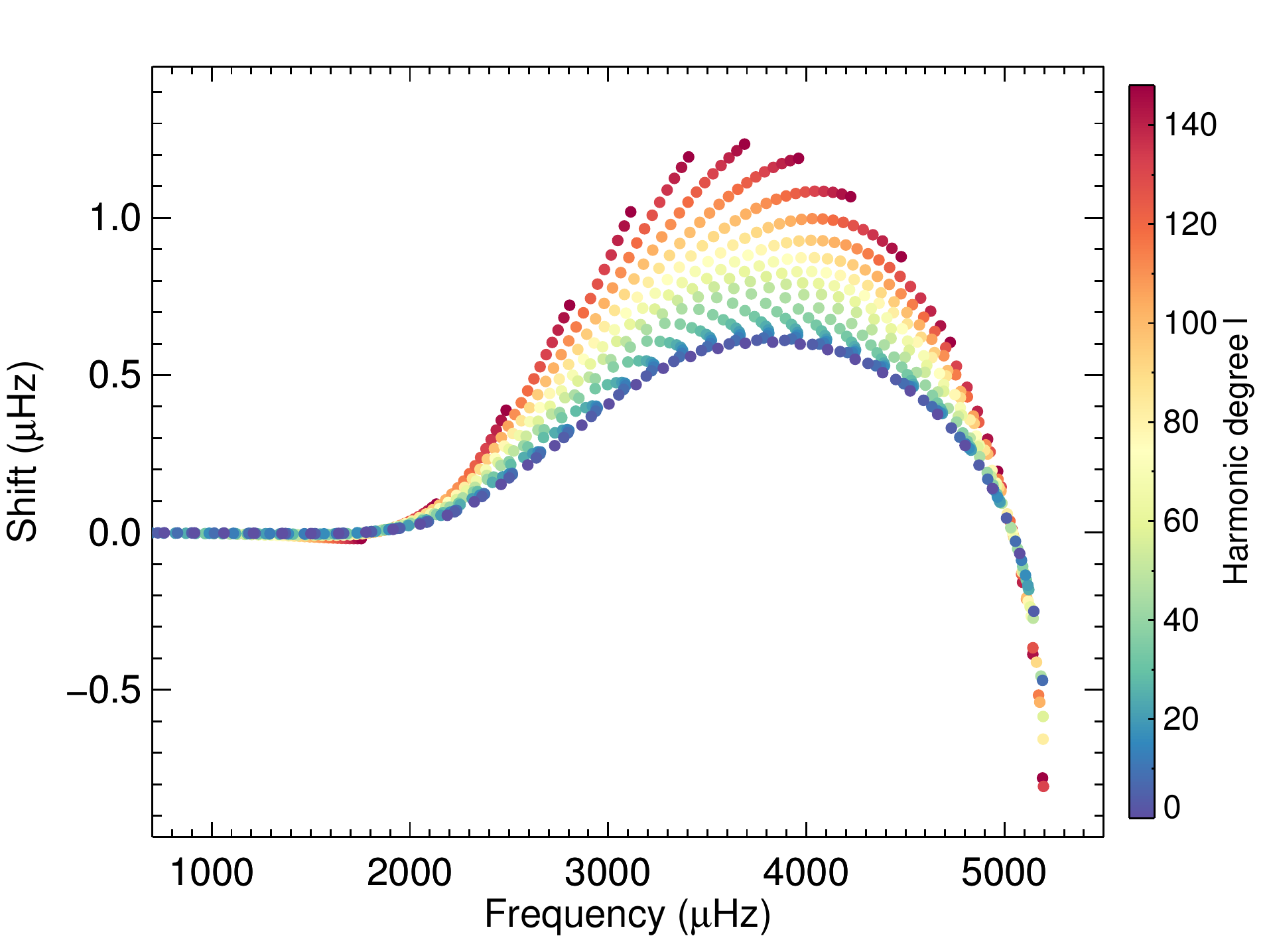}
		\caption{Difference of the multiplet frequency shifts for model B and model A (in the sense $B-A$) as a function of unperturbed mode frequency. Every fourth harmonic degree is shown.}
		\label{fig:5.1}}
\end{figure}
In Figure~\ref{fig:5.1}, the difference of the multiplet frequency shifts for model B and model A is shown as a function of unperturbed mode frequency. We show differences between the shifts of models B and A in the sense model B - model A, that is, the model with the weaker magnetic field minus the model with the stronger field. On the Sun, the frequency shifts are correlated with the level of magnetic activity. Hence, model B with a maximum field strength of $\unit[40]{kG}$ would correspond to the activity maximum and model A with $B_{\text{max}}=\unit[50]{kG}$ would correspond to the activity minimum. In this picture, the toroidal magnetic field gets weaker in the ascending part of the cycle and reaches its minimum strength at the activity maximum. It then gets built up again by the solar dynamo and reaches its maximum strength at the activity minimum. The large field strength then causes the activity to increase again as magnetic flux starts to rise to the surface.

\begin{figure}
	\centering{
		\includegraphics[width=0.6\textwidth]{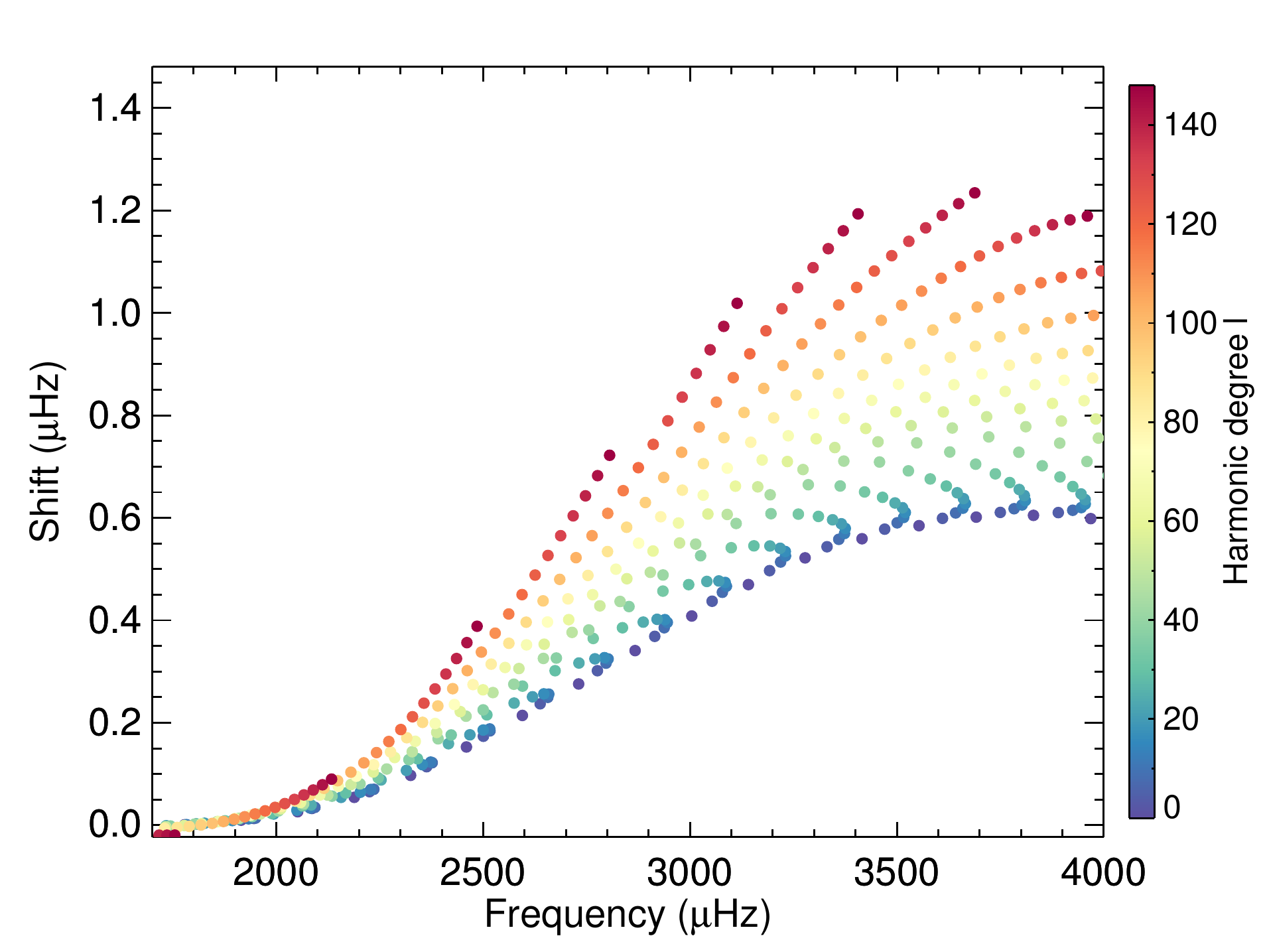}\\
		\vspace{2em}
		\includegraphics[width=0.6\textwidth]{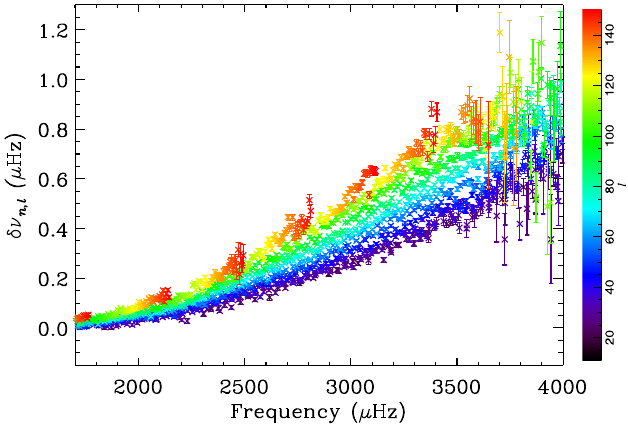}
		\caption{\textit{Top panel:} same as Figure~\ref{fig:5.1} but for a restricted frequency range. \textit{Bottom panel: } observed mean multiplet frequency shifts between the maximum of solar cycle 23 and the minimum of cycle 23. Figure adapted from \citet{Broomhall2017}. Note the different color coding in the two panels.}
		\label{fig:5}}
\end{figure}

In the top panel of Figure~\ref{fig:5}, we show the same as in Figure~\ref{fig:5.1} but for a restricted frequency range of 1700--4000$\unit{\mu Hz}$. In the bottom panel of Figure~\ref{fig:5}, we show the results of \citet[][called B17 hereafter]{Broomhall2017}. She investigated the mean multiplet frequency shifts between the maximum of solar cycle 23 and the minimum of the same cycle from Global Oscillation Network Group (GONG) data. As can be seen, the differences between our two model magnetic field produce shifts in the multiplet frequencies that are strikingly similar to those reported by \citetalias{Broomhall2017}. Our results have the correct order of magnitude, while our frequency shifts are about $\unit[0.1]{\mu Hz}$ higher than the solar shifts. This can, however, easily be corrected by adjusting the field strength of either model A or B. We find that the shifts of modes of low harmonic degree have a maximum at mode frequencies of $\nu\approx\unit[3800]{\mu Hz}$. As can be seen in Figure~\ref{fig:5.1}, they decrease for higher frequencies and even turn negative at about $\unit[5000]{\mu Hz}$. This behavior is not apparent in the result of \citetalias{Broomhall2017}. An extension of the measurement of the observed frequency shifts to mode frequencies above $\unit[4000]{\mu Hz}$ may help to discern the model fields that best reproduce the shifts.

As can be seen in the last two columns of Table~\ref{table:1}, the $\beta$ values of the six models we consider here, cover a wide range. The tachocline field, model C, has the smallest $\beta$ at its location of maximum magnetic field strength, but the highest value in the photosphere. Overall, the multiplet shifts for model C are very small, as can be seen in Figure~\ref{fig:4}. The shifts of model F, for which $\beta(\text{photosphere})=154$, are also only of the order of $\unit{nHz}$. The other four models have $\beta$ values at the photospheric level that are $<1$. These models produce multiplet frequency shifts that are of the order of $\unit{\mu Hz}$. From this, we gather that the magnetic field needs to have an appreciable field strength in the near-surface layers if the mean multiplet shifts shall reach the $\unit{\mu Hz}$ level as is observed for the Sun. 

For the model field E, which is a superposition of two fields, we find that the shifts are positive for most modes, see Figure~\ref{app:fig:3.2} in Appendix~\ref{app:sec:figures}. It is noteworthy that each of the two components of this model would lead to negative shifts on their own as can be seen for model A and model D. Also, the shifts are significantly different from zero already for mode frequencies below $\unit[2]{mHz}$.

%%%%%%%%%%%%%%%%%%%%%%%%%%%%%%%%%%%%%%%%%%%%%%%%%%%%%%%%%%%%%%%%%%%%%%%%%%%%%%%%%%%%%%%%%%%%%%%%%%%%%%%%%%%%%%%%%%%%%%%%%%%%%%%%%%%%%%%%%%%%%%%%%%%%
%%%%%%%%%%%%%%%%%%%%%%%%%%%%%%%%%%%%%%%%%%%%%%%%%%%%%%%%%%%%%%%%%%%%%%%%%%%%%%%%%%%%%%%%%%%%%%%%%%%%%%%%%%%%%%%%%%%%%%%%%%%%%%%%%%%%%%%%%%%%%%%%%%%%
\section{Conclusion}\label{sec:6}
The search for the layers where strong magnetic fields are located in the interior of the Sun can be conducted with helioseismic forward calculations and inversions. In this article, we described our efforts to carry out forward calculations of the effect of toroidal magnetic fields on solar multiplet frequencies of acoustic oscillations. The theoretical framework we present here is also applicable to stellar models.

We investigated six models for the magnetic field. A strong tachocline field, model C, produced only small shifts of the order of a few \unit{nHz}. The maximum field strength of \unit[300]{kG} is already at \citep{Antia2000} or even high above the expected limit of magnetic fields in this region of the Sun \citep{Arlt2007}. With this, we can safely state that toroidal magnetic fields in the solar tachocline are not responsible for the observed frequency shifts over the solar cycle.

\cite{Basu1997} put an upper limit of \unit[300]{kG} on the strength of toroidal magnetic fields concentrated below the convection zone base. This is the strength of magnetic field model C. Hence, the effect of fossil fields in the solar radiative zone with comparable strengths can also only be of the order of $\unit{nHz}$. Since we did not investigate poloidal magnetic field configurations, we cannot rule out poloidal fossil fields of this strength. We note again that the effect of near-surface magnetic field concentrations on acoustic oscillations are stronger by orders of magnitude and will thus likely impede the detection of fossil fields in the frequencies of solar acoustic oscillations. An investigation of the perturbation of eigenfunctions of low harmonic degree due to fossil fields is worthwhile and might put tighter constraints on the strength of such fields.

Shifts in multiplet frequencies of the Sun are measured between two differently magnetized states of the Sun. Commonly, these shifts are reported between an activity minimum and a following or preceding activity maximum. The difference of the multiplet shifts caused by our models A and B are found to be of the same order of magnitude as the observed shifts on the Sun. Also, the behavior of the shifts as a function of frequency and harmonic degree strongly resembles the solar values reported by \citetalias{Broomhall2017}. This might indicate that there are magnetic field distributions of the geometry and depth of our models A and B in the Sun. The location of the maximum field strength is at $\unit[0.9]{R_{\odot}}$ and they both have a field strength in the photosphere, which gives a plasma beta $\beta$ that is smaller than unity. In general, we found that the $\beta$ value in the very shallow layers must be of the order of unity to produce multiplet shifts that reach the observed $\unit{\mu Hz}$ level.

A detailed study of splitting coefficients with the theory we presented here is expedient and will be conducted shortly. This will yield further insight into the depth, shape, and strength of the magnetic field concentration necessary to produce the observed frequency shifts and splitting coefficients caused by toroidal magnetic fields. It was shown by \citet{Schad2013a} that analyses of perturbed eigenfunctions can be used to infer the solar meridional circulation. For this, the cross-coupling of modes has to be included, as self-coupling alone does not change the eigenfunction. Future integrated investigations of perturbed eigenfunctions, frequency shifts, and splitting coefficients, which can all be determined with the general matrix elements presented by \citet{Kiefer2017b} and in this article, are a promising tool to shed more light on the workings of the solar dynamo. 
\paragraph{Acknowledgments}
We are grateful to J{\o}rgen Christensen-Dalsgaard for supplying the extended version of Solar Model S. We also thank Kolja Glogowski for computing the full set of solar oscillation eigenmodes as well as for helpful discussions, Vincent B\"oning for important comments and fruitful discussions, and our institute's internal referee Wolfgang Schmidt for reading the initial manuscript and his remarks, which helped to improve this article. We thank the anonymous referee for taking the time to review this paper. The research leading to these results received funding from the European Research Council under the European Union’s Seventh Framework Program (FP/2007-2013)/ERC Grant Agreement no. 307117.

\appendix
\section{Decomposition of a Real Valued Vector Field}\label{app:sec:decompose}
In order to expand the Lorentz force in terms of spherical harmonics, we use that every real vector field $\mathbf{u}$ can be expanded in terms of vector spherical harmonics \citep{Dahlen1998}:
\begin{align}
\mathbf{u} = \sum_{l=0}^{\infty}\sum_{m=-l}^{l}\mathcal{R}_l^m\mathbf{Y}_l^m + \mathcal{S}_l^m\mathbf{\Psi}_l^m - \mathcal{T}_l^m\mathbf{\Phi}_l^m\label{app:vsh:1}.
\end{align}
The vector spherical harmonics, which are complete, are defined by
\begin{align}
\mathbf{Y}_l^m &= \sharm{l}{m}\mathbf{e}_r,\label{app:vsh1}\\
\mathbf{\Psi}_l^m &= r\boldsymbol{\nabla}\sharm{l}{m} = \left(\mathbf{e}_{\theta}\delth{}+\mathbf{e}_{\phi}\frac{1}{\sinth}\delphi{}\right)\sharm{l}{m},\label{app:vsh2}\\
\mathbf{\Phi}_l^m &= \mathbf{r}\times\boldsymbol{\nabla}\sharm{l}{m} = \mathbf{e}_r \times \boldsymbol{\Psi}_l^m = \left(-\mathbf{e}_{\theta}\frac{1}{\sinth}\delphi{}+\mathbf{e}_{\phi}\delth{}\right)\sharm{l}{m}\label{app:vsh3},	
\end{align}
where $l$ is the harmonic degree, $m$ is the azimuthal order with $-m\le l \le m$, $\sharm{l}{m}$ is a spherical harmonic function, and we work in spherical geometry with the coordinates $\left(r,\theta, \phi\right)$ radius, colatitude, and azimuth. The vector spherical harmonic coefficients in Equation~(\ref{app:vsh:1}) are given by
\begin{align}
\mathcal{R}_l^m &= \int\overline{\mathbf{Y}_l^m}\cdot\mathbf{u} \D\Omega,\label{app:vsh:2}\\
\mathcal{S}_l^m &= \frac{1}{l\left(l+1\right)}\int\overline{\boldsymbol{\Psi}_l^m}\cdot\mathbf{u} \D\Omega,\label{app:vsh:3}\\
\mathcal{T}_l^m &= -\frac{1}{l\left(l+1\right)}\int\overline{\boldsymbol{\Phi}_l^m}\cdot\mathbf{u} \D\Omega.\label{app:vsh:4}
\end{align}
With Equations~(\ref{app:vsh:1})--(\ref{app:vsh:4}), the components of the vector field $\mathbf{u}$ are calculated as
\begin{align}
u_{r}(r,\theta,\phi) &= \sum_{l=0}^{\infty}\sum_{m=-l}^{l}\mathcal{R}_l^m(r){Y}_l^m(\theta,\phi) ,\label{app:vsh:5}\\
u_{\theta}(r,\theta,\phi) &= \sum_{l=0}^{\infty}\sum_{m=-l}^{l} \mathcal{S}_l^m(r)\delth{}Y_l^m(\theta,\phi) + \mathcal{T}_l^m(r)\frac{1}{\sinth}\delphi{} Y_l^m(\theta,\phi),\label{app:vsh:6}\\
u_{\phi}(r,\theta,\phi) &= \sum_{l=0}^{\infty}\sum_{m=-l}^{l}\mathcal{S}_l^m(r)\frac{1}{\sinth}\delphi{}Y_l^m(\theta,\phi) - \mathcal{T}_l^m(r)\delth{}Y_l^m(\theta,\phi),\label{app:vsh:7}
\end{align}

%%%%%%%%%%%%%%%%%%%%%%%%%%%%%%%%%%%%%%%%%%%%%%%%%%%%%%%%%%%%%%%%%%%%%%%%%%%%%%%%%%%%%%%%%%%%%%%%%%%%%%%%%%%%%%%%%%%%%%%%%%%%%%%%%%%%%%%%%%%%%%%%%%%%
%%%%%%%%%%%%%%%%%%%%%%%%%%%%%%%%%%%%%%%%%%%%%%%%%%%%%%%%%%%%%%%%%%%%%%%%%%%%%%%%%%%%%%%%%%%%%%%%%%%%%%%%%%%%%%%%%%%%%%%%%%%%%%%%%%%%%%%%%%%%%%%%%%%%
\section{Expansion of the Lorentz Force}\label{app:sec:project}
The Lorentz force, see Equation~(\ref{eq:sec2:1}), for the toroidal configuration defined in Equation~(\ref{eq:sec2:2}), is given by
\begin{align}
\mathbf{F}_{\text{tor}}=& -\frac{1}{4\pi}\left[\frac{a\hat{a}}{r}\delth{}\sharm{s}{0}\delth{}\sharm{s\p}{0}+\hat{a}\delr{a}\delth{}\sharm{s}{0}\delth{}\sharm{s\p}{0}\right]\mathbf{e}_r\notag\\
&-\frac{1}{4\pi}\left[\frac{a\hat{a}}{r}\cotth\delth{}\sharm{s}{0}\delth{}\sharm{s\p}{0}+\frac{a\hat{a}}{r}\delthsq{}\sharm{s\p}{0}\delth{}\sharm{s\p}{0}\right]\mathbf{e}_{\theta},\label{app:lorentz}
\end{align}
where $a$ and $\hat{a}$ are the radial profiles of the two magnetic field component, which are superposed. All dependencies were dropped. The Lorentz force (\ref{app:lorentz}) will now be projected onto the vector spherical harmonics, as defined in Equations~(\ref{app:vsh:5})--(\ref{app:vsh:7}). Each vector component is treated separately. 
\subsection{The Radial Component}
We concentrate on one magnetic field configuration with indices $s,s'$, which is indicated by including the indices $s,s'$ to the notation of the vector spherical harmonic coefficients, e.g., $\mathcal{R}_{s,s',l}^m$. Calculating the sum over different configurations in Equation~(\ref{app:project:result}) then yields the total effect of the superposition. 

With Equation~(\ref{app:vsh:5}), the radial component of the Lorentz force can be written as
\begin{align}
\mathbf{F}_{\text{tor},r}(r,\theta,\phi) &= \sum_{\lambda=0}^{\infty}\sum_{\mu=-\lambda}^{\lambda}\mathcal{R}_{s,s',\lambda}^\mu(r){Y}_{\lambda}^{\mu}(\theta,\phi)\mathbf{e}_r ,\label{app:project:1}
\end{align}
where the vector spherical harmonic coefficient are
\begin{align}
\mathcal{R}_{s,s\p,\lambda}^{\mu}(r) &\enspace= \int \overline{\boldsymbol{Y}_{\lambda}^{\mu}} \cdot \mathbf{F} \D \Omega\notag\\
&\stackrel{\text{(\ref{app:vsh1})}}{=} \int \overline{\sharm{\lambda}{\mu}}\mathbf{e}_{r}\cdot \mathbf{F} \D \Omega\notag\\
&\stackrel{\text{(\ref{app:lorentz})}}{=} \int \overline{\sharm{\lambda}{\mu}} \left(-\frac{1}{4\pi}\left[\frac{a\hat{a}}{r}\delth{}\sharm{s}{0}\delth{}\sharm{s\p}{0}+\hat{a}\delr{a}\delth{}\sharm{s}{0}\delth{}\sharm{s\p}{0}\right]\right) \D \Omega\notag\\
&\enspace= -\frac{1}{4\pi}\left(\frac{a\hat{a}}{r}+\hat{a}\delr{a}\right)\int \overline{\sharm{\lambda}{\mu}}\delth{}\sharm{s}{0}\delth{}\sharm{s\p}{0} \D \Omega.\label{app:project:2}
\end{align}
We define the angular kernel 
\begin{align}
\mathcal{A}_{1} &= \overline{\gsharm{\lambda}{0}{\mu}}\delth{}\gsharm{s}{0}{0}\delth{}\gsharm{s\p}{0}{0},\label{app:project:3}
\end{align}
where we used that $\gsharm{l}{0}{m}=\sharm{l}{m}$, see \citet{Dahlen1998}. The relations for the generalized spherical harmonics we use here can be found in Appendix~(D) of \citet{Kiefer2017b}. With Equation~(D18) from \citet{Kiefer2017b}, we can write 
\begin{align}
\mathcal{A}_{1}&= \frac{1}{2}\om{0}{s}\om{0}{s'}
\left(\overline{\gsharm{\lambda}{0}{\mu}}\gsharm{s}{-1}{0}\gsharm{s'}{-1}{0}-\overline{\gsharm{\lambda}{0}{\mu}}\gsharm{s}{1}{0}\gsharm{s'}{-1}{0}-\overline{\gsharm{\lambda}{0}{\mu}}\gsharm{s}{-1}{0}\gsharm{s'}{1}{0}+\overline{\gsharm{\lambda}{0}{\mu}}\gsharm{s}{1}{0}\gsharm{s'}{1}{0}\right).\label{app:project:4}
\end{align}
The angular integral over the product of three generalized spherical harmonics can be calculated with help of Equation~(C.198) of \citet{Dahlen1998}. Making use of the properties of the Wigner 3j symbols (E30a)--(E30c) in \citet{Kiefer2017b}, we find that $\mu = 0$, as otherwise the angular integral vanishes. It can also be seen that the integrals over the first and last term in the bracket in Equation~(\ref{app:project:4}) always vanish due to Equation (E30a) in \citet{Kiefer2017b}. We thus find
\begin{align}
\mathcal{R}_{s,s\p,\lambda}(r) &=   \om{0}{s}\om{0}{s'}\gamma_{\lambda}\gamma_{s}\gamma_{s'}\left(\frac{a\hat{a}}{r}+\hat{a}\delr{a}\right)
\wtj{s}{s'}{\lambda}{0}{0}{0}\wtj{s}{s'}{\lambda}{1}{-1}{0},\label{app:project:R}
\end{align}
where we made use of Equations~(E27) and (E28) from \citet{Kiefer2017b} and we dropped the upper index on $\mathcal{R}$ as $\mu=0$ always holds.

\subsection{The Azimuthal and Colatitudinal Components}
Carrying out the same procedure for the azimuthal component as for the radial part leads to $\mu=0$. With Equations~(\ref{app:vsh3}), (\ref{app:vsh:7}), and (\ref{app:lorentz}), we find that $\mathcal{T}_{\lambda}^{0}=0$ and hence $\mathbf{F}_{\text{tor},\phi}=0$, as can be expected from Equation~(\ref{app:lorentz}). It remains to explicitly calculate the colatitudinal component of the Lorentz force:
\begin{align}
\mathbf{F}_{\text{tor},\theta} = \sum_{\lambda=0}^{\infty} \mathcal{S}_{s,s\p,\lambda}^{0}\delth{}Y_{\lambda}^{0}\mathbf{e}_{\theta},
\end{align}
where the vector spherical harmonic coefficient is given by
\begin{align}
\mathcal{S}_{s,s\p,\lambda}^0(r) &\enspace= \frac{1}{\lambda\left(\lambda+1\right)}\int \overline{\boldsymbol{\Psi}} \cdot \mathbf{F} \D \Omega\notag\\
&\stackrel{\text{(\ref{app:vsh2})}}{=} \frac{1}{\lambda\left(\lambda+1\right)}\int \left(\mathbf{e}_{\theta}\delth{}+\mathbf{e}_{\phi}\frac{1}{\sinth}\delphi{}\right)\overline{\sharm{\lambda}{0}}\cdot \mathbf{F} \D \Omega\notag\\
&\stackrel{\text{(\ref{app:lorentz})}}{=}\frac{1}{\lambda\left(\lambda+1\right)}\int \delth{} \overline{\sharm{\lambda}{0}}\left(-\frac{1}{4\pi}\left[\frac{a\hat{a}}{r}\cotth\delth{}\sharm{s}{0}\delth{}\sharm{s\p}{0}+\frac{a\hat{a}}{r}\delthsq{}\sharm{s}{0}\delth{}\sharm{s\p}{0}\right]\right) \D \Omega\notag\\
&\enspace= -\frac{a\hat{a}}{4\pi\lambda\left(\lambda+1\right) r} \int \left(\cotth\delth{} \overline{\sharm{\lambda}{0}}\delth{}\sharm{s}{0}\delth{}\sharm{s\p}{0} +\delth{} \overline{\sharm{\lambda}{0}}\delthsq{}\sharm{s}{0}\delth{}\sharm{s\p}{0}\right)\D \Omega.\label{app:project:theta:1}
\end{align}
We define the two angular kernels
\begin{align}
\mathcal{A}_{2} &= \cotth\delth{}\overline{\sharm{\lambda}{0}}\delth{}\sharm{s}{0}\delth{}\sharm{s'}{0}, \\
\mathcal{A}_{3} &= \delth{}\overline{\sharm{\lambda}{0}}\delthsq{}\sharm{s}{0}\delth{}\sharm{s'}{0}.
\end{align}
The following properties and equations are then used to obtain the vector spherical harmonic coefficient: Equation~(D18) from \citet{Kiefer2017b} to evaluate the colatitudinal derivatives; Equation~(D16) from \citet{Kiefer2017b} with $m=0$ to absorb the factor $\cotth$ in the kernel $\mathcal{A}_2$; Equation~(C.198) from \citet{Dahlen1998} to evaluate the angular integral in Equation~(\ref{app:project:theta:1}); properties (E27)--(E30c) of the Wigner 3j symbols in \citet{Kiefer2017b}. With all of this, we find
\begin{align}
\mathcal{S}_{s,s\p,\lambda}(r) =&	
-\om{0}{s}\om{0}{s}\om{0}{s'}\gamma_{\lambda}\gamma_{s}\gamma_{s'}\frac{a\hat{a}}{\sqrt{2}r}\wtj{s}{s'}{\lambda}{0}{0}{0}\wtj{s}{s'}{\lambda}{0}{-1}{1},
\end{align}
where we dropped the upper index on $\mathcal{S}$ as $\mu=0$ always holds.

The complete Lorentz force vector for a superposition of toroidal magnetic fields can thus be written as
\begin{align}
\mathbf{F}_{\text{tor}}(r,\theta)=& \sum_{s,s'}\sum_{\lambda=0}^{s+s'}\left(\mathcal{R}_{s,s',\lambda}(r){Y}_{\lambda}^{0}(\theta)\mathbf{e}_r + \mathcal{S}_{s,s\p,\lambda}(r)\delth{}Y_{\lambda}^{0}(\theta)\mathbf{e}_{\theta}\right),\label{app:project:result}
\end{align}
where the second summation extends over all even values between $0$ and $s+s'$ due to properties of the Wigner 3j symbols (E28) and (E30c) from \citet{Kiefer2017b}.

%%%%%%%%%%%%%%%%%%%%%%%%%%%%%%%%%%%%%%%%%%%%%%%%%%%%%%%%%%%%%%%%%%%%%%%%%%%%%%%%%%%%%%%%%%%%%%%%%%%%%%%%%%%%%%%%%%%%%%%%%%%%%%%%%%%%%%%%%%%%%%%%%%%%
%%%%%%%%%%%%%%%%%%%%%%%%%%%%%%%%%%%%%%%%%%%%%%%%%%%%%%%%%%%%%%%%%%%%%%%%%%%%%%%%%%%%%%%%%%%%%%%%%%%%%%%%%%%%%%%%%%%%%%%%%%%%%%%%%%%%%%%%%%%%%%%%%%%%
\section{The Sensitivity Kernels}\label{app:sec:kernels}
We reproduce the sensitivity kernels presented in \citet{Lavely1992}, Section (6c), which appear in the general matrix element for the indirect effect given in Equation~(\ref{matrixelement_indirect}):\footnote{We corrected two mistakes compared to \citet{Lavely1992}: In Equation~(\ref{app:kernels:6}) we added a factor $\xr$ to the second term in the second square bracket and in Equation~(\ref{app:kernels:7}) we corrected the factor before the first integral from $r^2$ to $r^{\lambda}$.}
\begin{align}
F(r) &= r^{-1}\left(2\xr-l\left(l+1\right)\xh\right),\label{app:kernels:1}\\
F'(r)&= r^{-1}\left(2\xrh-l'\left(l'+1\right)\xhh\right),\label{app:kernels:2}\\
K_{\lambda}(r) &= \left(\delr{\xrh}+F\p\right)\left(\delr{\xr}+F\right)B_{l\p \lambda l}^{(0)+},\label{app:kernels:3}\\
G_{\lambda}^{(1)}(r) &= \frac{1}{2}\rho_0r^{-1}\left(\xr\delr{\xhh}+r^{-1}\xr\xhh-\delr{\xr}\xhh-2F\xhh\right)B_{l\p l\lambda}^{(1)+}\notag\\
&\hspace{1em}+\frac{1}{2}\rho_0r^{-1}\left(\xrh\delr{\xh}+r^{-1}\xrh\xh-\delr{\xrh}\xh-2F'\xh\right)B_{ll\p \lambda}^{(1)+} +\rho_0 r^{-2}\xr\xrh \lambda\left(\lambda+1\right)B_{l\p \lambda l}^{(0)+},\label{app:kernels:4}\\
G_{\lambda}^{(2)}(r) &=\frac{1}{2}\rho_0 r^{-1}\xr\xhh B_{l\p l\lambda}^{(1)+} +\frac{1}{2}\rho_0 r^{-1}\xrh\xh B_{ll'\lambda}^{(1)+}-\rho_0\left(F'\xr+\xrh F\right)B_{l' \lambda l}^{(0)+},\label{app:kernels:5}\\
R_{\lambda}^{(1)}(r) &= \left[-\omega_{\text{ref}}^2\xh\xhh+r^{-1}\left(\delta\varphi'\xh+\delta\varphi\xhh\right)+\frac{1}{2}g_0r^{-1}\left(\xrh\xh+\xhh\xr\right)\right]B_{l\p \lambda l}^{(1)+}\notag\\
&\hspace{1em}+\left[8\pi G\rho_0\xr\xrh+\delr{\delta\varphi'}\xr+\delr{\delta\varphi}\xrh-\omega_{\text{ref}}^2\xr\xrh-\frac{1}{2}g_0\left(4r^{-1}\xr\xrh+\xrh F+\xr F'\right)\right]B_{l\p \lambda l}^{(0)+},\label{app:kernels:6}\\
R_{\lambda}^{(2)}(r) &= R_{\lambda}^{(1)}(r) + \frac{4\pi G}{2\lambda+1}\left\{r^{\lambda}\int\limits_{r}^{R_{\odot}}r^{-\lambda}\left[\left(\lambda+1\right)G_{\lambda}^{(2)}(r)-rG_{\lambda}^{(1)}(r)\right]\D r\right.\notag\\
&\left.\hspace{10em} -r^{-\lambda-1}\int\limits_0^r r^{\lambda+1}\left[\lambda G_{\lambda}^{(2)}(r)+rG_{\lambda}^{(1)}(r)\right]\D r\right\},\label{app:kernels:7}
\end{align}
where $K_{\lambda}(r)$ is the bulk modulus perturbation kernel and $R_{\lambda}^{(2)}(r)$ is the density perturbation kernel.

The Woodhouse coefficients \citep{Woodhouse1980} are given by 
\begin{align}
B_{l'l''l}^{(N)\pm} = \frac{1}{2}\left(1\pm\left(-1\right)^{l'+l''+l}\right)\left[\frac{\left(l'+N\right)!\left(l+N\right)!}{\left(l'-N\right)!\left(l-N\right)!}\right]^{\frac{1}{2}}\left(-1\right)^N\wtj{l'}{l''}{l}{-N}{0}{N}.
\end{align}
The cases, which occur in the sensitivity kernels, are 
\begin{align}
B_{l\p sl}^{(0)\pm}&= \frac{1}{2}\left(1\pm\left(-1\right)^{l'+s+l}\right)\wtj{l'}{s}{l}{0}{0}{0},\\
B_{l\p sl}^{(1)\pm} &= -\frac{1}{2}\left(1\pm\left(-1\right)^{l'+s+l}\right)\left[l'\left(l'+1\right)l\left(l+1\right)\right]^{\frac{1}{2}}\wtj{l'}{s}{l}{-1}{0}{1}.
\end{align}
For useful identities of the Woodhouse coefficients, the reader is referred to the Appendix of \citet{Woodhouse1980} and Appendix D.2.3 of \citet{Dahlen1998}.

%%%%%%%%%%%%%%%%%%%%%%%%%%%%%%%%%%%%%%%%%%%%%%%%%%%%%%%%%%%%%%%%%%%%%%%%%%%%%%%%%%%%%%%%%%%%%%%%%%%%%%%%%%%%%%%%%%%%%%%%%%%%%%%%%%%%%%%%%%%%%%%%%%%%
%%%%%%%%%%%%%%%%%%%%%%%%%%%%%%%%%%%%%%%%%%%%%%%%%%%%%%%%%%%%%%%%%%%%%%%%%%%%%%%%%%%%%%%%%%%%%%%%%%%%%%%%%%%%%%%%%%%%%%%%%%%%%%%%%%%%%%%%%%%%%%%%%%%%
\section{Perturbations in Structural Quantities}\label{app:sec:perturb}
In Section~\ref{sec:3} we found for the perturbations in the structural quantities:
\begin{align}
\delta\rho_{s,s\p}^{\lambda}(r) &= \frac{1}{g_0}\left[\Dr{\rho_0}\delta\phi_{s,s\p}^{\lambda}(r)+\mathcal{R}_{s,s\p,\lambda}(r)+\Dr{}\left(r\mathcal{S}_{s,s\p,\lambda}(r)\right)\right]\label{app:just:1},\\
\delta p_{s,s\p}^{\lambda}(r) &= -\rho_0\delta\phi_{s,s\p}^{\lambda}(r)-r\mathcal{S}_{s,s\p,\lambda}(r)\label{app:just:2},\\
\delta {c^2}_{s,s\p}^{\lambda}(r) &= \left[\left(\frac{\partial \ln\Gamma_1}{\partial\ln p}\right)_\rho+1\right]\frac{\delta p_{s,s\p}^{\lambda}(r)}{p_0}c_0^2+ \left[\left(\frac{\partial \ln\Gamma_1}{\partial\ln \rho}\right)_p-1\right]\frac{\delta \rho_{s,s\p}^{\lambda}(r)}{\rho_0}c_0^2 \label{app:just:3},
\end{align}
where $\delta\phi_{s,s\p}^{\lambda}(r)$ is found by integrating Equation~(\ref{eq:sweet}) numerically. Equation~(\ref{app:just:1}) can be obtained from Equation~(\ref{eq:sweet}) by using the Poisson equation
\begin{align}
\boldsymbol{\nabla}^2 \phi(r) = 4\pi G \rho(r)\label{app:eq:poisson}
\end{align}
to solve for the perturbation in the density $\delta\rho_{s,s\p}^{\lambda}(r)$. The derivation of Equation~(\ref{app:just:1}) can also be found in \citet{Mathis2004}.

Introducing perturbations $Q \rightarrow Q + \delta Q$ with $Q=p,\mathbf{g},\rho$ to the equation of hydrostatic support $\D p / \D r=-\mathbf{g}\rho$ yields
\begin{align}
\dr{\delta p_{s,s\p}^{\lambda}(r)} = -\mathbf{g}_0\delta \rho_{s,s\p}^{\lambda}(r)-\delta g_{s,s\p}^{\lambda}(r)\rho_0,\label{app:just:perturbhsp}
\end{align}
where the equilibrium equation has been subtracted and only terms linear in the perturbations are retained. The gravitational acceleration is expanded, like the structural quantities, according to Equation~(\ref{expand:perturbation}). The expansion coefficient of the aspherical perturbation to the gravitational acceleration is given by 
\begin{align}
\delta g_{s,s\p}^{\lambda}(r) = \dr{\delta \phi_{s,s\p}^{\lambda}(r) }-\frac{\mathcal{R}_{s,s\p,\lambda}(r)}{\rho_0},\label{app:just:g}
\end{align}
which is derived in \citet{Mathis2004}, Section 5.2. By inserting Equation~(\ref{app:just:1}) and (\ref{app:just:g}) into Equation~(\ref{app:just:perturbhsp}), we find
\begin{align}
\dr{\delta p_{s,s\p}^{\lambda}(r)} = -\Dr{\rho_0}\delta\phi_{s,s\p}^{\lambda}(r)-\Dr{}\left(r\mathcal{S}_{s,s\p,\lambda}(r)\right)-\dr{\delta \phi_{s,s\p}^{\lambda}(r) }\rho_0\label{app:just:perturbhsp2}.
\end{align}
Applying the inverse chain rule and eliminating the radial derivative yields Equation~(\ref{app:just:2}).

The kernel for the indirect effect as given in Equation~(90) of \citet{Lavely1992} includes perturbations of the bulk modulus $\kappa_0$ and density $\rho_0$. It can be transformed to perturbations in squared sound speed $c_0^2$ and density $\rho_0$. The transformation is obtained by introducing perturbations in the defining equation of the squared sound speed $c_0^2 = \kappa_0/\rho_0$, subtracting the unperturbed equation, and linearizing in the perturbations. This yields
\begin{align}
\delta \kappa_{s,s\p}^{\lambda}(r) = \rho_0(r)\delta {c^2}_{\,s,s\p}^{\lambda}(r) + c_0^2\delta \rho_{s,s\p}^{\lambda}(r),\label{app:transform:perturbations}
\end{align}
which was used to obtain Equation~(\ref{matrixelement_indirect}) from Equation~(90) in \citet{Lavely1992}. The aspherical perturbation to the bulk modulus can be expanded according to Equation~(\ref{expand:perturbation}). The perturbation in squared sound speed (Equation~(\ref{app:just:3})) is derived in \citet{Aerts2010}, Section 3.6. Compared to \citet{Aerts2010}, we neglect perturbations to the chemical abundances.

\section{Additional Figures}\label{app:sec:figures}

\begin{figure}[b!]
	\centering{
		\includegraphics[width=0.32\textwidth]{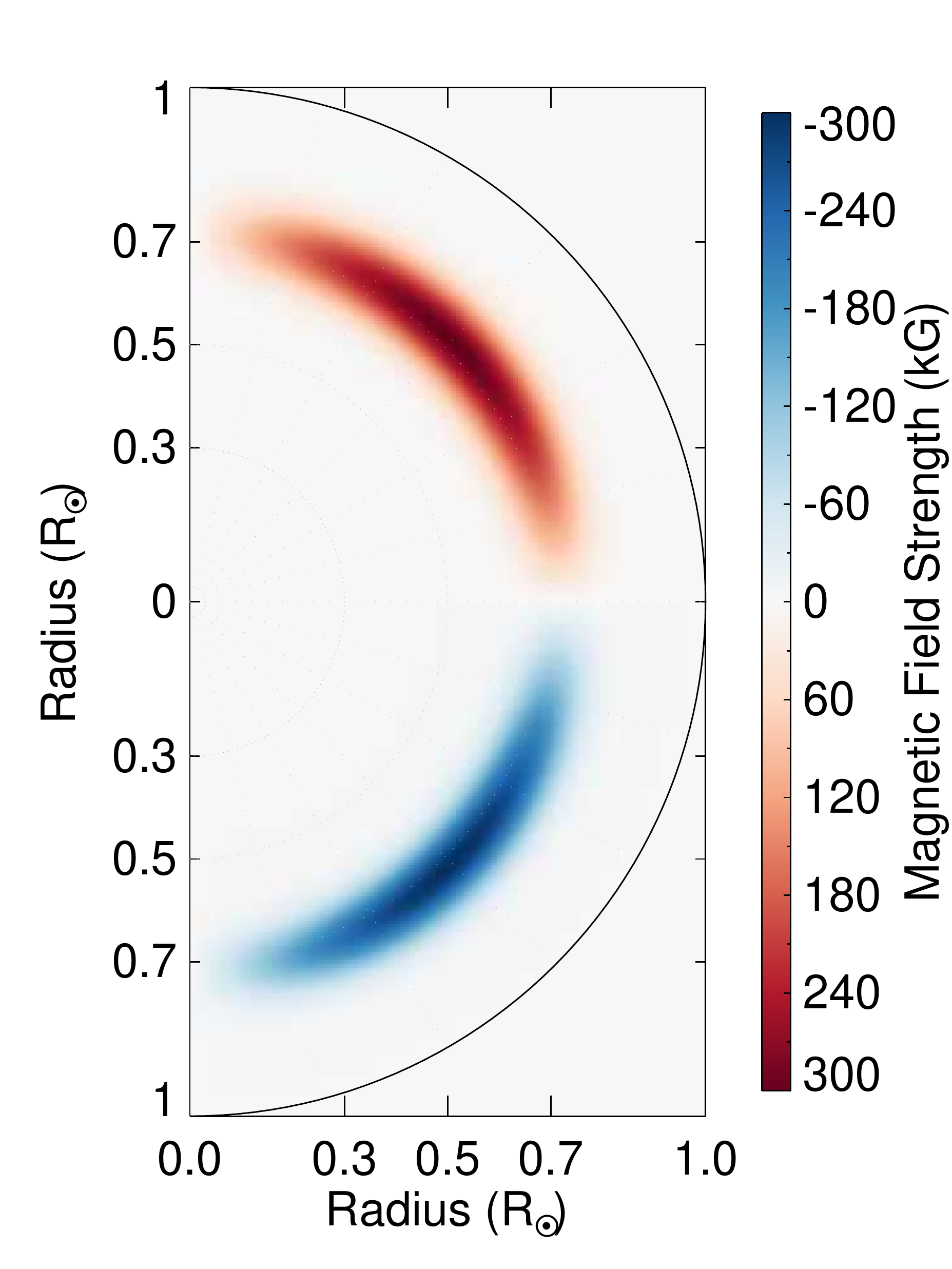}
		\caption{Visualization of magnetic field model C.\label{app:fig:1.1}}
		\includegraphics[width=0.47\textwidth]{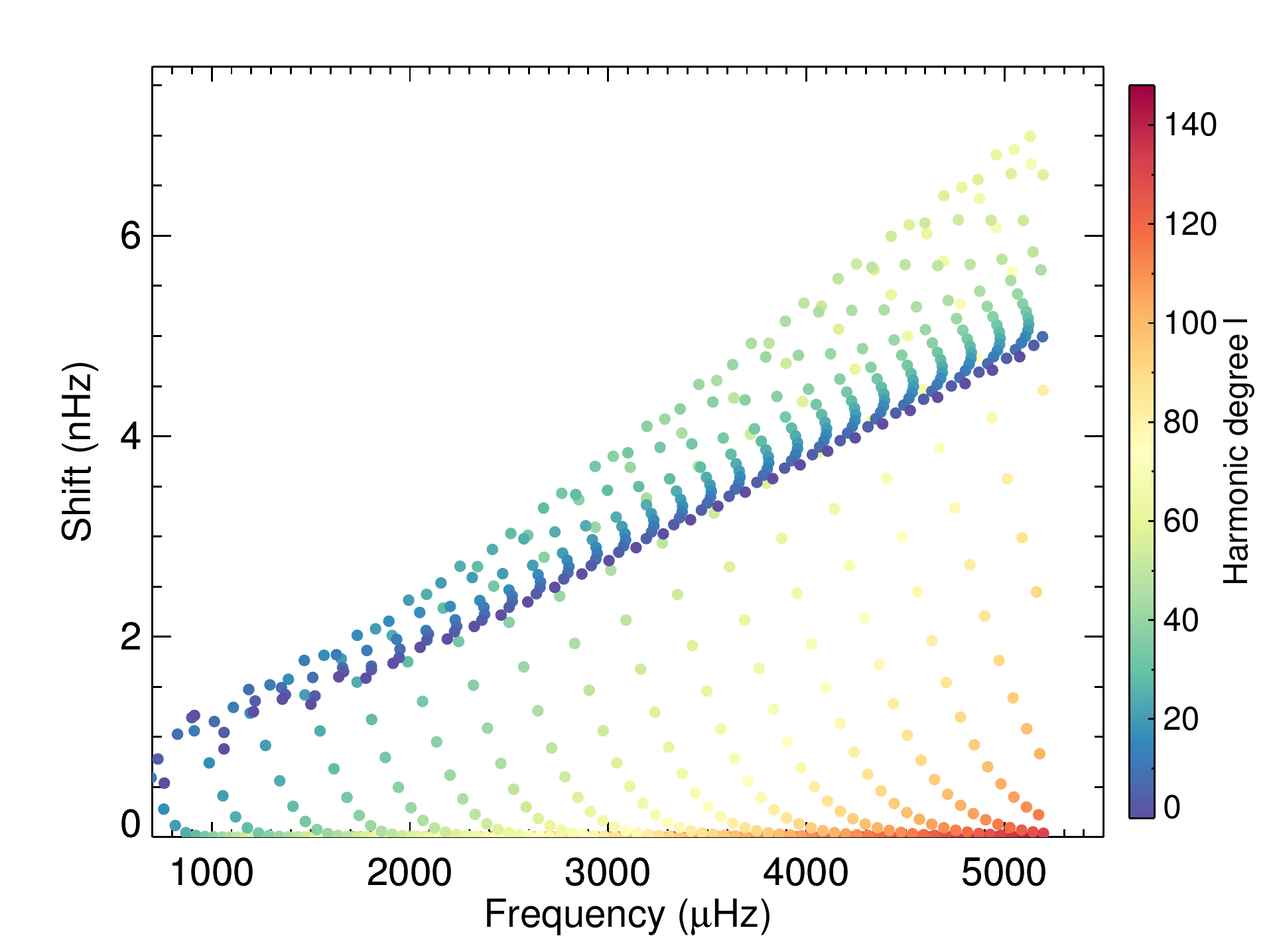}
		\includegraphics[width=0.47\textwidth]{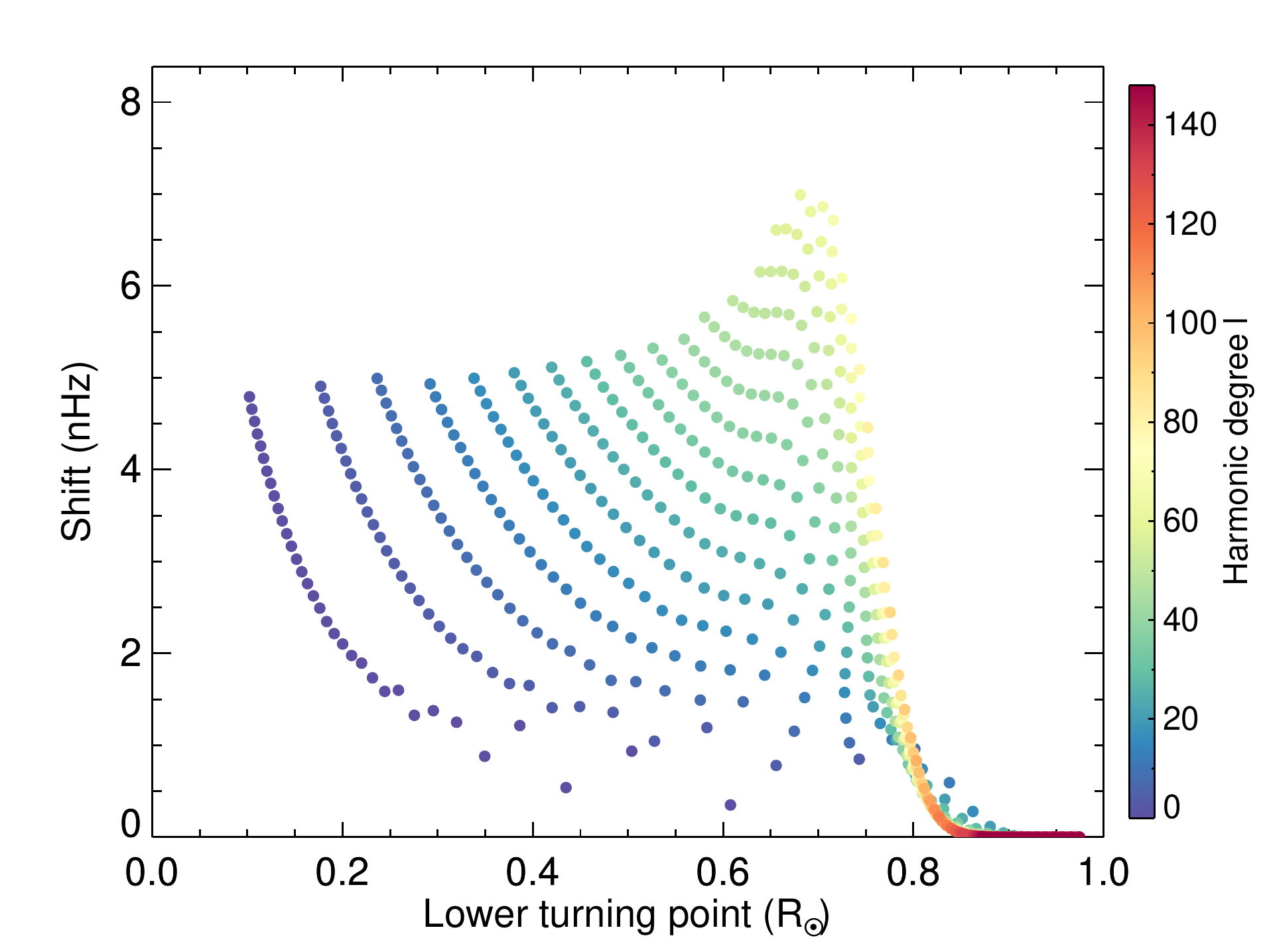}
		\includegraphics[width=0.47\textwidth]{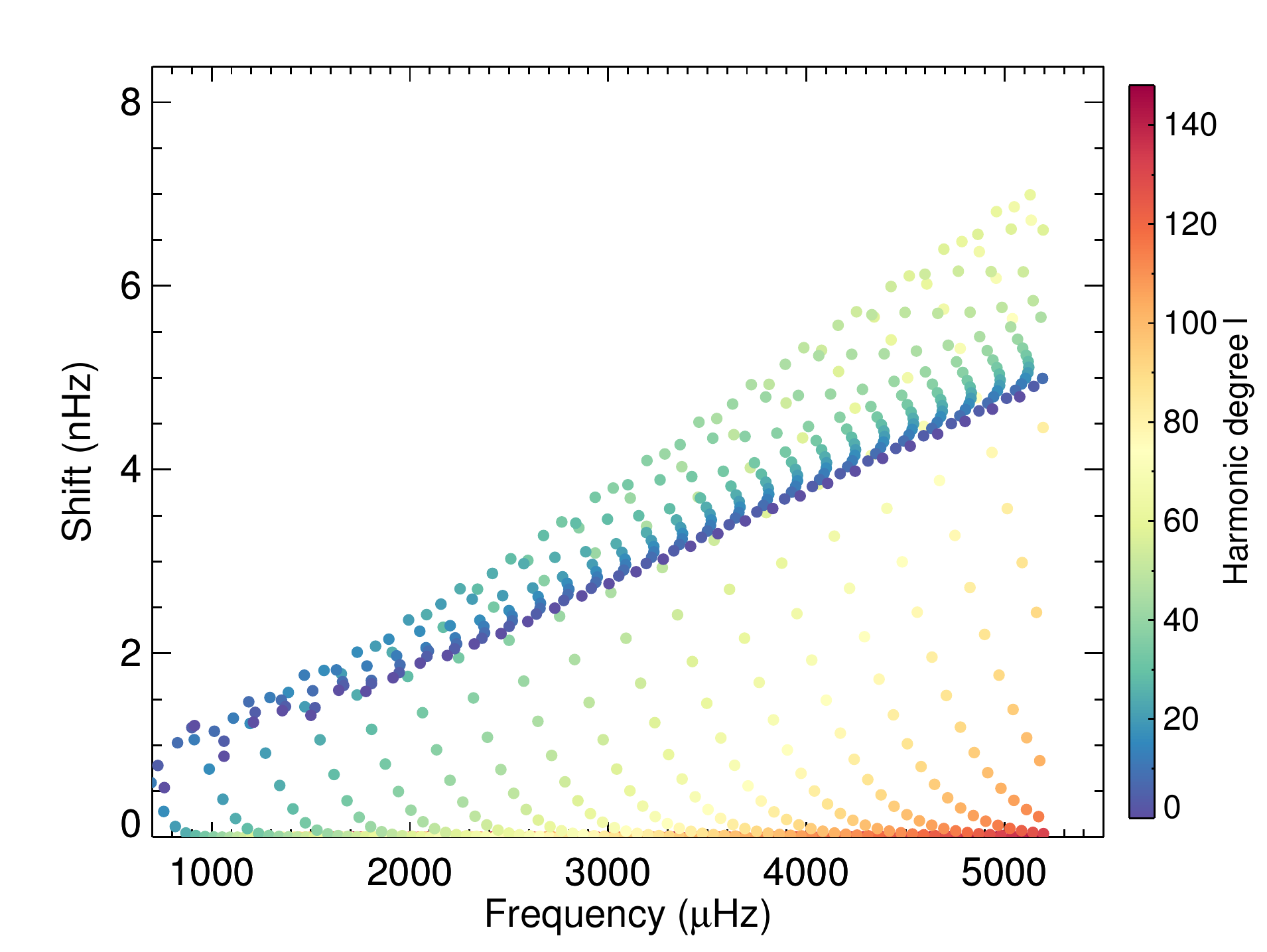}
		\includegraphics[width=0.47\textwidth]{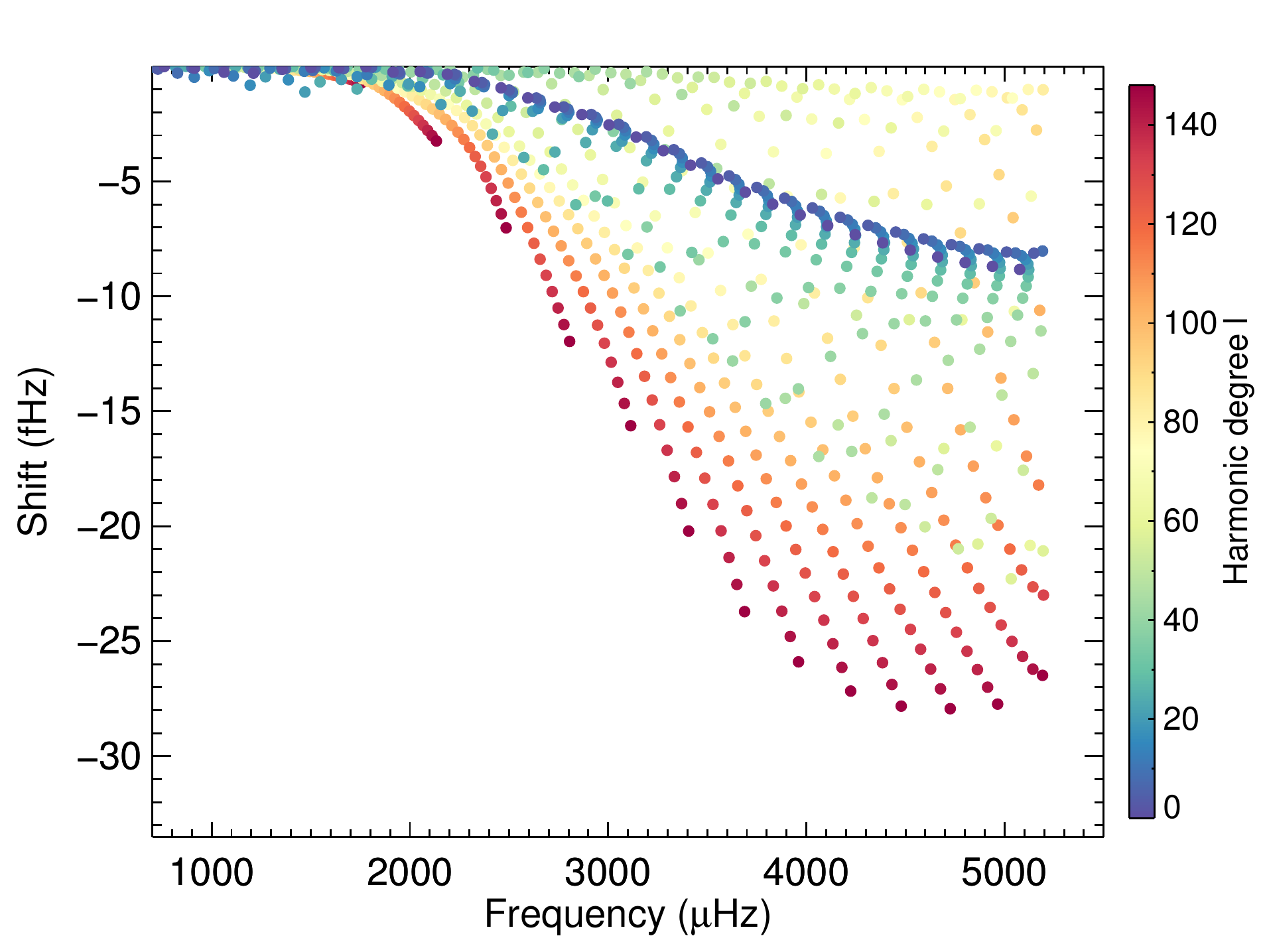}
		\caption{\textit{Top row:} multiplet frequency shifts for model C as a function of unperturbed mode frequency (\textit{left panel}) and as a function of lower turning point (\textit{right panel}). \textit{Bottom row:} multiplet frequency shifts for model C as a function of unperturbed mode frequency for the direct and indirect effect in the left and right panels, respectively. Notice the different magnitudes of the shifts in the left and right panels. Every fourth harmonic degree is shown in all plots.}
		\label{app:fig:1.2}}
\end{figure} 

\begin{figure}
	\centering{
		\includegraphics[width=0.32\textwidth]{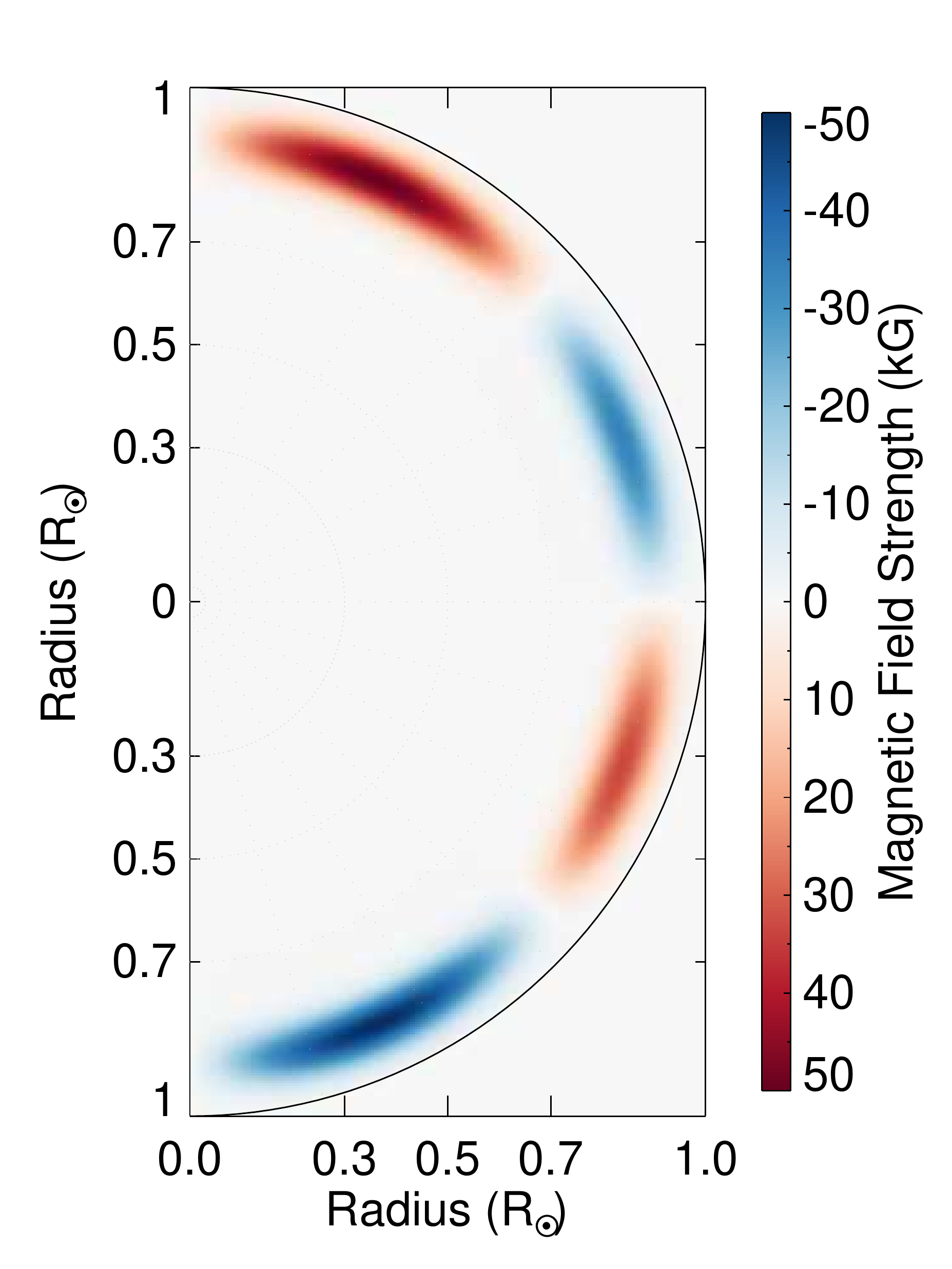}
		\caption{Visualization of magnetic field model D.\label{app:fig:2.1}}
		\includegraphics[width=0.47\textwidth]{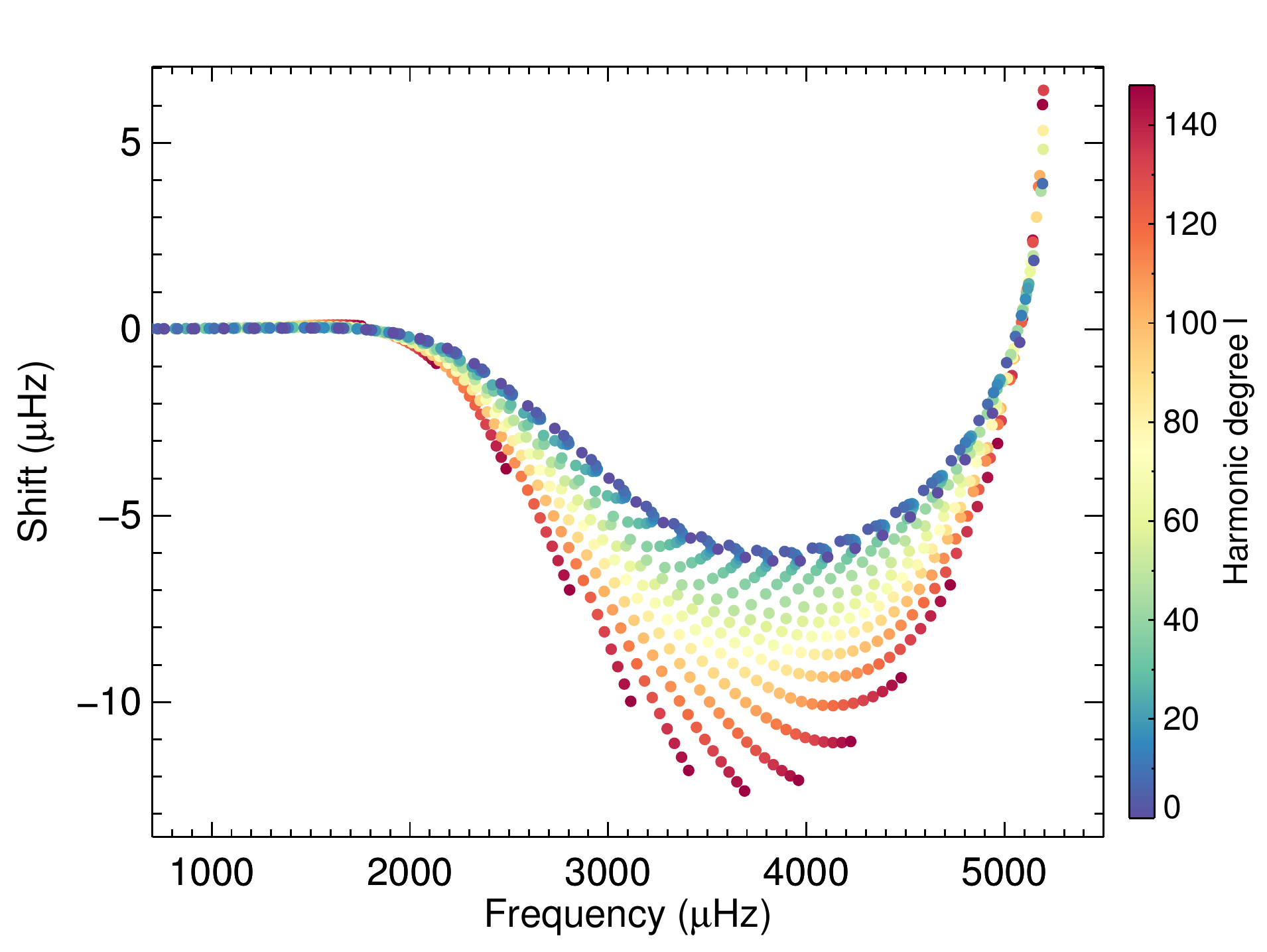}
		\includegraphics[width=0.47\textwidth]{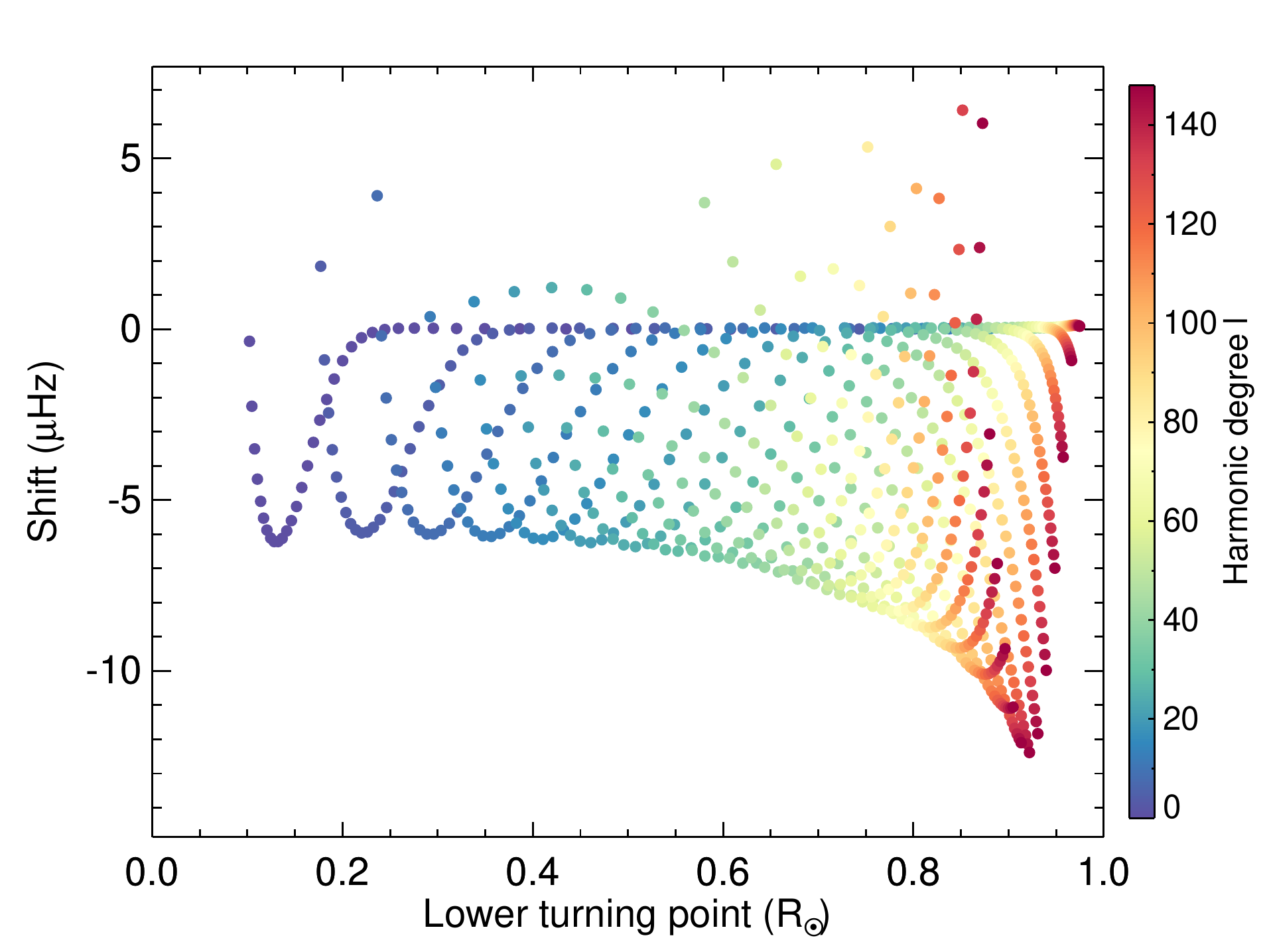}
		\includegraphics[width=0.47\textwidth]{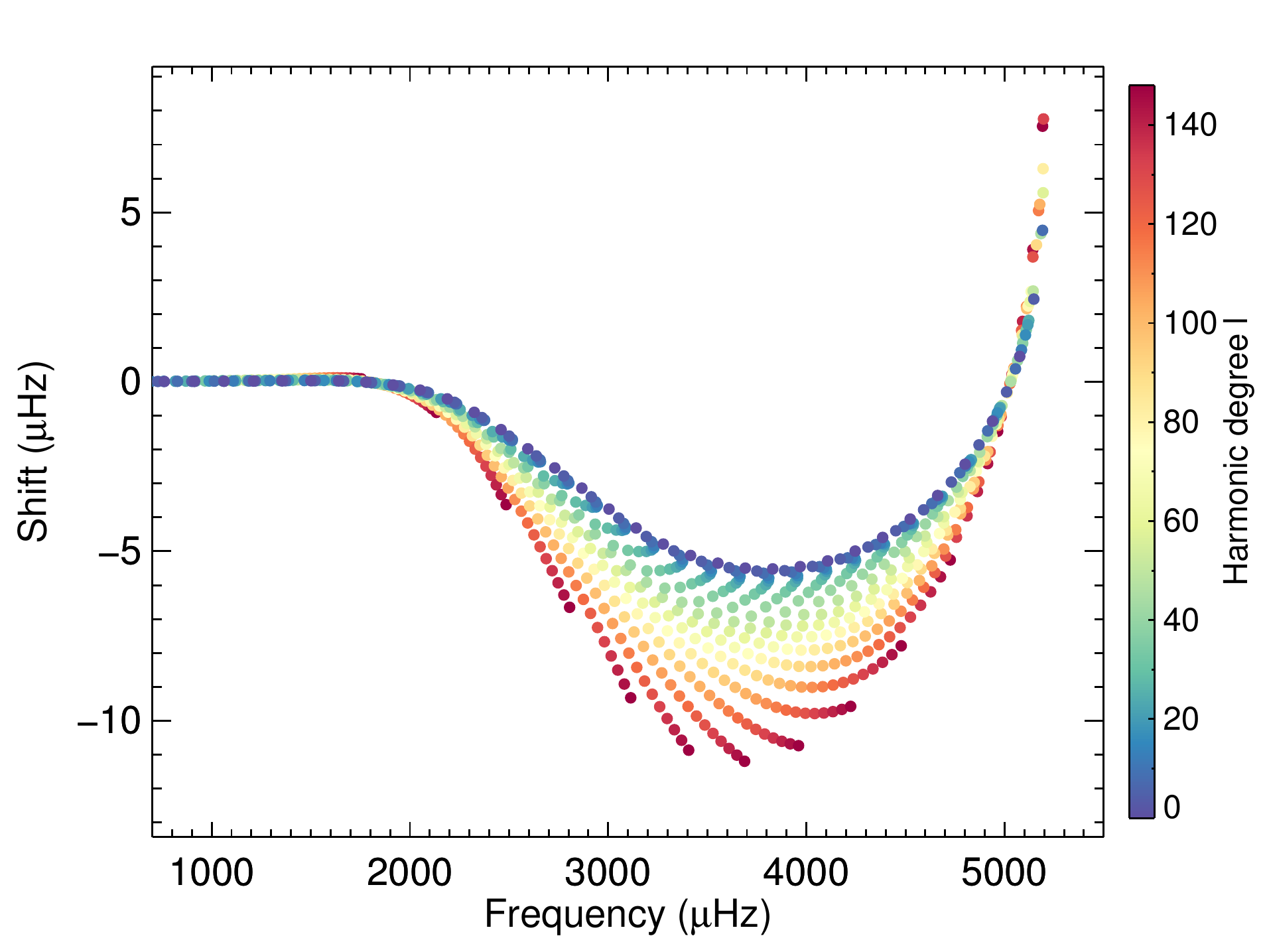}
		\includegraphics[width=0.47\textwidth]{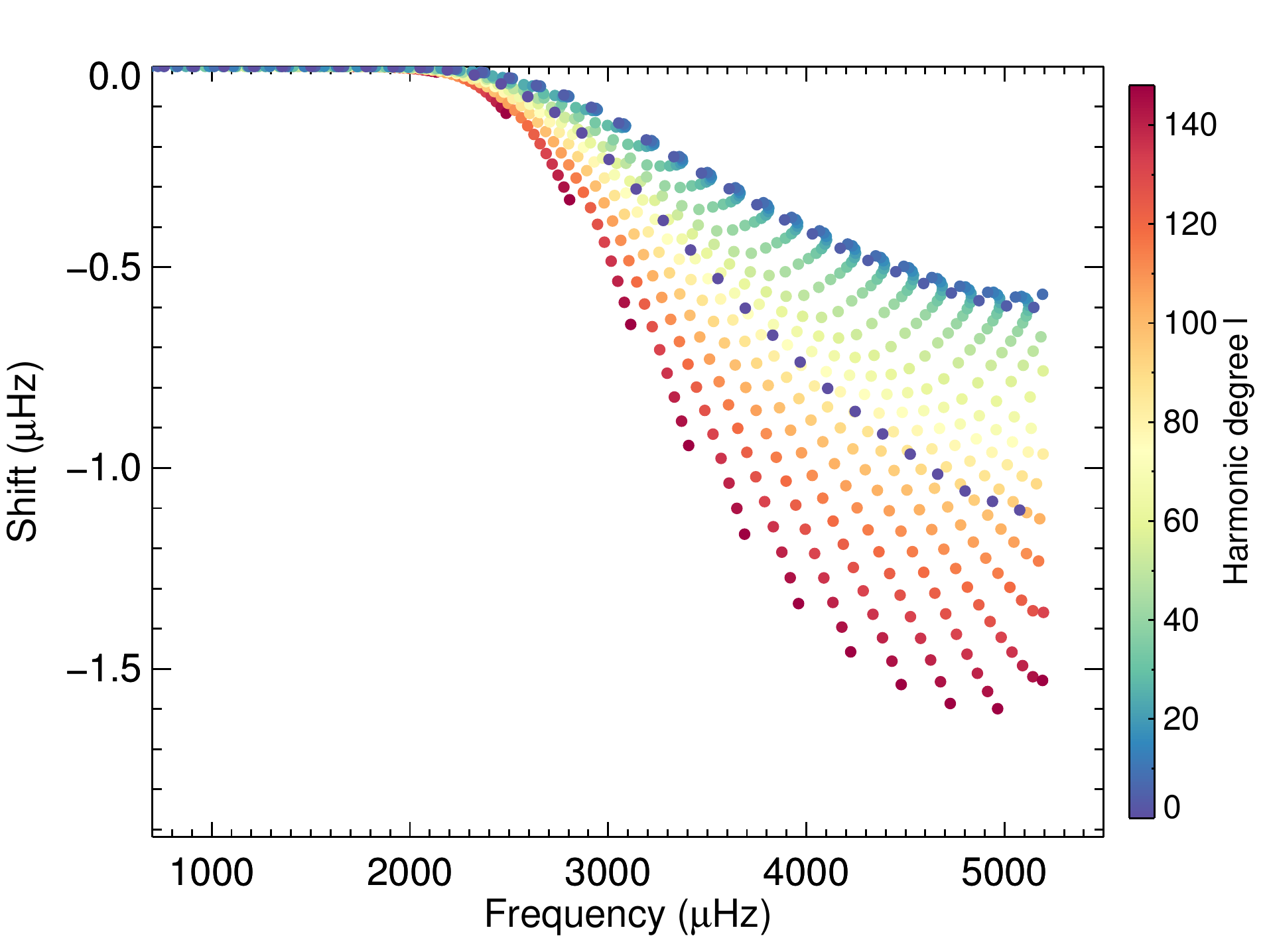}
		\caption{\textit{Top row:} multiplet frequency shifts for model D as a function of unperturbed mode frequency (left panel) and as a function of lower turning point (right panel). \textit{Bottom row:} multiplet frequency shifts for model D as a function of unperturbed mode frequency for the direct and indirect effect in the left and right panels, respectively. Every fourth harmonic degree is shown in all plots.}
		\label{app:fig:2.2}}
\end{figure} 

\begin{figure}
	\centering{
		\includegraphics[width=0.32\textwidth]{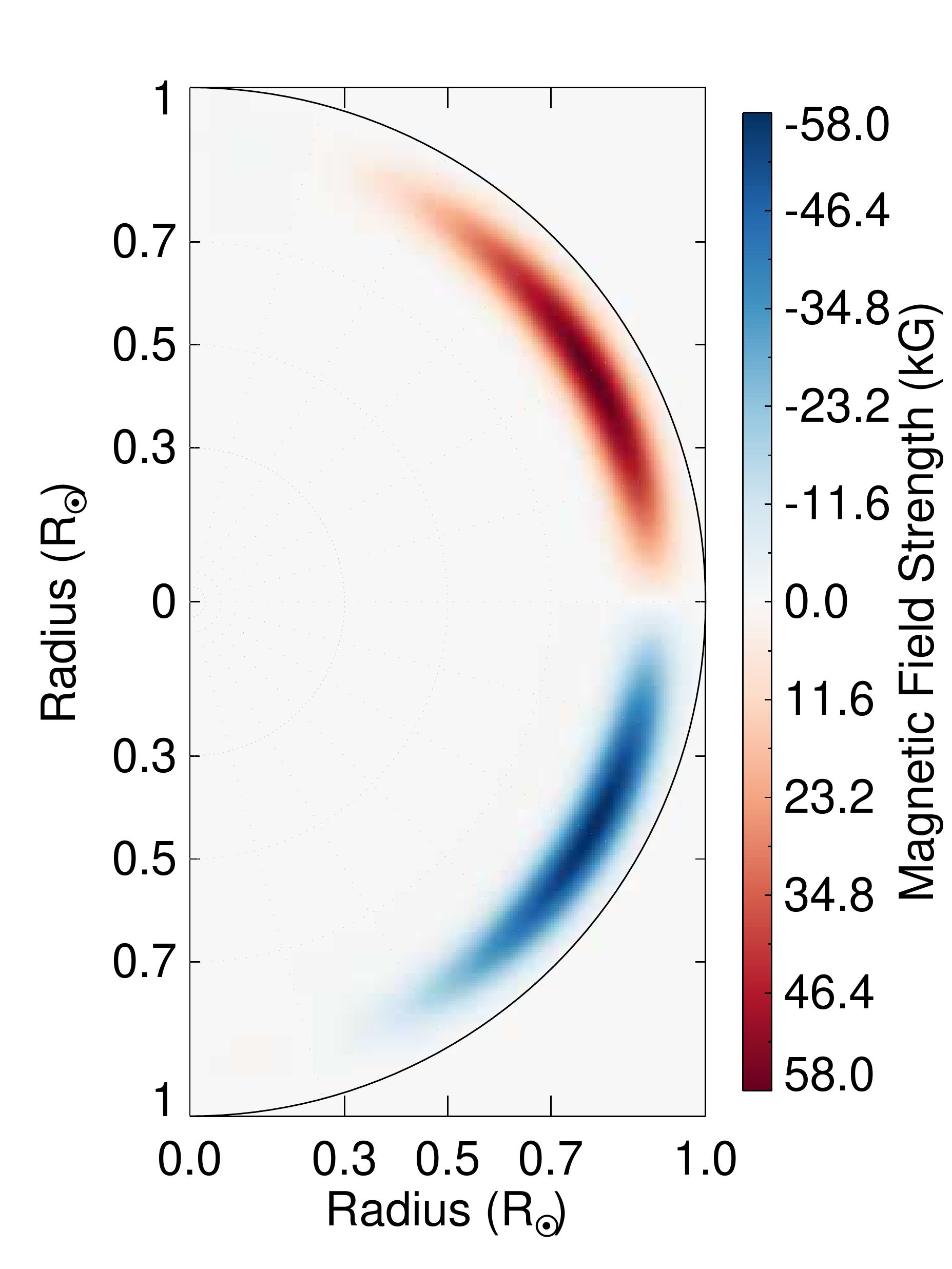}
		\caption{Visualization of magnetic field model E.\label{app:fig:3.1}}
		\includegraphics[width=0.47\textwidth]{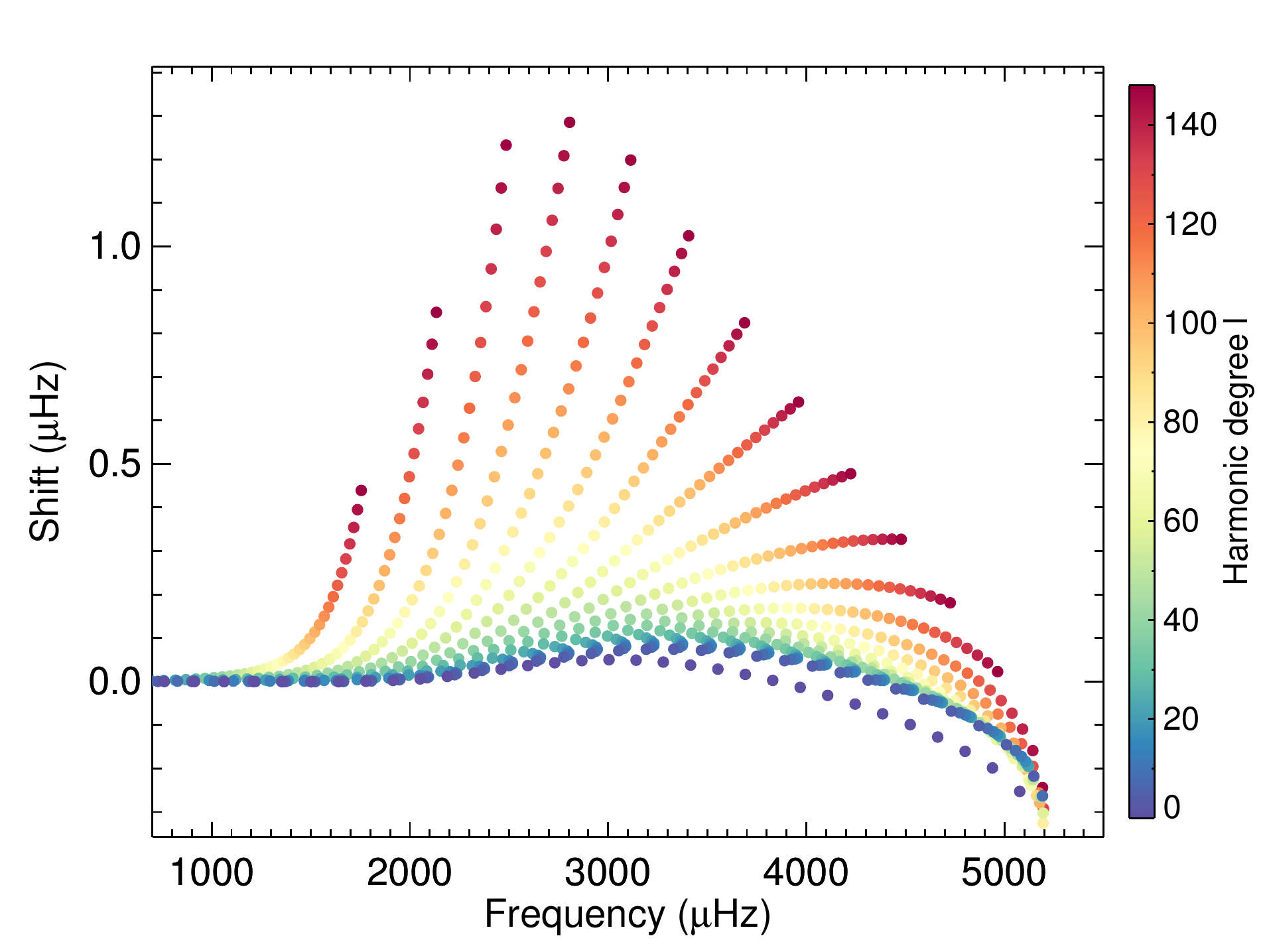}
		\includegraphics[width=0.47\textwidth]{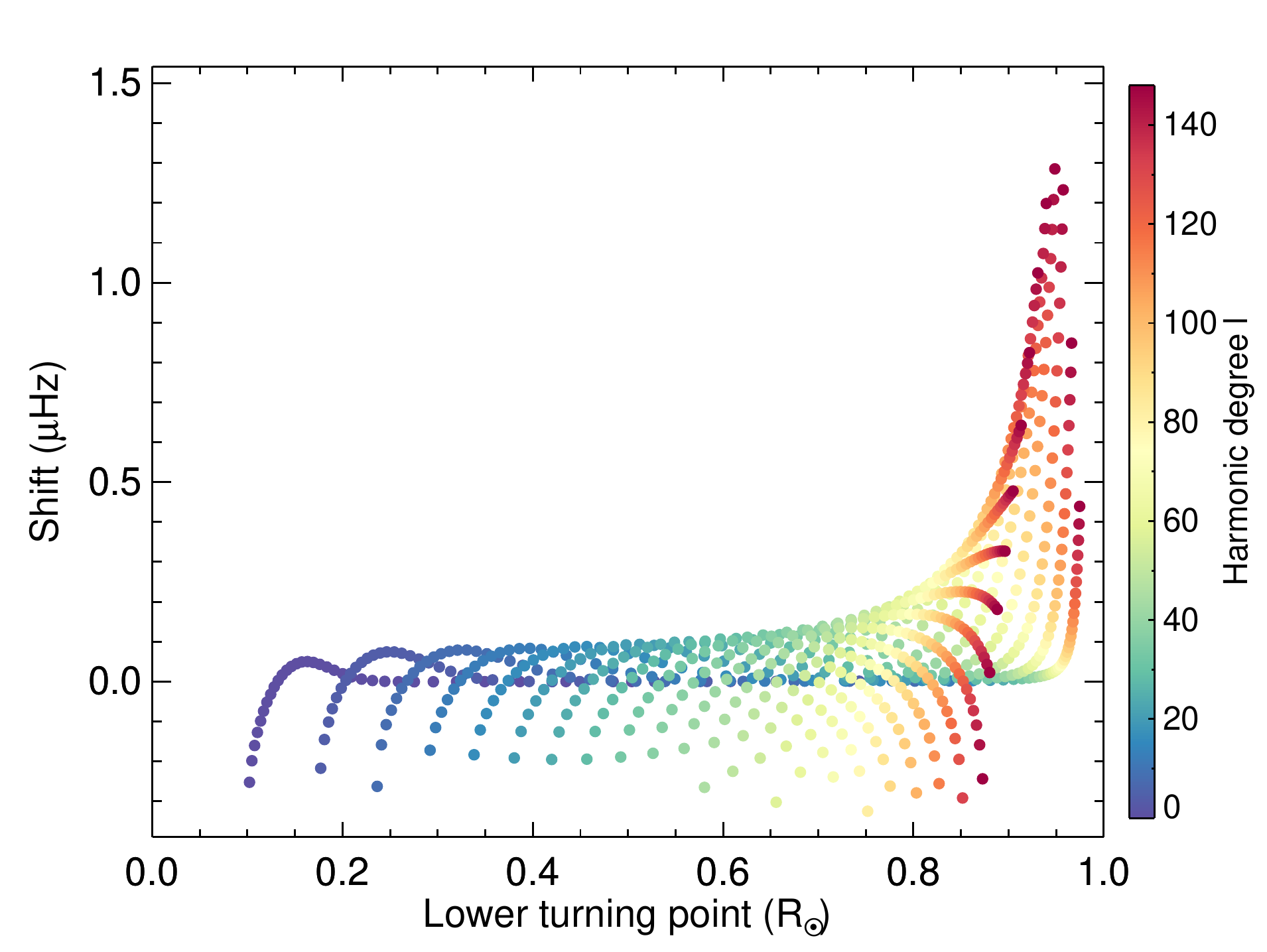}
		\includegraphics[width=0.47\textwidth]{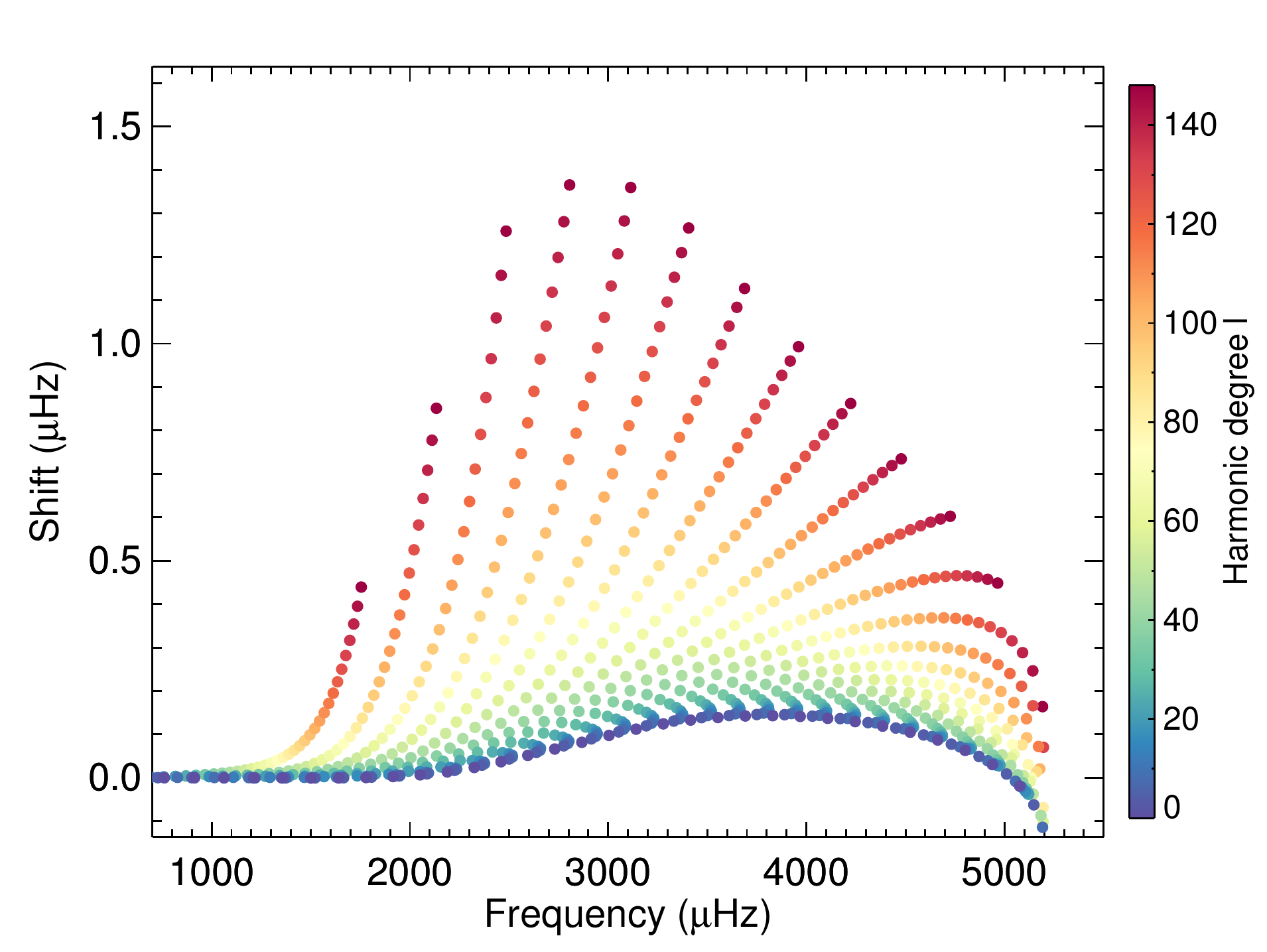}
		\includegraphics[width=0.47\textwidth]{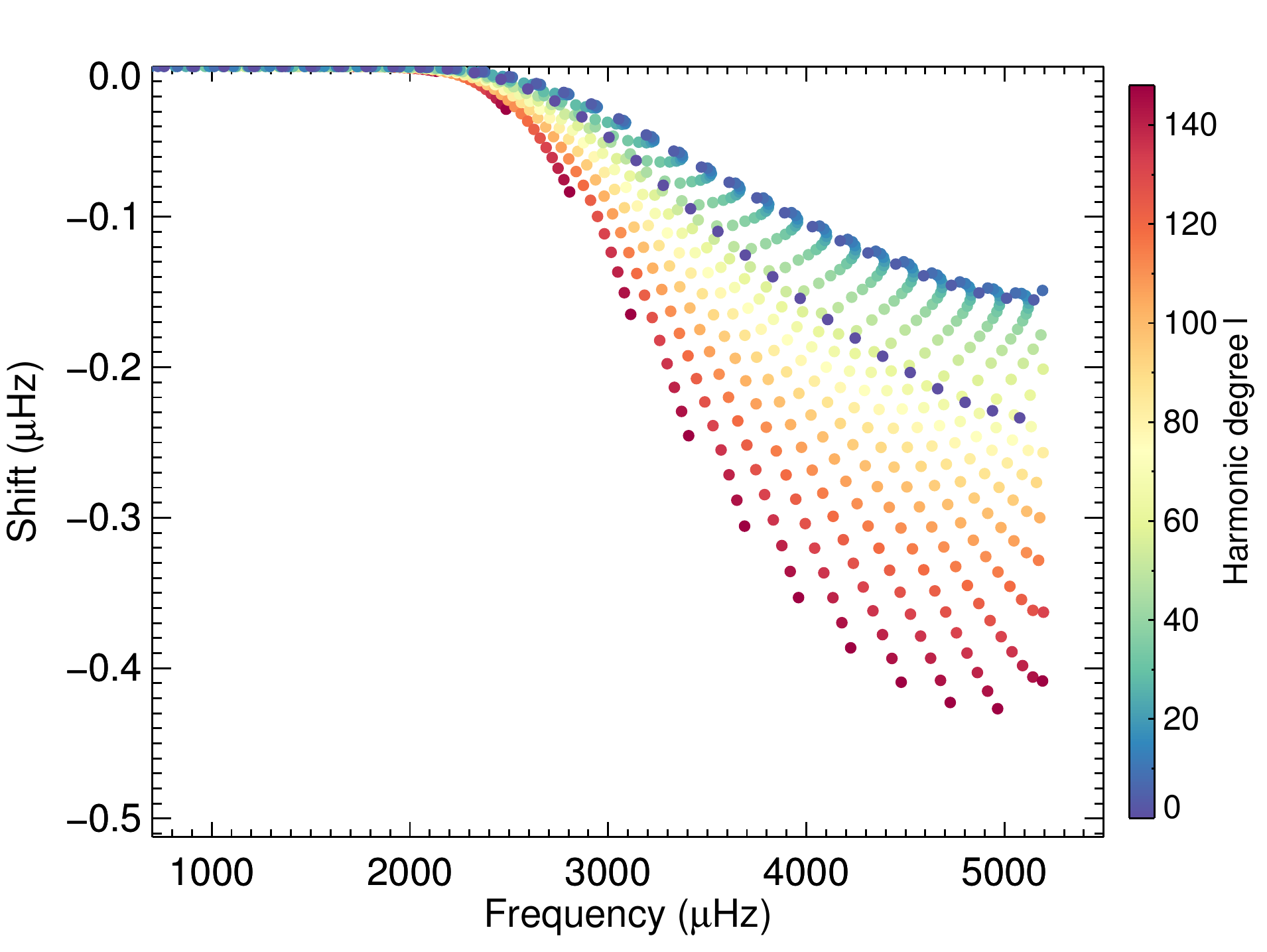}
		\caption{\textit{Top row:} multiplet frequency shifts for model E as a function of unperturbed mode frequency (left panel) and as a function of lower turning point (right panel). \textit{Bottom row:} multiplet frequency shifts for model E as a function of unperturbed mode frequency for the direct and indirect effect in the left and right panels, respectively. Every fourth harmonic degree is shown in all plots.}
		\label{app:fig:3.2}}
\end{figure} 

\begin{figure}
	\centering{
		\includegraphics[width=0.32\textwidth]{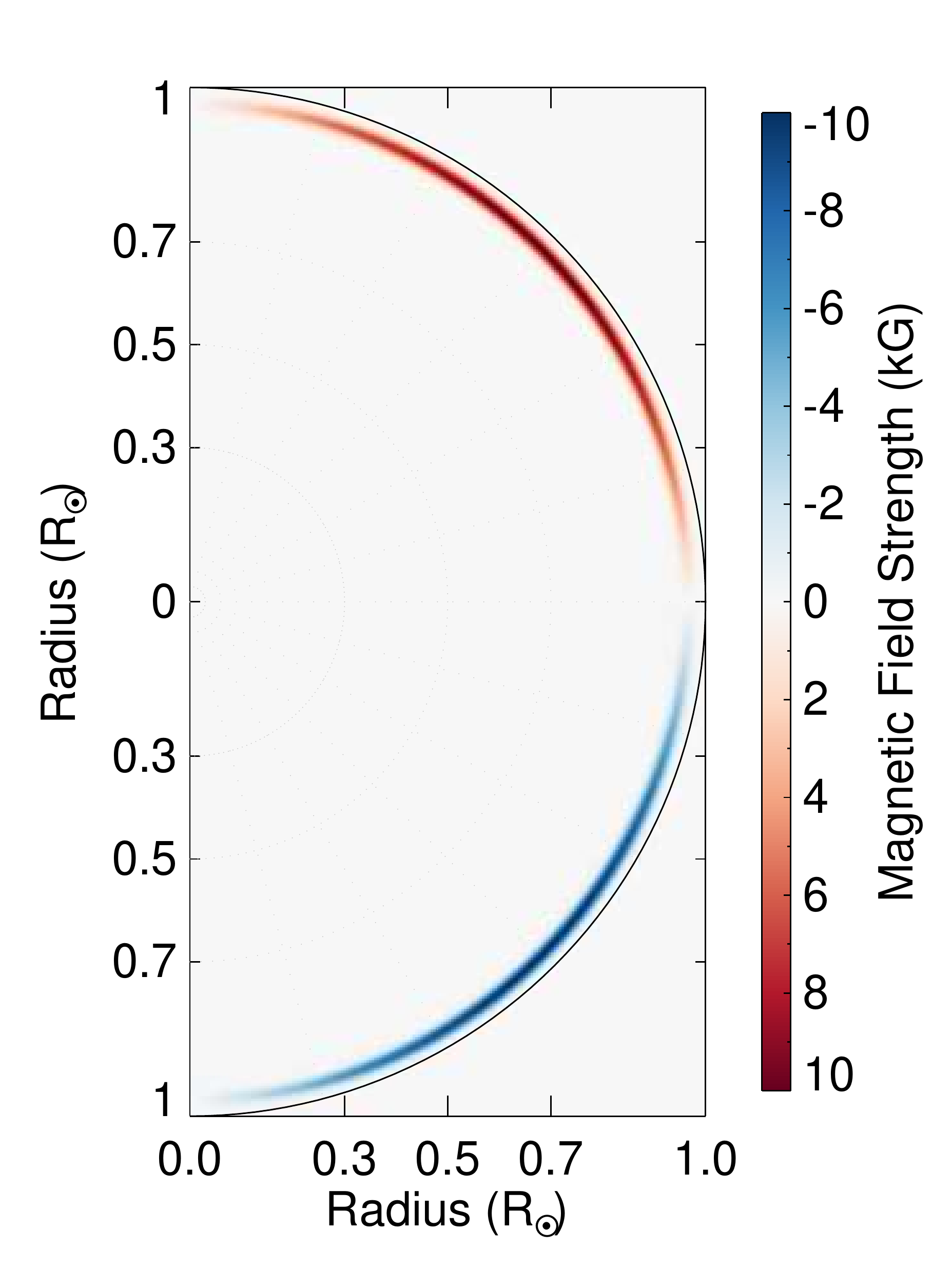}
		\caption{Visualization of magnetic field model F.\label{app:fig:4.1}}
		\includegraphics[width=0.47\textwidth]{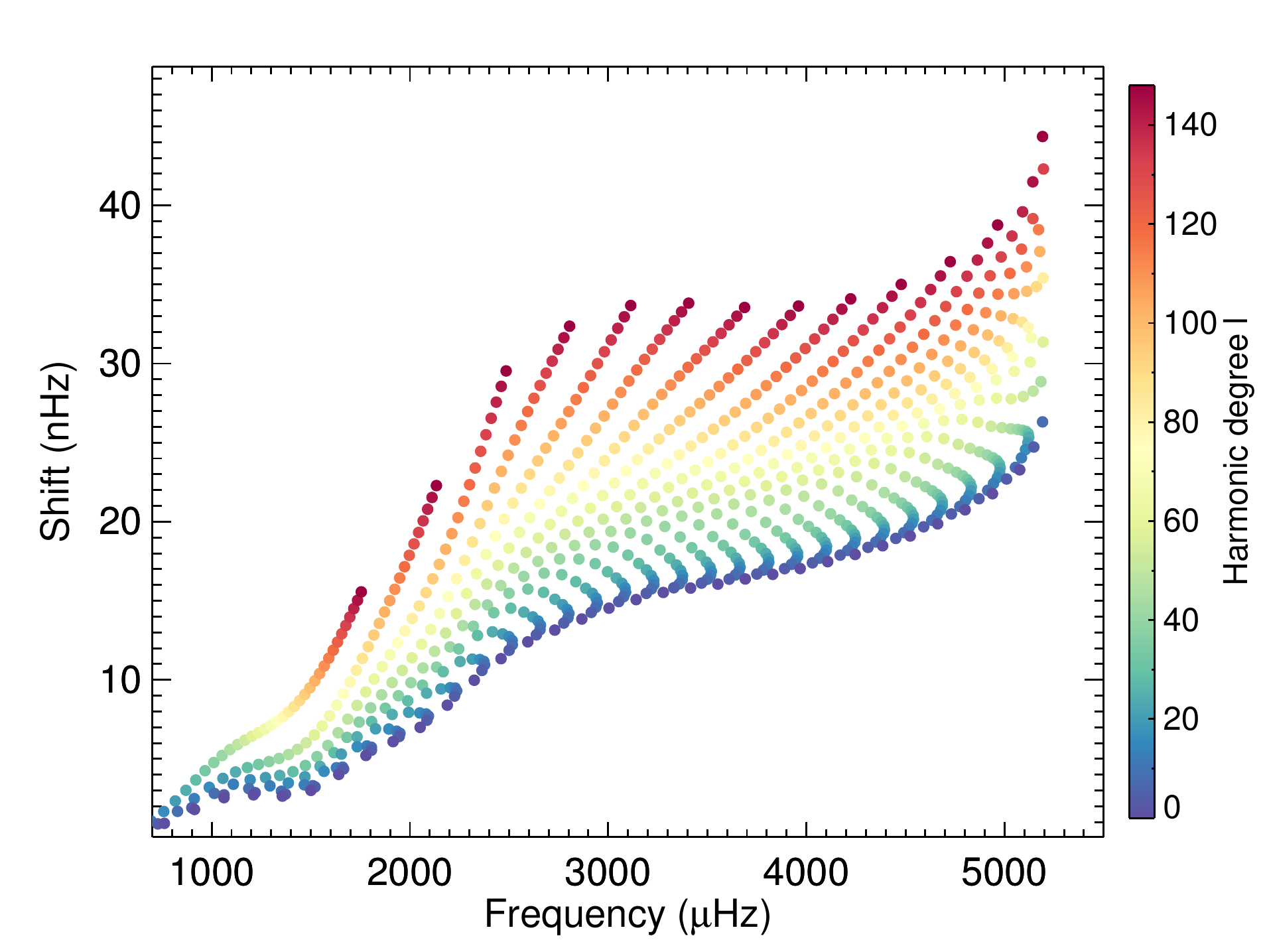}
		\includegraphics[width=0.47\textwidth]{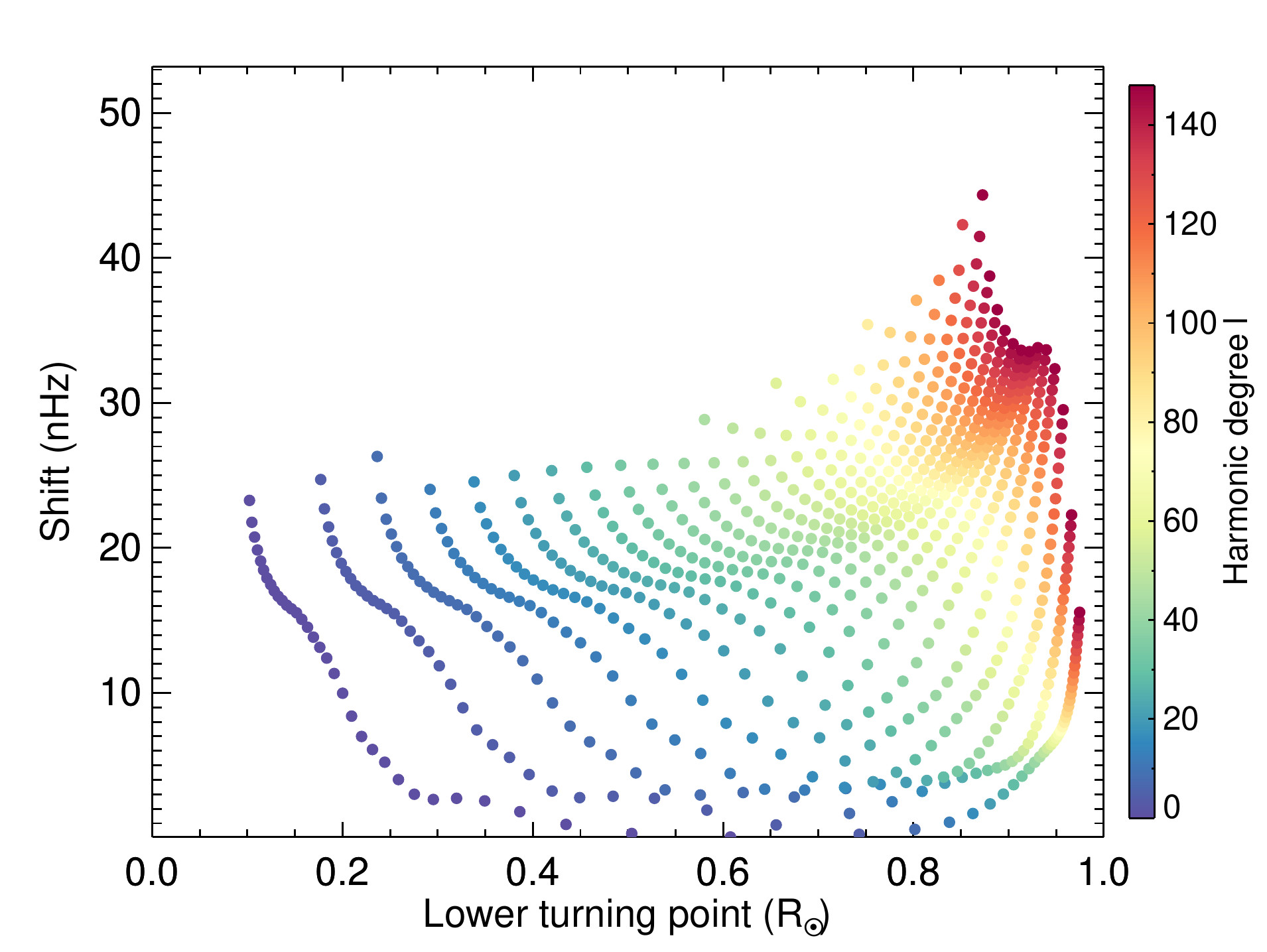}
		\includegraphics[width=0.47\textwidth]{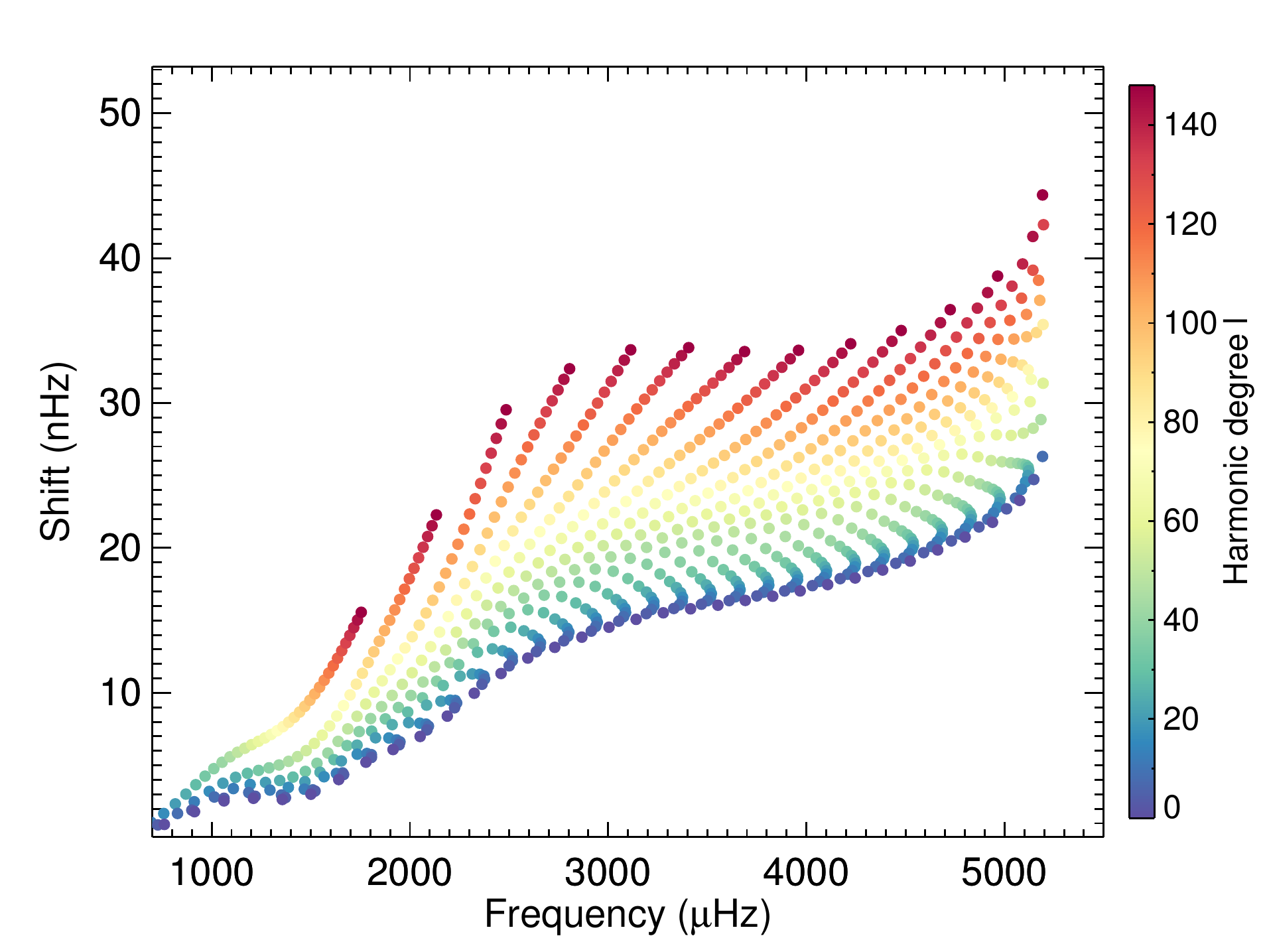}
		\includegraphics[width=0.47\textwidth]{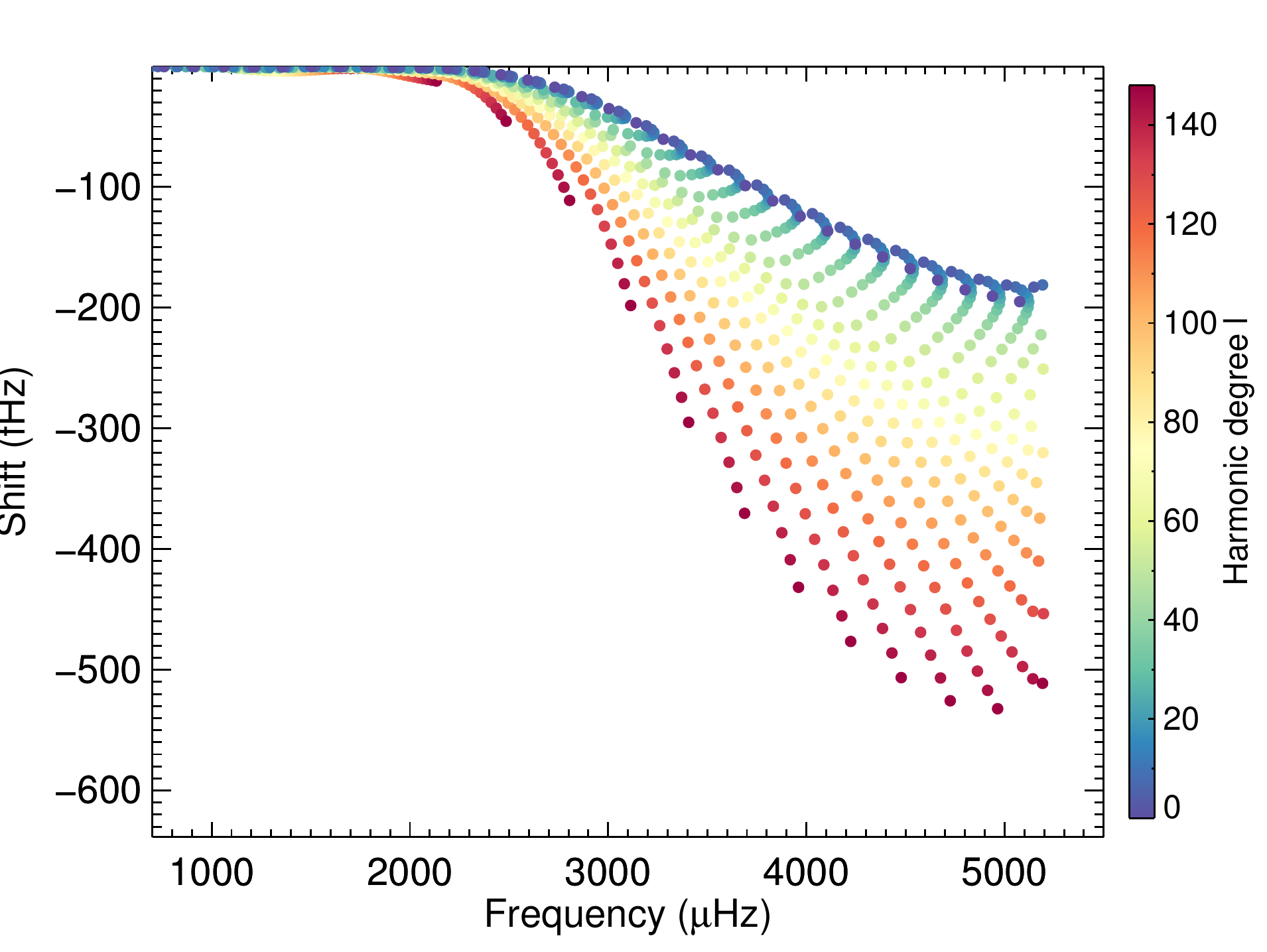}
		\caption{\textit{Top row:} multiplet frequency shifts for model F as a function of unperturbed mode frequency (left panel) and as a function of lower turning point (right panel). \textit{Bottom row:} multiplet frequency shifts for model F as a function of unperturbed mode frequency for the direct and indirect effect in the left and right panels, respectively. Notice the different magnitudes of the shifts in the left and right panels. Every fourth harmonic degree is shown in all plots. }
		\label{app:fig:4.2}}
\end{figure}

\clearpage

\bibliographystyle{astronurl_ads}
\bibliography{references}

\begin{thebibliography}{}

\bibitem[\protect\astroncite{Aerts et~al.}{2010}]{Aerts2010}
Aerts, C., Christensen-Dalsgaard, J., and Kurtz, D.~W.: 2010,
\newblock {\em {Asteroseismology}},
\newblock Springer, Dordrecht, 1 edition,
\newblock
  {\small[\href{http://adsabs.harvard.edu/abs/2010aste.book.....A}{URL}]},
\newblock {\small[\href{http://dx.doi.org/10.1007/978-1-4020-5803-5}{DOI}]}

\bibitem[\protect\astroncite{Antia et~al.}{2000}]{Antia2000}
Antia, H.~M., Chitre, S.~M., and Thompson, M.~J.: 2000,
\newblock {\em Astronomy \& Astrophysics} {\bf 360}, 335,
\newblock
  {\small[\href{http://adsabs.harvard.edu/abs/2000A&A...360..335A}{URL}]}

\bibitem[\protect\astroncite{Arlt et~al.}{2007}]{Arlt2007}
Arlt, R., Sule, A., and R{\"{u}}diger, G.: 2007,
\newblock {\em Astronomy \& Astrophysics} {\bf 461(1)}, 295,
\newblock
  {\small[\href{http://adsabs.harvard.edu/abs/2007A%26A...461..295A}{URL}]},
\newblock {\small[\href{http://dx.doi.org/10.1051/0004-6361:20065192}{DOI}]}

\bibitem[\protect\astroncite{Baldner et~al.}{2009}]{Baldner2009}
Baldner, C.~S., Antia, H.~M., Basu, S., and Larson, T.~P.: 2009,
\newblock {\em The Astrophysical Journal} {\bf 705(2)}, 1704,
\newblock
  {\small[\href{http://adsabs.harvard.edu/abs/2009ApJ...705.1704B}{URL}]},
\newblock {\small[\href{http://dx.doi.org/10.1088/0004-637X/705/2/1704}{DOI}]}

\bibitem[\protect\astroncite{Ball and Gizon}{2014}]{Ball2014}
Ball, W.~H. and Gizon, L.: 2014,
\newblock {\em Astronomy \& Astrophysics} {\bf 568}, A123,
\newblock
  {\small[\href{http://adsabs.harvard.edu/abs/2014A%26A...568A.123B}{URL}]},
\newblock {\small[\href{http://dx.doi.org/10.1051/0004-6361/201424325}{DOI}]}

\bibitem[\protect\astroncite{Basu}{1997}]{Basu1997}
Basu, S.: 1997,
\newblock {\em Monthly Notices of the Royal Astronomical Society} {\bf 288(3)},
  572,
\newblock
  {\small[\href{http://adsabs.harvard.edu/abs/1997MNRAS.288..572B}{URL}]},
\newblock {\small[\href{http://dx.doi.org/10.1093/mnras/288.3.572}{DOI}]}

\bibitem[\protect\astroncite{Basu}{2016}]{Basu2016}
Basu, S.: 2016,
\newblock {\em Living Reviews in Solar Physics} {\bf 13}, 2,
\newblock
  {\small[\href{http://adsabs.harvard.edu/abs/2016LRSP...13....2B}{URL}]},
\newblock {\small[\href{http://dx.doi.org/10.1007/s41116-016-0003-4}{DOI}]}

\bibitem[\protect\astroncite{Basu et~al.}{2012}]{Basu2012}
Basu, S., Broomhall, A.-M., Chaplin, W.~J., and Elsworth, Y.: 2012,
\newblock {\em The Astrophysical Journal} {\bf 758(1)}, 43,
\newblock
  {\small[\href{http://adsabs.harvard.edu/abs/2012ApJ...758...43B}{URL}]},
\newblock {\small[\href{http://dx.doi.org/10.1088/0004-637X/758/1/43}{DOI}]}

\bibitem[\protect\astroncite{Broomhall}{2017}]{Broomhall2017}
Broomhall, A.-M.: 2017,
\newblock {\em Solar Physics} {\bf 292}, 67,
\newblock
  {\small[\href{http://adsabs.harvard.edu/abs/2017SoPh..292...67B}{URL}]},
\newblock {\small[\href{http://dx.doi.org/10.1007/s11207-017-1068-5}{DOI}]}

\bibitem[\protect\astroncite{Charbonneau}{2010}]{Charbonneau2010}
Charbonneau, P.: 2010,
\newblock {\em Living Reviews in Solar Physics} {\bf 7}, 3,
\newblock
  {\small[\href{http://adsabs.harvard.edu/abs/2010LRSP....7....3C}{URL}]},
\newblock {\small[\href{http://dx.doi.org/10.12942/lrsp-2010-3}{DOI}]}

\bibitem[\protect\astroncite{Christensen-Dalsgaard}{2008}]{Christensen-Dalsgaard2008}
Christensen-Dalsgaard, J.: 2008,
\newblock {\em Astrophysics and Space Science} {\bf 316}, 113,
\newblock
  {\small[\href{http://adsabs.harvard.edu/abs/2008Ap&SS.316..113C}{URL}]},
\newblock {\small[\href{http://dx.doi.org/10.1007/s10509-007-9689-z}{DOI}]}

\bibitem[\protect\astroncite{Christensen-Dalsgaard
  et~al.}{1996}]{Christensen-Dalsgaard1996}
Christensen-Dalsgaard, J., D{\"{a}}ppen, W., Ajukov, S.~V., Anderson, E.~R.,
  Antia, H.~M., Basu, S., Baturin, V.~A., Berthomieu, G., Chaboyer, B., Chitre,
  S.~M., Cox, A.~N., Demarque, P., Donatowicz, J., Dziembowski, W.~A., Gabriel,
  M., Gough, D.~O., Guenther, D.~B., Guzik, J.~A., Harvey, J.~W., Hill, F.,
  Houdek, G., Iglesias, C.~A., Kosovichev, A.~G., Leibacher, J.~W., Morel, P.,
  Proffitt, C.~R., Provost, J., Reiter, J., Rhodes, E.~J., Rogers, F.~J.,
  Roxburgh, I.~W., Thompson, M.~J., and Ulrich, R.~K.: 1996,
\newblock {\em Science} {\bf 272(5266)}, 1286,
\newblock
  {\small[\href{http://adsabs.harvard.edu/abs/1996Sci...272.1286C}{URL}]},
\newblock {\small[\href{http://dx.doi.org/10.1126/science.272.5266.1286}{DOI}]}

\bibitem[\protect\astroncite{Cowling}{1941}]{Cowling1941}
Cowling, T.~G.: 1941,
\newblock {\em Monthly Notices of the Royal Astronomical Society} {\bf 101(8)},
  367,
\newblock
  {\small[\href{http://adsabs.harvard.edu/abs/1941MNRAS.101..367C}{URL}]},
\newblock {\small[\href{http://dx.doi.org/10.1093/mnras/101.8.367}{DOI}]}

\bibitem[\protect\astroncite{Dahlen and Tromp}{1998}]{Dahlen1998}
Dahlen, F.~A. and Tromp, J.: 1998,
\newblock {\em {Theoretical Global Seismology}},
\newblock Princeton University Press, Princeton, New Jersey, 1 edition

\bibitem[\protect\astroncite{Duez et~al.}{2008}]{Duez2008}
Duez, V., Mathis, S., Brun, A.~S., and Turck-Chi{\`{e}}ze, S.: 2008,
\newblock in K.~G. Strassmeier, A.~G. Kosovichev, and J.~E. Beckman (eds.),
  {\em Cosmic Magnetic Fields: From Planets, to Stars and Galaxies}, Vol.~4 of
  {\em Proceedings of the International Astronomical Union}, pp 177--184,
\newblock
  {\small[\href{http://adsabs.harvard.edu/abs/2009IAUS..259..177D}{URL}]},
\newblock {\small[\href{http://dx.doi.org/10.1017/S1743921309030427}{DOI}]}

\bibitem[\protect\astroncite{Duez et~al.}{2010}]{Duez2010}
Duez, V., Mathis, S., and Turck-Chi{\`{e}}ze, S.: 2010,
\newblock {\em Monthly Notices of the Royal Astronomical Society} {\bf 402(1)},
  271,
\newblock
  {\small[\href{http://adsabs.harvard.edu/abs/2010MNRAS.402..271D}{URL}]},
\newblock
  {\small[\href{http://dx.doi.org/10.1111/j.1365-2966.2009.15955.x}{DOI}]}

\bibitem[\protect\astroncite{Dziembowski and Goode}{2005}]{Dziembowski2005}
Dziembowski, W.~A. and Goode, P.~R.: 2005,
\newblock {\em The Astrophysical Journal} {\bf 625(1)}, 548,
\newblock
  {\small[\href{http://adsabs.harvard.edu/abs/2005ApJ...625..548D}{URL}]},
\newblock {\small[\href{http://dx.doi.org/10.1086/429712}{DOI}]}

\bibitem[\protect\astroncite{Edmonds}{1960}]{Edmonds}
Edmonds, A.~R.: 1960,
\newblock {\em {Angular momentum in quantum mechanics}},
\newblock Princeton University Press, Princeton, New Jersey, 2 edition,
\newblock
  {\small[\href{http://adsabs.harvard.edu/abs/1960amqm.book.....E}{URL}]}

\bibitem[\protect\astroncite{Fan}{2009}]{Fan2009}
Fan, Y.: 2009,
\newblock {\em Living Reviews in Solar Physics} {\bf 6}, 4,
\newblock
  {\small[\href{http://adsabs.harvard.edu/abs/2009LRSP....6....4F}{URL}]},
\newblock {\small[\href{http://dx.doi.org/10.12942/lrsp-2009-4}{DOI}]}

\bibitem[\protect\astroncite{Garc{\'{i}}a et~al.}{2010}]{Garcia2010}
Garc{\'{i}}a, R.~A., Mathur, S., Salabert, D., Ballot, J., R{\'{e}}gulo, C.,
  Metcalfe, T.~S., and Baglin, A.: 2010,
\newblock {\em Science} {\bf 329(5995)}, 1032,
\newblock
  {\small[\href{http://adsabs.harvard.edu/abs/2010Sci...329.1032G}{URL}]},
\newblock {\small[\href{http://dx.doi.org/10.1126/science.1191064}{DOI}]}

\bibitem[\protect\astroncite{Gough and Thompson}{1990}]{Gough1990}
Gough, D.~O. and Thompson, M.~J.: 1990,
\newblock {\em Monthly Notices of the Royal Astronomical Society} {\bf 242(1)},
  25,
\newblock
  {\small[\href{http://adsabs.harvard.edu/abs/1990MNRAS.242...25G}{URL}]},
\newblock {\small[\href{http://dx.doi.org/10.1093/mnras/242.1.25}{DOI}]}

\bibitem[\protect\astroncite{Hanasoge}{2017}]{Hanasoge2017}
Hanasoge, S.~M.: 2017,
\newblock {\em Monthly Notices of the Royal Astronomical Society} {\bf 470(3)},
  2780,
\newblock
  {\small[\href{http://adsabs.harvard.edu/abs/2017MNRAS.470.2780H}{URL}]},
\newblock {\small[\href{http://dx.doi.org/10.1093/mnras/stx1342}{DOI}]}

\bibitem[\protect\astroncite{Hathaway}{2015}]{Hathaway2015}
Hathaway, D.~H.: 2015,
\newblock {\em Living Reviews in Solar Physics} {\bf 12}, 4,
\newblock
  {\small[\href{http://adsabs.harvard.edu/abs/2015LRSP...12....4H}{URL}]},
\newblock {\small[\href{http://dx.doi.org/10.1007/lrsp-2015-4}{DOI}]}

\bibitem[\protect\astroncite{Herzberg}{2016}]{Herzberg2016}
Herzberg, W.: 2016,
\newblock {\em Ph.D. thesis}, Albert-Ludwigs Universit{\"{a}}t Freiburg,
\newblock {\small[\href{https://freidok.uni-freiburg.de/data/11692}{URL}]},
\newblock {\small[\href{http://dx.doi.org/10.6094/UNIFR/11692}{DOI}]}

\bibitem[\protect\astroncite{Herzberg and Roth}{2018}]{Herzberg2017}
Herzberg, W. and Roth, M.: 2018,
\newblock {\em in prep.}

\bibitem[\protect\astroncite{Jim{\'{e}}nez-Reyes
  et~al.}{1998}]{Jimenez-Reyes1998}
Jim{\'{e}}nez-Reyes, S.~J., R{\'{e}}gulo, C., Pall{\'{e}}, P.~L., and {Roca
  Cortes}, T.: 1998,
\newblock {\em Astronomy \& Astrophysics} {\bf 329}, 1119,
\newblock
  {\small[\href{http://adsabs.harvard.edu/abs/1998A&A...329.1119J}{URL}]}

\bibitem[\protect\astroncite{Kiefer et~al.}{2017a}]{Kiefer2017}
Kiefer, R., Schad, A., Davies, G., and Roth, M.: 2017a,
\newblock {\em Astronomy \& Astrophysics} {\bf 598}, A77,
\newblock
  {\small[\href{http://adsabs.harvard.edu/abs/2017A&A...598A..77K}{URL}]},
\newblock {\small[\href{http://dx.doi.org/10.1051/0004-6361/201628469}{DOI}]}

\bibitem[\protect\astroncite{Kiefer et~al.}{2017b}]{Kiefer2017b}
Kiefer, R., Schad, A., and Roth, M.: 2017b,
\newblock {\em The Astrophysical Journal} {\bf 846(2)}, 162,
\newblock
  {\small[\href{http://adsabs.harvard.edu/abs/2017ApJ...846..162K}{URL}]},
\newblock {\small[\href{http://dx.doi.org/10.3847/1538-4357/aa8634}{DOI}]}

\bibitem[\protect\astroncite{Lavely and Ritzwoller}{1992}]{Lavely1992}
Lavely, E.~M. and Ritzwoller, M.~H.: 1992,
\newblock {\em Philosophical Transactions of the Royal Society of London A:
  Mathematical, Physical and Engineering Sciences} {\bf 339(1655)}, 431,
\newblock
  {\small[\href{http://adsabs.harvard.edu/abs/1992RSPTA.339..431L}{URL}]},
\newblock {\small[\href{http://dx.doi.org/10.1098/rsta.1992.0048}{DOI}]}

\bibitem[\protect\astroncite{Libbrecht and Woodard}{1990}]{Libbrecht1990}
Libbrecht, K.~G. and Woodard, M.~F.: 1990,
\newblock {\em Nature} {\bf 345(6278)}, 779,
\newblock
  {\small[\href{http://adsabs.harvard.edu/abs/1990Natur.345..779L}{URL}]},
\newblock {\small[\href{http://dx.doi.org/10.1038/345779a0}{DOI}]}

\bibitem[\protect\astroncite{Mathis and Zahn}{2004}]{Mathis2004}
Mathis, S. and Zahn, J.-P.: 2004,
\newblock {\em Astronomy \& Astrophysics} {\bf 425(1)}, 229,
\newblock
  {\small[\href{http://adsabs.harvard.edu/abs/2004A&A...425..229M}{URL}]},
\newblock {\small[\href{http://dx.doi.org/10.1051/0004-6361:20040278}{DOI}]}

\bibitem[\protect\astroncite{Mathis and Zahn}{2005}]{Mathis2005}
Mathis, S. and Zahn, J.-P.: 2005,
\newblock {\em Astronomy \& Astrophysics} {\bf 440(2)}, 653,
\newblock
  {\small[\href{http://adsabs.harvard.edu/abs/2005A&A...440..653M}{URL}]},
\newblock {\small[\href{http://dx.doi.org/10.1051/0004-6361:20052640}{DOI}]}

\bibitem[\protect\astroncite{Mestel and Moss}{1977}]{Mestel1977}
Mestel, L. and Moss, D.~L.: 1977,
\newblock {\em Monthly Notices of the Royal Astronomical Society} {\bf 178(1)},
  27,
\newblock
  {\small[\href{http://adsabs.harvard.edu/abs/1977MNRAS.178...27M}{URL}]},
\newblock {\small[\href{http://dx.doi.org/10.1093/mnras/178.1.27}{DOI}]}

\bibitem[\protect\astroncite{Miesch and Teweldebirhan}{2016}]{Miesch2016}
Miesch, M. and Teweldebirhan, K.: 2016,
\newblock {\em Advances in Space Research} {\bf 58(8)}, 1571,
\newblock
  {\small[\href{http://adsabs.harvard.edu/abs/2016AdSpR..58.1571M}{URL}]},
\newblock {\small[\href{http://dx.doi.org/10.1016/j.asr.2016.02.018}{DOI}]}

\bibitem[\protect\astroncite{Regge}{1958}]{Regge1958}
Regge, T.: 1958,
\newblock {\em Il Nuovo Cimento} {\bf 10(3)}, 544,
\newblock {\small[\href{http://dx.doi.org/10.1007/BF02859841}{DOI}]}

\bibitem[\protect\astroncite{Roth}{2002}]{Roth2002}
Roth, M.: 2002,
\newblock {\em Ph.D. thesis}, Albert-Ludwigs-Universit{\"{a}}t Freiburg,
\newblock {\small[\href{https://freidok.uni-freiburg.de/data/512}{URL}]}

\bibitem[\protect\astroncite{Sakurai and Napolitano}{2014}]{Sakurai2014}
Sakurai, J.~J. and Napolitano, J.~J.: 2014,
\newblock {\em {Modern Quantum Mechanics}},
\newblock Pearson Education Limited, Harlow, UK, 2 edition

\bibitem[\protect\astroncite{Salabert et~al.}{2016}]{Salabert2016a}
Salabert, D., R{\'{e}}gulo, C., Garc{\'{i}}a, R.~A., Beck, P.~G., Ballot, J.,
  Creevey, O.~L., {P{\'{e}}rez Hern{\'{a}}ndez}, F., {do Nascimento Jr.},
  J.-D., Corsaro, E., Egeland, R., Mathur, S., Metcalfe, T.~S., Bigot, L.,
  Ceillier, T., and Pall{\'{e}}, P.~L.: 2016,
\newblock {\em Astronomy \& Astrophysics} {\bf 589}, A118,
\newblock
  {\small[\href{http://adsabs.harvard.edu/abs/2016A&A...589A.118S}{URL}]},
\newblock {\small[\href{http://dx.doi.org/10.1051/0004-6361/201527978}{DOI}]}

\bibitem[\protect\astroncite{Schad}{2013}]{Schad2013}
Schad, A.: 2013,
\newblock {\em Ph.D. thesis}, Albert-Ludwigs-Universit{\"{a}}t Freiburg

\bibitem[\protect\astroncite{Schad et~al.}{2011}]{Schad2011}
Schad, A., Timmer, J., and Roth, M.: 2011,
\newblock {\em The Astrophysical Journal} {\bf 734(2)}, 97,
\newblock
  {\small[\href{http://adsabs.harvard.edu/abs/2011ApJ...734...97S}{URL}]},
\newblock {\small[\href{http://dx.doi.org/10.1088/0004-637X/734/2/97}{DOI}]}

\bibitem[\protect\astroncite{Schad et~al.}{2013}]{Schad2013a}
Schad, A., Timmer, J., and Roth, M.: 2013,
\newblock {\em The Astrophysical Journal Letters} {\bf 778(2)}, L38,
\newblock
  {\small[\href{http://adsabs.harvard.edu/abs/2013ApJ...778L..38S}{URL}]},
\newblock {\small[\href{http://dx.doi.org/10.1088/2041-8205/778/2/L38}{DOI}]}

\bibitem[\protect\astroncite{Sweet}{1950}]{Sweet1950}
Sweet, P.~A.: 1950,
\newblock {\em Monthly Notices of the Royal Astronomical Society} {\bf 110(6)},
  548,
\newblock
  {\small[\href{http://adsabs.harvard.edu/abs/1950MNRAS.110..548S}{URL}]},
\newblock {\small[\href{http://dx.doi.org/10.1093/mnras/110.6.548}{DOI}]}

\bibitem[\protect\astroncite{Woodard and Noyes}{1985}]{Woodard1985}
Woodard, M.~F. and Noyes, R.~W.: 1985,
\newblock {\em Nature} {\bf 318(6045)}, 449,
\newblock
  {\small[\href{http://adsabs.harvard.edu/abs/1985Natur.318..449W}{URL}]},
\newblock {\small[\href{http://dx.doi.org/10.1038/318449a0}{DOI}]}

\bibitem[\protect\astroncite{Woodhouse}{1980}]{Woodhouse1980}
Woodhouse, J.~H.: 1980,
\newblock {\em Geophysical Journal International} {\bf 61(2)}, 261,
\newblock
  {\small[\href{http://adsabs.harvard.edu/abs/1980GeoJ...61..261W}{URL}]},
\newblock
  {\small[\href{http://dx.doi.org/10.1111/j.1365-246X.1980.tb04317.x}{DOI}]}

\bibitem[\protect\astroncite{Woodhouse and Dahlen}{1978}]{Woodhouse1978}
Woodhouse, J.~H. and Dahlen, F.~A.: 1978,
\newblock {\em Geophysical Journal International} {\bf 53(2)}, 335,
\newblock
  {\small[\href{http://adsabs.harvard.edu/abs/1978GeoJ...53..335W}{URL}]},
\newblock
  {\small[\href{http://dx.doi.org/10.1111/j.1365-246X.1978.tb03746.x}{DOI}]}

\end{thebibliography}

\end{document}